\documentstyle[aps,epsfig,psfig,citesort]{revtex}

\begin{document}

\def \lesssim{\leavevmode{\raisebox{-.5ex}{ $\stackrel{<}{\sim}$ } } }
\def \gtrsim{ \leavevmode{\raisebox{-.5ex}{ $\stackrel{>}{\sim}$ } } }
\def \subarr{  \begin{array}{c} \mbox{\scriptsize{$ \{ n_{\ell}\}$ }} \\ 
        \mbox{\scriptsize{ $\{m_{\ell}\}$}}             \\ 
        \mbox{\scriptsize{ $\{p_{\ell}\}$ }}  \end{array} }
\def \subun{ \leavevmode{\raisebox{-1.6ex}{$\stackrel{
	\textstyle{\ell_i,\ell_j}}
	{\mbox{\scriptsize{unconstrained}}} $}} }
\def \maxj{\begin{array}{c} \mbox{max} \\ 
        \mbox{\scriptsize{ $j$}}             \\ 
        \mbox{\scriptsize{ $ 0 \leq j \leq \jmax $ }}  \end{array} }
\def \minF{\begin{array}{c} \mbox{min} \\ 
        \mbox{\scriptsize{ $F$}}  \end{array} }
\def \minNfo{\begin{array}{c} \mbox{min} \\ 
        \mbox{\scriptsize{ $\Nfo$}}  \end{array} }
\def \nl{\textstyle{\mbox{\scriptsize{$\{n_{\ell}\}$}}} }
\def \ml{\mbox{\scriptsize{$\{m_{\ell}\}$}} }
\def \pl{\mbox{\scriptsize{$\{p_{\ell}\}$}} }
\def \be{\begin{equation}}
\def \ee{\end{equation}}
\def \bea{\begin{eqnarray}}
\def \eea{\end{eqnarray}}
\def \zn{z_{\mbox{\tiny N}}}
\def \jmax{j_{\mbox{\tiny MAX}}}
\def \ei{\epsilon_i}
\def \ej{\epsilon_j}
\def \Qj{Q_j}
\def \Qk{Q_k}
\def \Qi{Q_i}
\def \qi{q_i}
\def \Qu{Q_{\mbox{\tiny U}}}
\def \Qf{Q_{\mbox{\tiny F}}}
\def \Qsurf{Q_{\mbox{\tiny SURF}}}
\def \eio{\epsilon_{i o}}
\def \ejo{\epsilon_{j o}}
\def \eabn{\epsilon_{\alpha\beta}^{\mbox{\tiny N}}}
\def \eab{\epsilon_{\alpha\beta}}
\def \Dab{\Delta_{\alpha\beta}}
\def \Dabn{\Delta_{\alpha\beta}^{\mbox{\tiny N}}}
\def \setDab{\{ \Delta_{\alpha\beta} \} }
\def \setDabn{\{ \Delta_{\alpha\beta}^{\mbox{\tiny N}} \} }
\def \li{\ell_i}
\def \leff{\ell_{\mbox{\tiny EFF}}}
\def \lj{\ell_j}
\def \si{s_i}
\def \sj{s_j}
\def \sm{s_m}
\def \be{\begin{equation}}
\def \ee{\end{equation}}
\def \bea{\begin{eqnarray}}
\def \eea{\end{eqnarray}}
\def \aQ{\alpha_{\mbox{\tiny Q}}}
\def \muQ{\mu_{\mbox{\tiny Q}}}
\def \lQ{\lambda_{\mbox{\tiny Q}}}
\def \lT{\lambda T}
\def \ldag{\lambda^{\ddag}}
\def \D{\partial}
\def \DF{\Delta F}
\def \DFdag{\Delta F^{\ddag}}
\def \Fdag{F^{\ddag}}
\def \Fbardag{\overline{F}^{\ddag}}
\def \fbar{\overline{f}}
\def \dfi{\delta f_i}
\def \Nfdag{N_{\mbox{\tiny F}}^{\ddag}}
\def \Nfo{N_{\mbox{\tiny F}}^{o}}
\def \Nf{N_{\mbox{\tiny F}}}
\def \nf{n_{\mbox{\tiny F}}}
\def \nfdag{n_{\mbox{\tiny F}}^{\ddag}}
\def \Fbar{\overline{F}}
\def \Fstar{F^{\star}}
\def \dij{\delta_{ij}}
\def \di{\delta_{i}}
\def \dj{\delta_{j}}
\def \d{\delta}
\def \s{\sigma}
\def \Qdag{Q^{\dag}}
\def \Qddag{Q^{\ddag}}
\def \Qo{Q^{o}}
\def \En{E_{\mbox{\tiny N}}}
\def \dli{\delta\ell_i}
\def \dl{\delta\ell}
\def \dlj{\delta\ell_j}
\def \dei{\delta\epsilon_i}
\def \dsi{\delta s_i}
\def \de{\delta\epsilon}
\def \dej{\delta\epsilon_j}
\def \dQi{\delta Q_i}
\def \dQj{\delta Q_j}
\def \dQ{\delta Q}
\def \ds{\delta s}
\def \setei{\left\{ \epsilon_i \right\}}
\def \eistar{\ei^{\star}}
\def \seteistar{\{ \ei^{\star} \} }
\def \seteio{\{ \ei^o \} }
\def \setebar{\{ \ebar \} }
\def \seteir{\{ \ei^{\mbox{\tiny R}} \} }
\def \ejstar{\ej^{\star}}
\def \setsi{\left\{ s_i \right\}}
\def \setej{\left\{ \epsilon_j \right\}}
\def \setli{\left\{ \ell_i \right\}}
\def \setlj{\left\{ \ell_j \right\}}
\def \setQi{\left\{ Q_i \right\}}
\def \setQiQ{\left\{\Qi\left(Q\right)\right\}}
\def \setQiQstar{\left\{\Qi^{\star}\left(Q\right)\right\}}
\def \setQistarQ{\left\{ Q_i^{\star} \left(Q\right)\right\}}
\def \setQistar{\left\{ Q_i^{\star} \right\}}
\def \setQj{\left\{ Q_j \right\}}
\def \setQk{\left\{ Q_k \right\}}
\def \Ebar{\overline{E}}
\def \ebar{\overline{\epsilon}}
\def \eibar{\overline{\epsilon}_i}
\def \libar{\overline{\ell_i}}
\def \ebarN{\overline{\epsilon}_{\mbox{\tiny N}}}
\def \en{\epsilon_{\mbox{\tiny N}}}
\def \lbar{\overline{\ell}}
\def \Tg{T_{\mbox{\tiny G}}}
\def \Tf{T_{\mbox{\tiny F}}}
\def \Td{T_{\mbox{\tiny D}}}
\def \FH{F_{\mbox{\tiny HOMO}}}
\def \kH{k_{\mbox{\tiny HOMO}}}
\def \kF{k_{\mbox{\tiny F}}}
\def \kB{k_{\mbox{\tiny B}}}
\def \wq{\omega_{\mbox{\tiny Q}}}
\def \Qast{Q^{\ast}}
\def \JQ{ {\cal J}^{\star}\left( Q\right)}
\def \cN{c_{\mbox{\tiny N}}}
\def \Ecore{E_{\mbox{\tiny C}}}
\def \Fcore{F_{\mbox{\tiny C}}}
\def \Fbarcore{\overline{F}_{\mbox{\tiny C}}}
\def \Shalo{S_{\mbox{\tiny H}}}
\def \pcore{p_{\mbox{\tiny C}}}
\def \Sroute{S_{\mbox{\tiny ROUTE}}}
\def \Qical{{\cal Q}_i}
\def \Ei{E_i}
\def \spini{\sigma_i}
\def \spinj{\sigma_j}
\def \setspini{\left\{ \sigma_i \right\}}
\def \sistar{\sigma_i^{\star}}
\def \hM{h_{\mbox{\tiny M}}}
\def \tauF{\tau_{\mbox{\tiny F}}}
\def \ai{\alpha_i}
\def \asim{\alpha_{\mbox{\tiny SIM}}}
\def \athry{\alpha_{\mbox{\tiny THRY}}}

\renewcommand{\thesection}{\arabic{section}}
\renewcommand{\thesubsection}{\arabic{section}.\Alph{subsection}}
\renewcommand{\thesubsubsection}{\arabic{section}.\Alph{subsection}.\arabic{subsubsection}}
\renewcommand{\theequation}{\thesection.\arabic{equation}}

\title{\Large\bf 
Structural and energetic heterogeneity in protein folding
}
\author{{\large Steven S.~Plotkin and Jos\'{e} N. Onuchic \\
Department of Physics, University of California, San Diego
}}

\maketitle
\par
\bigskip
\noindent
\begin{abstract}
A general theoretical framework is developed using free energy
functional methods to understand the effects of heterogeneity in
the folding of a well-designed protein. 
Native energetic heterogeneity arising from
non-uniformity in native stability, as well as entropic heterogeneity
intrinsic to the topology of the native structure are both
investigated as to their impact on the folding free energy landscape
and resulting folding mechanism. Given a minimally frustrated
protein, both structural and
energetic heterogeneity lower the
thermodynamic barrier to folding, and designing in sufficient
heterogeneity can eliminate the barrier at the folding transition
temperature. Sequences with different distributions of stability
throughout the protein and correspondingly different folding
mechanisms may still be good folders to the same structure. This
theoretical framework allows for a systematic study of the coupled
effects of energetics and topology in protein folding, and provides
interpretations and predictions for future experiments which may
investigate these effects.
\end{abstract}


\section{Introduction}
\setcounter{equation}{0}

Theories of protein folding currently focus primarily on predicting
properties of the folding mechanism given the native structure and/or
energy function is known {\it a priori}. The most powerful approach to
this end has been the energy landscape theory, used in one form or
another in most descriptions of
folding~\cite{WolynesPG92,BryngelsonJD95,DillKA95:ps,BallKD96,DillKA97,VeitshansT97,Onuchic97,PandeVS97:bj,Dobson98,GarelT:rev98,BrooksCL98:pnas,FershtAR99:book,GruebeleM99,WalesDJ99,OnuchicJN00:apc}.
This approach takes advantage of the huge number of conformational
states available to a protein by treating the energetics of those
conformations statistically, just as the description of a phase
transition from a liquid to a crystal is understood through the
application of statistical mechanics to the numerous degrees of
freedom in the problem.  However in understanding the
self-organization of proteins and biological systems in general, it is
necessary to study properties particular to finite-sized systems,
e.g. barrier heights and corresponding rates.  For a finite system
such as a protein, characteristic features present in the amino acid
sequence give rise to residual signatures in thermodynamic and kinetic
properties. For example, although the four-helix proteins Im7 and Im9
are structural homologs, Im9 folds by a two-state mechanism while Im7
folds through an {\it en route} intermediate~\cite{FergusonN99}: the
free energy landscape may fluctuate sequence to sequence for aa chains
that fold to the same native structure.  Other experiments also
indicate that rates and/or intermediates may differ for structural
homologs~\cite{KimDE98,DalessioPM00}.

Since a knowledge of the native structure does not completely
determine the free energy profile, we might ask what information does,
and also what parameters must be known to predict other properties of
the folding mechanism, such as the specificity or diffusivity of the
folding nucleus, for example~\cite{AbkevichVI94,Onuchic96,Klimov98,LiL00}.
By analyzing the energetic statistics of ensembles of states,
landscape theory provides a framework to distinguish folding processes
common to an ensemble of sequences from those peculiar to individual sequences.
A particular
property, for example folding transition 
temperature $\Tf$, is not strongly dependent on the detailed
Hamiltonian of the protein, but only on 
a few thermodynamic parameters such as overall native stability, chain
stiffness, overall variance in energies (if for example the sequence
were thread through random structures), and perhaps also 
on chain length and overall hydrophobicity which induces generic collapse.  
The folding temperature may be expected to be a universal or
self-averaging property for the ensemble of
sequences having these parameters. Such  quantities are
important since they are amenable to analysis: the transition
temperature may be expressed quantifiably through the parameters
mentioned above as 
\be
\Tf = \frac{z |\ebar|}{2 s_o} \left( 1 + \sqrt{1-\frac{2 s_o b^2}{z
\ebar^2} } \right)
\label{eqtf}
\ee
where $\ebar$ is the average native stability per residue ($\ebar <0$),
$s_o$ is the 
entropy gain per residue in unfolding which depends on chain
stiffness and also net hydrophobicity, 
$b$ is the non-native energetic variance per residue, and
$z$ is the average number of neighbors per residue which is weakly a
function of the chain length $N$. Equation~(\ref{eqtf}) further
gives the folding temperature for new sequences if these parameters
can be determined, or conversely eq.~(\ref{eqtf}) may help to determine
intrinsic protein parameters from a measurement of the transition
temperature. Equations for self-averaging properties provide a simple 
framework for  understanding the phenomena; for example if the
variance is weak compared to the 
stability (the protein is well-designed) then the folding temperature
is to the first approximation proportional to the stability gap over
the entropy of unfolding: $\Tf \approx -\En/S_o$. So for example a
well-designed stiff 
protein has a higher transition temperature than a flexible one, since
it has less conformations per residue in the unfolded state. 
Moreover, by eq.~(\ref{eqtf}) $\Tf$ is always below $z |\ebar|/s_o$
when non-native strength $b=0$,
so non-native interactions always lower the folding temperature by
stabilizing the unfolded state. As $b$ is increases, $\Tf$ decreases
but remains real so long as~\cite{GoldsteinRA-AMH-92}
\be 
\sqrt{\frac{z}{2 s_o}} \frac{|\ebar|}{b} \geq 1 \: ,
\label{tftgeq}
\ee
the equality in~(\ref{tftgeq}) holding when $\Tf$ reaches its minimum
value of $\sqrt{z/2 s_o} \, b$ which is precisely the
glass temperature $\Tg$, defined as the
temperature where the entropy of a random aa sequence is no longer
extensive, i.e. where a random system becomes trapped in one of a
small number of low energy states. Because
$\Tf$ is found by equating the total free energy of the folded and
unfolded  states, it
should be only weakly sensitive to the details of the actual
distribution of native 
stability within the protein. On the other hand, properties such as
the folding barrier and its corresponding rate may depend strongly on
the distribution of native stability throughout the protein.

To theoretically treat the thermodynamics of folding and unfolding, we
quantify a model protein in terms of the full native Hamiltonian
$\setei$, as well as the full distribution of native contact lengths
$\setli$, under the assumption that the protein under consideration is
well-designed with $\Tf$ reasonably larger than $\Tg$, i.e. the
inequality~(\ref{tftgeq}) is easily satisfied. Native heterogeneity is
retained explicitly, while non-native interactions are treated through
an average background field- the scalar quantity $b$ in
eq.s~(\ref{eqtf}) and~(\ref{tftgeq}).  We are thus isolating the
effects of native heterogeneity on the folding mechanism.

Certainly if the entropy around the transition state were small, as
when the inequality~(\ref{tftgeq}) is not well-satisfied, the position
and height of the rate determining barrier(s) would fluctuate wildly
sequence to sequence.  However, since proteins have evolved native
stabilities larger than their RMS non-native energy
scale~\cite{BryngelsonJD87,GoldsteinRA-AMH-92,LeopoldPE92,ShakhnovichEI93a,OnuchicJN95:pnas,Bornberg99:pnas,Buchler99:jcp}
as manifested in inequality~(\ref{tftgeq}), the temperature $\Tf$
where the native state is stable is sufficiently higher than $\Tg$,
such that there is an extensive amount of residual entropy left under
folding conditions. Nevertheless, we find that even though there is a
large entropy present in the system, there are still in fact strong
dependencies of the barrier and folding mechanism on the distribution
of native stability and distribution of native contact lengths. In
fact local fluctuations in native stability and structure do not
average out, but contribute extensively in determining properties of
the folding mechanism!

For a property which is not self-averaging over a given ensemble of
sequences, the parameters specified to determine the ensemble are
either not sufficiently accurate or are incomplete.  
For example, the folding transition
temperature $\Tf$ is not self-averaging over the ensemble of sequences
that fold to a particular native structure, since these different
sequences may have different native stability, flexibility, etc. 
Nonetheless, a quantity such as folding temperature or folding barrier
which fluctuates over an incompletely specified ensemble may have a
mean that is still useful in characterizing trends as a function
of one or more parameters. An example  is the
increase on average in folding rate, or decrease in folding barrier,
as mean native contact length $\lbar$ (more specifically $\lbar/N$) is 
decreased~\cite{Plaxco98}, for which several models have been
proposed~\cite{MunozV99,FershtAR00},
and which we consider here within our theoretical framework as well
(see section~\ref{sect:mean}). The observed correlation between rates
and $\lbar$ implies that many proteins are sufficiently well-designed
such that native topology plays an important if not dominant role in
governing folding mechanism, a topic recently investigated by several
authors~\cite{AlmE99,MunozV99,ShoemakerWang99,FinkelsteinAV99:pnas,SheaJE99,RiddleDS99,DuR99:jcp,FershtAR00,ClementiC00:jmb,ClementiC00:pnas,TavernaDM00,MaritanA00}.

Our intention here is to go beyond the first moment of the contact
length distribution $\lbar$ or stability distribution $\ebar$. We
investigate how the full distributions of energetics and topology as
well as correlations between them affects the free energy profile,
corresponding barrier, folding rate, and overall folding
mechanism. For one example, we find in section~\ref{sect:structvar}
that there is a net correlation between structural variance defined
through the second moment of the contact length distribution and
folding rate for well-designed proteins with a given mean contact
length $\lbar$.  Expanding on our previous
work~\cite{PlotkinSS00:pnas}, we find that native heterogeneity, both
entropic and energetic, plays an important role in quantifying protein
folding mechanisms.  We show how one can extend the analysis of
thermodynamic quantities by using functionals to describe folding
properties which are not necessarily self-averaging but which may
depend on distributions of coupling parameters.  To this end we derive
a simple field theory with a nonuniform order parameter to study
fluctuations away from uniform ordering, through free energy
functional methods introduced earlier by Wolynes and
collaborators~\cite{ShoemakerBA97,ShoemakerBA99,ShoemakerWang99}. The
theory is in good agreement with lattice simulations also performed in
this paper. Similar effects have also been observed in Monte Carlo
simulations of sequence evolution for lattice protein models, when the
selection criteria involves fast folding
rate~\cite{GutinAM95:pnas}. Here we see how, from general
considerations of the energy landscape theory, selecting for rate can
induce heterogeneity in the transition state ensemble. The folding
barrier for a well-designed protein is maximized when the nucleus is
the most diffuse.  For typical values of native energies,
well-designed proteins have heterogeneous funneled folding
mechanisms~\cite{Radford92,BaiY95,BoczkoEM95,Lazaridis97,Brookspnas98}.

Our results are also supported by several
experiments in the literature as described in the conclusions section,
and suggest experiments to be performed.  For example the reduction of
barrier height with folding heterogeneity should be experimentally
testable by measuring the dependence of folding rate for a
well-designed protein on the dispersion of
$\phi$-values~\cite{FershtAR92}. It is important that before and after the
mutation(s) the protein remains fast-folding to the native structure
without ``off-pathway'' intermediates, and that its native state
stability remain approximately the same, perhaps by tuning
environmental variables.

In the arguments below we associate reductions in the
free energy barrier 
$\DFdag$ to increases in the folding rate $\kF$. 
This is true as long as the 
prefactor  $k_o$ in the expression for the rate
\be
\kF = k_o \mbox{e}^{-\DFdag/T}
\label{eq:k1}
\ee
is more weakly affected than the barrier height under redistribution
of native stability. 
While the distribution of native stability must indeed couple with the
specific distribution of non-native interactions, for well-designed
proteins with large transition-state entropy, it is more likely that
the effect on the prefactor comes from the coupling of the transition
temperature $\Tf$ to the distribution of native stability, as long as
the protein still folds to the same native structure. In other words
we must consider the effect on the prefactor as the ratio of the transition
temperature to glass temperature $\Tf/\Tg$ changes, i.e. as the left
hand side of~(\ref{tftgeq}) changes and the 
protein becomes more or less well-designed.
However we find
(see fig.s~\ref{fig:sdplot} and~\ref{fig:TTFF}) that 
even for perturbations in native energy large enough to kill the
barrier, the folding
temperature varies only weakly compared to the changes in barrier
height, so that the ratio 
$\Tf/\Tg$ should also not vary significantly compared to the changes in
barrier height.  Thus the barrier height is the strongest determinant
of folding rate in well-designed proteins.
For G\={o}-like models with native interactions alone, the energy distribution
does not strongly affect the prefactor: rates obtained  from
simulations are well fit with those predicted from the change in
barrier height alone (see fig.~\ref{fig:tau}), i.e. the energy distribution
does not strongly affect the reconfiguration kinetics appearing in the
prefactor, compared to its effect on the barrier height.
For a full discussion of the above effects see
section~\ref{sect:kinetics}.

When the Hamiltonians consists of pair interactions alone,
redistributions of native stability can eliminate the barrier entirely
at the folding temperature, see figures~\ref{fig:Fplot}
and~\ref{fig:FvsR}.  It is worth noting that many-body interactions
which are believed to be present in real protein
interactions~\cite{HorovitzA92,HaoM97,SorensonJM98,Klimov98:fd,LumK99,TakadaS99:jcp,PrinceRB99}
tend to increase the barrier
height~\cite{KolinskiA96:prot,PlotkinSS97,EastwoodMP00}, and in their
presence the barrier may be more robust to redispersement of native
stability.

A funnel folding mechanism consisting of a large number of routes to
the native structure is preserved for a wide variety of folding
scenarios and barrier heights, including folding through on-pathway
intermediates. For the distributions of native energy
necessary to induce folding 
through one or a few routes, the folding temperature drops by about a
factor of six, which indicates that for realistic energy functions
which are also composed of non-native interactions, folding would be
exceedingly slow at the low temperatures where the native state would
be stable. However for proteins which are large and multi-domain, it
is possible that entropic or energetic heterogeneity may induce
significantly route-like folding near the transition temperature.

The paper is organized as follows. First in
section~\ref{sect:strat} we outline the
strategy of the calculation and illustrate it with a simple example in
section~\ref{sect:simple}. Next we anticipate and explain several of
the results with
physical arguments  in section~\ref{sec:phys}; its 
reading is not essential for the rest of the paper, but should be very
helpful to the reader. In section~\ref{sect:f} the full expression for
the free energy functional is derived, and the functional is implemented in
section~\ref{section:opt} where comparison is made to the results from
lattice simulations. Finally we conclude and suggest future research.

\section{The General Strategy}
\setcounter{equation}{0}
\label{sect:strat}

It is first necessary to
characterize the generic properties of the native state.
We adopt a coarse grained description for the native
structure, and describe it by its distributions of native contact
energies $\setei$ and 
native loop lengths or contact lengths $\setli$ (see figure~\ref{fig1}).
Here $\ei$ is the solvent averaged effective energy of contact $i$, and
$\li$ is the sequence length pinched off by contact $i$ (see
fig.~\ref{fig1})~\cite{note:backtop}.
We use a single subscript for the labeling index $i$
because we are only considering
effects on the particular set of native contacts for a given native
structure. Non-native interactions are treated by an average field,
since the protein is assumed to be well-designed to its native
structure, and native interactions are then most important in
determining the folding mechanism. 
The index $i$ runs from  $1$ to $M$, where $M$ is the
number of residue pair contacts  in the lowest energy native
structure. $M$ scales roughly extensively, i.e. $M = z N$, with $N$ the number
of residues  in the polymer chain. Here $z$ is the mean number of contacts
per residue:  a function of either the lattice
coordination number or the off-lattice cut-off length. It is of order
$1$, with surface area corrections dying away as
$N^{-1/3}$~\cite{Douglas95}. 
We can quantify nativeness in  the first approximation by the fraction
of native contacts $Q$, with $0 < Q < 1$. Other parameters 
are also reasonable for stratifying the landscape:
the fraction of correct (native) dihedral angles~\cite{OnuchicJN95:pnas},
coarse grained position in 
space in the native structure~\cite{BryngelsonJD89,PortmanJprl98}, or
even the ensembles having 
a given probability to fold before unfolding~\cite{DuR98:jcp}.
But $Q$ is the most suitable for calculation in the present theory.
At partial degrees of nativeness the  probability to form
contact $i$ is  defined as $\Qi (Q)$, and we define $\Qi^{\star} (Q)$
as the fraction 
of time contact $i$ is formed at equilibrium in the ensemble with $M
Q$ native contacts, or equivalently the fraction of proteins in a
macroscopic sample with a given degree of nativeness having that
contact formed.
Non-uniformity in $\Qi^{\star}$ values at partial degrees of nativeness would
indicate that the protein prefers to fold some regions over others.

Following the formalism used in inhomogeneous fluids~\cite{PercusJK82,EvansR92}
and the theory of first order transitions~\cite{Gunton83}
we write a free energy functional $F(\setQi, \setei , \setli)$ to
characterize 
the effects of structural and energetic heterogeneity superimposed on
the overall 
folding funnel. This approach has been used earlier by Bohr and Wolynes 
to describe domain growth
in proteins~\cite{BohrHG92} and more recently as a calculational
tool for experimental 
$\phi$-values~\cite{ShoemakerBA97,ShoemakerBA99,ShoemakerWang99}.

The free energy functional is first interpreted as depending upon the
local contact probabilities $Q_i (Q) = \left< \Theta(r_i -
r_i^{\mbox{\tiny N}} ) \right>_{\mbox{\tiny T}} (Q)$ 
where $i$ labels the native contact 
between two residues,  $r_i$ the distance between them, 
$\Theta$ is a function that measures proximity such as a step
function for off-lattice models or a Kronecker delta for non-bonded
nearest neighbor sites on-lattice, and $\left< \cdots \right>_{\mbox{\tiny T}}$
indicates the an average over the ensemble at $Q$. We will typically
take  $\left< \cdots \right>_{\mbox{\tiny T}}$ to be a Boltzmann
weighted average; then $\Qi^{\star}$
is the thermally averaged fraction of the time two parts of the
protein are in proximity (contact)~\cite{note:kb}:
\be
Q_i^{\star} (Q) = \left< \delta_i \right>_{\mbox{\tiny T}} =
\sum_{c \in Q}^{'}  \delta(i,c) \frac{\exp(-E_c/T)}{Z}
\label{Qdelta}
\ee
where $\delta(i,c) = 1$ if contact $i$ is made in configuration $c$,
and $\delta(i,c) = 0$ otherwise.
The sum may be taken over any ensemble of theoretical interest. Here
we have in mind the ensemble defined as having a given degree of
overall order $Q = (1/M) \sum_i \Qi$~\cite{note:avg}.

In the functional method, 
all the contact energies $\setei$ and loop lengths $\setli$ for
a protein are initially assumed as given, and the thermal (most probable) 
distribution of contact probabilities $\left\{ Q_i^{\star} \left(\setei,
\setli, Q \right) \right\}$
is found by minimizing the free energy functional $F(\setQiQ | \setei,
\setli )$ with respect to the distribution of occupation
probabilities, subject to the constraint that the average probability is $Q$,
i.e. $\sum_i \Qi = M Q$ ($Q$ then parameterizes the values of the $\Qi's$).
Examples of the functions $\Qi^{\star}(Q)$ are plotted in
figures~\ref{fig:QiQ} and~\ref{fig:QivsQroute}. 
This procedure is analogous to finding the most probable distribution of
occupation numbers, and thus the thermodynamics, by maximizing the
microcanonical entropy for a system of particles obeying a given
occupation statistics. Here the 
effective particles (the contacts) obey Fermi-Dirac statistics,
(see eq.~(\ref{Qstar})), since no more than one bond can ``occupy'' a
contact.
The system can be understood to have a set of free energy levels
obeying a distribution governed by the native structure and energies
of the protein, and we 
seek the fraction of time (the probability) those levels are occupied
given a fixed overall number of levels filled.

The free energy  for a
system obeying the thermal (most probable) distribution $\left\{ Q_i^\star
\left(Q, \setei,\setli\right) \right\}$
is then considered a
function of the contact energies for a {\it fixed} native
structure: $F(Q, \setei | \setli)$. That is, we consider the folding free
energy barrier as a functional of the interaction energies $\setei$ 
for a {\em given} native topology. The free energy depends on the
energies $\setei$ both explicitly and implicitly through the thermal
contact probabilities $ \left\{ Q_i^\star\left(\setei | Q,
\setli\right) \right\}$. 
Then we can seek the special distribution of contact
energies $\{ \ei^\star (\li ) \}$ that extremizes (minimizes or
maximizes, depending on the second derivative) the
thermodynamic folding barrier to a particular structure by
finding the extremum of $F^\dag(\setei | \setli)$ with
respect to the contact energies $\ei$, subject to the
constraint of fixed total native energy:
$\sum_i \ei = M\ebar=\En $, i.e. while
maintaining the same overall stability of the native structure.
Thus we are isolating the effect of heterogeneity on the folding mechanism.
This distribution when substituted into the free energy 
gives in principle the extremum free energy barrier as a function of native
structure $F^\dag (\setli )$, which might then be optimized for the
fastest/slowest folding structure and its corresponding barrier.
However we found that in fact the only distribution of energies for
which the 
free energy was an extremum is in fact the distribution which {\it
maximizes} the barrier by tuning all the contact probabilities to the
same value: $\Qi(\Qdag) = \Qdag$. In this case the coupling energies
would be tuned to eliminate 
any information contained in the native structure, except for the mean
loop length $\lbar = (1/M)\sum_i \li$. Any perturbation away from
this scenario lowers the free energy barrier.
We also examine the effects of structural dispersion on the barrier,
i.e. a free energy for variable loop distribution but fixed coupling
energies $F(Q,\setli | \setei )$,
and arrive at the same conclusion: for fixed energies, increasing
structural variance (at fixed average loop length) lowers the barrier
and thus speeds the rate, as 
long as the protein is sufficiently well-designed that the rate is
governed by the free energy barrier.

\subsection{An example}
\label{sect:simple}

We illustrate the procedure by applying it to a more trivial system- an
Ising paramagnet in a non-uniform external field.
The Hamiltonian for this system is
\be
{\cal H} = - \sum_{i=1}^N \ei \spini \: ,
\ee
where $\spini = \pm 1$ is the $i$th spin and $\ei$ is its local field energy. 
To obtain the free energy functional we need an expression for the
entropy in terms of the spin degrees of freedom. If the
field was uniform, the entropy per spin $s(q)$ could be written in terms the
fraction of up spins $N_{+}/N \equiv q$ as
\be
s(q) = \frac{S(q)}{N} = \frac{1}{N} \ln \frac{N!}{(Nq)! \,  [N(1-q)]!}
\cong  \left[-q\ln q -(1-q) \ln (1-q) \right] \: .
\ee
Here $q = (1+\overline{\sigma})/2$ where $\overline{\sigma} =
(1/N)\sum_i \spini$ is the average magnetization per site.
However if the field varies from site to site so will the equilibrium
value of the spin. To allow for this the entropy per spin must be
written in terms of $\qi = (1 + \spini)/2$, and the total entropy is
then a functional  $S(\{ \qi \})$. The free energy functional is then
\be
F\left( \setspini , \setei \right) = \sum_{i=1}^{N} \left[
- \ei \spini + T \left( \frac{1+\spini}{2} \ln \frac{1+\spini}{2} +
\frac{1-\spini}{2} \ln \frac{1-\spini}{2} \right) \right] \: .
\label{Fpara}
\ee
The equilibrium values of the spins $\sistar$ are obtained by finding
the extremum of the free energy, perhaps subject to a fixed overall
magnetization~\cite{note:mag} $M = \sum_i \spini$:
\be
\frac{\d}{\d \spini} \left( F + \hM \sum_j \spinj \right) = 0 \: .
\label{eq:dfdspin}
\ee
This leads to the equation 
\be
\sistar = \tanh \left( \frac{\ei + \hM}{T} \right)
\label{eq:sistar}
\ee
for the equilibrium values of the spins. Each spin follows its local
field according to the well-known Brillion function (for spin $1/2$). 
The Lagrange
multiplier $\hM$ is determined from the sum $\sum_i \sistar = M$.
For a uniform field $\epsilon$, $\hM = T \tanh^{-1} (M/N) -
\epsilon$.
The second variation of $F$ is positive indicating the free energy is
minimized and $\sistar$ are the thermal equilibrium values:
\be
\left.\frac{\d^2 F}{\d \spini \d \spinj}\right|_{ \{\sistar \}} = \dij
\frac{T}{1-\sigma_i^{\star \, 2}} > 0 \: .
\ee
Substituting $\sistar(\ei)$ (eq.~(\ref{eq:sistar})) back into the free energy
functional~(\ref{Fpara}) gives the free energy in terms of the
coupling energies $\{ \ei \}$:
\be
\frac{F\left(\setei \right)}{T} = - N \ln 2 - \sum_{i=1}^{N} \ln
\left[ \cosh \left( \frac{\ei + \hM}{T}\right) \right] + \frac{\hM
M}{T} \: .
\ee
This is equivalent to the form obtained from the partition function in
the canonical ensemble.

Now we can seek the set of fields $\eistar$ that extremizes the free
energy subject to a given total coupling energy $E = \sum_i \ei$:
\be
\frac{\d}{\d \ei} \left( F + p \sum_j \ej \right) = 0 \: .
\ee
This yields the condition that all the spins have the same value and
thus that the field be uniform:
\bea
\spini (\eistar) &=& p  \\
\eistar &=& \epsilon
\eea
However, second variation of the free energy gives
\be
\left.\frac{\d^2 F}{\d \ej \d \ei}\right|_{\{\ei \} = \{ \epsilon \} }
= - \dij \frac{1}{T} \,{\mbox{sech}}^2 \left( \frac{\epsilon +
\hM}{T}\right)
\label{2ndising}
\ee
which is negative, indicating this choice of coupling energies
{\it maximizes} the free energy. Thus any perturbations away from the 
uniform field will lower the free energy.
Although the entropy functional is much more complicated for proteins,
we find that there too the only free energy extremum is a maximum. 

\section{Physical Arguments for the effects}
\setcounter{equation}{0}
\label{sec:phys}

Before we present the full free energy functional theory, 
we include here some physically motivated arguments for the effects
to give the reader a
deeper intuitive feel for the results derived later within the more
general framework. This section is fairly 
independent of the rest of the text; it is broken up into subsections
which may be skipped or read in any order.
The first subsection concerns the effect on rates by changing the
interaction energies of contacts that are likely or unlikely to begin
with. The second subsection consists of various proofs describing how
heterogeneity lowers the barrier and the consequences of this
phenomenon, and in the third subsection we show a simple proof that a
protein with stability distributed uniformly has maximal
conformational entropy.

\subsection{Making early-forming 
native contacts relatively stronger will tend to speed 
folding more than making late-forming native contacts relatively
stronger stronger by the same
amount}
\label{sec:likely} 
The argument proceeds as follows~\cite{Nymeyer:priv}.
First notice that
\be
\frac{\D F}{\D \ei} = \frac{\D}{\D\ei} \left( -T \ln Z\right) =
\frac{1}{Z} \sum'_c \frac{\D E_c}{\D\ei} \mbox{e}^{-E_c/T} \: ,
\ee
where the sum is over all the states with a given similarity $Q$ to
the native. Since the energy of conformation $c$ is a sum of its contact
energies: $E_c = \sum_{j\in c} \ej$, $\D E_c/\D\ei = \delta(i,c)$ and
thus by eq.~(\ref{Qdelta})
\be
\frac{\D F}{\D \ei} = \Qi \: .
\label{dqde}
\ee
This result, derived below within the functional formalism
(c.f. eq.~(\ref{eqphi})), means that the change in free energy $\delta
F$ in 
perturbing a contact $i$ an amount $\delta \ei$ is equal to the amount
of that perturbation times the fraction of time that contact is formed.
An analogous equation holds for an inhomogeneous fluid, where the
density of fluid at position ${\bf x}$, $n({\bf x})$, plays the role
of contact probability and the external field at ${\bf x}$, $u({\bf
x})$, plays the role of the perturbation: $\delta F/\delta u({\bf x})
= n({\bf x})$.

Now imagine taking two contacts $i=1,2$ within the protein having
formation probabilities $Q_1, Q_2$, and making equal and opposite
energetic perturbations on them $\de >0$ (see fig.~\ref{fig:eqopp}).
Now by eq.~(\ref{dqde}), the total change in free energy to first
order is
\be
\delta F \cong -Q_1 \de + Q_2 \de = -\left(Q_1 - Q_2 \right) \de
\label{df1st}
\ee
so if $Q_1 > Q_2$ the change in free energy is negative and if $Q_2 >
Q_1$, $\delta F > 0$. Since contacts are {\em typically} unformed or
less formed in the unfolded state, we can say that if $Q_1(\Qdag) >
Q_2 (\Qdag)$, $\delta \DFdag < 0$ and {\it vice-versa}. 
Since for well-designed two-state folders the rate is controlled by the
free energy barrier, the assertion is then demonstrated (c.f. the
discussion in the introduction regarding barrier governed rates,
and also see section~\ref{sect:kinetics}).
Some obvious caveats include perturbations on a protein not
well-designed, 
perturbations of contacts involving
residues anomalously formed in the 
unfolded state, or situations where strengthening one of the contacts
lowers the free energy of an on-pathway intermediate; for these
exceptional cases the effect may not be observed.

\subsection{Adding native heterogeneity will always
lower the thermodynamic folding barrier in a well-designed protein.}
\label{sec:add}
A homogeneously ordering protein is equally likely to fold from
anywhere within it. As perturbations are made away from this scenario,
say in native interaction energies or through structural variance,
the folding barrier tends to decrease.
Several arguments detailed in the following subsections suggest this effect.
\subsubsection{Random Energy Model method}
\label{sect:remarg}
Consider making
random energetic perturbations on the contact energies of an initially
homogeneous idealized system (where all contact probabilities 
are the same: $\Qi =Q$) with free energy barrier $\FH$ and
folding rate $k_o \exp ( -\FH/T )$. Then the new rate is 
\be
k_f = k_o \exp \left(-\frac{\FH
+ \delta F\left(T\right) }{T} \right) = \kH \exp\left( - \frac{\delta
F\left( T \right)}{T} \right) \: .
\label{raterem}
\ee
If the total native (unconstrained) energetic variance $\sum_i \dei^2$
is $\Delta \En^2$, the variance 
at the transition state is approximately $\Delta \En^{2\: \ddag} =
\Qddag (1 - \Qddag) \Delta 
\En^2$, given the energies must sum to total native energy $\En$. The
variance vanishes at $Q=0$ since there are no 
contacts made there, and vanishes at $Q=1$ since all the $\sum_i \ei =
\En$, i.e. all the energies must sum to a fixed number and thus their sum
cannot vary. Approximating the transition state as an ensemble of states
with uncorrelated energies, i.e. a random energy
model~\cite{DerridaB81}, and considering only the effects of changing
native interactions, the energy will always decrease twice as much as
the entropy times the temperature under the influence of
heterogeneity. Thus the free energy barrier decreases:
\be
\delta F( T ) = \delta E( T ) - T \delta S(
T ) = -\frac{\Delta \En^{2\: \ddag}}{T} -
\frac{\Delta \En^{2\: \ddag}}{2 T} = 
-\frac{\Qddag\left(1-\Qddag\right)\Delta \En^2}{2 T} \: ,
\label{DFrem}
\ee
and the rate in eq.~(\ref{raterem}) increases as
\be
k_f \approx \kH \exp\left(\frac{\Qddag \left(1-\Qddag\right) \Delta
\En^2}{2 T^2}\right) \: . 
\label{rate1}
\ee
This crude argument yields essentially the same result as a much more
detailed functional analysis, c.f.  eq.~(\ref{rateFunc}) of the text;
an additional factor appears there to account for polymer effects on the
number of routes to the native state.
By this argument even for an initial unperturbed funnel which is fully
symmetric (an idealized case where all contacts are equally likely to
be formed), introducing arbitrary heterogeneity lowers the folding barrier.

As we will see below, since energies and entropies enter the
expression for $\Qi$ on the 
same footing, the above statements apply to the dispersion in loop
entropies inherent to a particular native structure, see
fig.~\ref{fig:sv}. 

\subsubsection{Argument from Transition state theory}
\label{sect:tst}
The result~(\ref{rate1}) is not surprising from the point of view of
transition 
state theory. Another way of describing it is to note that the time
rate of change of the population ${\cal N}$ in a metastable state is
proportional 
to the escape rate, and the escape rate is itself proportional to the
ratio of partition functions, transition state to reactant:
\be
\frac{\dot{{\cal N}}}{{\cal N}} = k \sim
\frac{Z^{\neq}}{Z^o} \: .
\ee
Now we imagine the transition state to be composed of an ensemble of 
$\Omega^{\neq}$
microstates of the system having energies $\Ei^{\neq} =
\overline{E}^{\neq} + \delta E_i$, where $\delta \Ei$ is distributed
state to state from a Gaussian distribution: $P(\delta \Ei) \sim \exp
(- (\delta \Ei)^2 / 2 Nb^2)$. Then 
\be
Z^{\neq} = \sum_i \mbox{e}^{-\beta \Ei^{\neq}} \sim \mbox{e}^{-\beta
\overline{E}^{\neq}} \Omega^{\neq} \left< \mbox{e}^{-\beta \delta E}
\right> = \overline{Z}^{\neq} \mbox{e}^{N b^2/2 T^2} \: .
\ee
So, neglecting changes in the prefactor, the rate increases
exponentially with the variance in the transition state, which scales
extensively with system size.

\subsubsection{Optimum fluctuation method}
\label{sec:of}
Applications of nucleation in disordered
media~\cite{Karpov94,Oxtoby96} to protein folding show 
similar trends in the folding rate. In a
system such as a protein there may be regions where nucleation of the
folded state is favored
due to local energetic or entropic inhomogeneities. These may speed
the rate of nucleation by decreasing the effective thermodynamic
nucleation barrier, referred to in the nucleation literature as the
optimum fluctuation (see fig.~\ref{opt}). In the theory of electronic
band tails in disordered systems the optimum fluctuation method has
been used to calculate the density of states at the mobility
edge~\cite{LifshitzIM64,HalperinBI66,LangerJS66}.

Let the folding rate for a given nucleation barrier $\Fdag$ be
$k(\Fdag) = k_o \exp\left(-\Fdag/T\right)$ and the probability
distribution of nucleation barriers be defined as $P(\Fdag) \equiv
\exp\left(-\phi (\Fdag)\right)$. Then the average rate in the presence
of a distribution of barriers is
\be
\overline{k} = \int \! dF \, P(F) k(F) \approx k_o \mbox{e}^{-G(T)}
\label{of1}
\ee
where 
\be
G(T) = \minF \left( \phi (F) +\frac{F}{T} \right)
\label{Gmin}
\ee
The decrease in activation barrier by the REM argument above amounts
to letting the free energy barriers be Gaussianly
distributed about a mean $\Fbardag$ with variance $\Delta^2$, so $\phi
(F) \approx (F-\Fbardag)^2/2 \Delta^2$. Then straightforwardly
from~(\ref{Gmin}) the effective barrier $\Fstar(T) = \Fbardag -
\Delta^2/T$ and thus
\be
\overline{k} \approx k_o \exp \left( -\frac{\Fbardag}{T} +
\frac{\Delta^2}{2 T^2} \right) = 
\kH \exp\left(\frac{ \Delta^2}{2 T^2}\right)  \: ,
\label{eq:optfrate}
\ee
which has  the same form as eq.~(\ref{rate1}), since this
is essentially the same argument as above.

We could have just as well found the disorder averaged time to
nucleate a folded structure, from $\overline{\tau} = \int P(F) \tau(F)
= \tau_o \exp ( \Fbardag/T + \Delta^2/2 T^2 )$. Large barriers
increase the mean time above zero-disorder value when the averaging is
done. This does not mean the rate observed in an experiment slows,
since the important quantity governing the decrease in reactant
population is the transition rate $k$, as described above in
section~\ref{sect:tst}.

{\it Example:  Free energy potentials observed in simulations.} $\;$
Consider for example the free energy profile 
obtained from off-lattice Monte Carlo simulations of a uniform energy
G\={o} model to the native structure of CheY (see fig.~\ref{chey}).
The profile is obtained by Boltzmann
sampling the states and partitioning them to different ensembles given
their overall number of native contacts. 
This profile is proportional to the function $G(T)$ above in
eq.~(\ref{Gmin}), i.e. it contains by  
construction the reduction in barrier due to structural (entropic)
fluctuations, and so should be a good
predictor of the rate.
The probability a particular
native core (illustrated in fig.~\ref{core}) is sampled is 
\be
\pcore \approx \mbox{e}^{-\Ecore/T + \Shalo} = \mbox{e}^{-\Fcore/T}
\ee
where $\Ecore$ is the energy of the native core ($\Ecore = \En Q$ for
all cores in the uniform G\={o} model) and $\Shalo$ is the entropy of
the polymer halo dressing the core. The role of native cores and
halos in calculating free energy profiles was discussed
in~\cite{PlotkinSS97}. 
So up to a $Q$ independent constant, the free energy at $Q$ is
obtained from the relative probabilities as
\bea
\frac{F}{T}(Q) &\sim& - \log \left[ \sum_{\mbox{\tiny CORES}}^Q
\mbox{e}^{-\Fcore/T} \right] \nonumber \\
&\sim& - \log \int^{'} \!\! d\Fcore \, n\left(\Fcore\right)
\mbox{e}^{-\Fcore/T}
\label{simof}
\eea
where the sum has been replaced by an integral over the number of
cores at $Q$ having free energy $\Fcore$. Equation~(\ref{simof}) is
analogous to eq.s~(\ref{of1}) and~(\ref{Gmin}) above obtained from the
optimum fluctuation method.
As a limiting case consider an idealized folding funnel with no
dispersion in energetics or entropics. Then $n(\Fcore) = \exp(\Sroute)
\d (\Fcore - \Fbarcore)$, where $\exp (\Sroute)$ is the number of
cores, or routes through the bottleneck. The free energy
in~(\ref{simof}) then becomes
\be
\frac{F}{T}(Q) \sim \frac{E(Q)}{T} - \overline{\Shalo}(Q) - \Sroute (Q)
\ee
which reproduces the mean-field free energy
profile~\cite{PlotkinSS97} in the absence of non-native
interactions. When there is  
dispersion in the free energies of a partially structured protein,
there is a distribution of core free energies. 
Heuristically we can approximate this by a
Gaussian distribution:
\be
n\left(\Fcore\right) = \mbox{e}^{\Sroute} P(\Fcore) \approx \exp \left(
\Sroute - \frac{\left( \Fcore - \Fbarcore \right)^2}{2 \Delta^2}\right) \: . 
\ee
Then 
\bea
\frac{F}{T}(Q) &\sim& - \Sroute (Q) - \log \left[ \int^{'} \!\! d\Fcore
\mbox{e}^{-\Fcore/T - (\Fcore - \Fbarcore)^2/2 \Delta^2} \right]
\nonumber \\
&\sim& - \Sroute (Q) + \frac{\Fbarcore}{T} - \frac{\Delta^2}{2 T^2} \:
.
\label{eq:hetcore}
\eea
So the barrier observed by sampling the Monte Carlo data or running
molecular dynamics is the
optimal fluctuation barrier, which includes in it the lowering effect
due to structural dispersion in fig.~(\ref{chey}) and both structural and
energetic dispersion in general.
The lowering of the barrier due to structural variance is further
investigated in section~\ref{sect:structvar} (see also fig.~\ref{fig:sv}).
Additionally, experiments which monitor equilibrium properties related
to relaxation rates or native structure formation measure the optimum
fluctuation, as mentioned above in the context of transition state
theory.
$\Sroute (Q)$ in eq.~(\ref{eq:hetcore}) is the log of the total number of
possible native cores at $Q$. This quantity will appear in the free
energy functional below, where after the functional is minimized it is
interpreted as the number of thermally accessible routes at a given
temperature.

\subsubsection{Thermodynamic perturbation theory}
\label{sec:landau}
Another argument for the lowering of the barrier 
makes use of thermodynamic perturbation
theory~\cite{Landau80}. 
Consider a G\={o} model with $M$ contacts, whose
configurational states are perturbed in energy by a random
contribution $V_c\equiv\delta E_c$ so that the new energy of state $c$
is $E_c = E_c^o + V_c$. Let the native energy be unchanged: $\delta E
= 0$ in the native state.
Then the change in free energy to second order in $V$ is 
\be
\delta \Delta F (Q) = \left<V\right> - \frac{1}{2T} \left< \left(
V-\left<V\right> \right)^2 \right>
\label{landau}
\ee
where 
\be
\left<V\right>=\frac{1}{Z}\sum'_{c \in Q} V_c \exp(-E_c^o/T) =
\left< \delta E\right>'_o
\ee
is calculated by summing over all configurations $c$ having $Q$ native
contacts.
Now since the change in a configuration's energy is the sum over
perturbations of native contacts made in that state, 
\bea
\left< \delta E\right>'_o &=& \sum'_{c \in Q} \delta E_c
\frac{\mbox{e}^{-E_c/T}}{Z} \nonumber \\
&=& \sum'_{c \in Q} \sum_{j\in \cN} \delta(j,c) \dej
\frac{\mbox{e}^{E_c/T}}{Z} 
= \sum_{j\in \cN} \left< \delta_j \dej \right> \nonumber \\
&=& \sum_{j=1}^{M}
\left< \delta_j \right> \dej = \sum_{j=1}^{M} \Qj \dej \: .
\label{number}
\eea
The last equality follows from eq.~(\ref{Qdelta}). Thus the first
order change in free energy is simply the sum of the perturbations
times the fraction of time those perturbations are felt, as in
eq.~(\ref{df1st}). However here the first order term is the sum of a large
number of random uncorrelated terms, and so is Gaussianly distributed
over realizations of the perturbation. The mean of this distribution 
is zero since the perturbation is randomly made contact to contact:
\be
\overline{\delta\Delta F^{\dag}} = \sum_i^M \overline{Q_i \dei} = M
Q \: \overline{\de} = 0 \: ,
\ee
i.e. $\overline{\de} = (1/M) \sum_i^M \dei = 0$, because the native
energy is unchanged~\cite{note:land}.
The standard deviation
\be
\sqrt{\overline{\left(\delta\Delta F^{\dag}\right)^2}} = 
\sqrt{M Q \left(1-Q\right)} \: b 
\ee
scales like $\sqrt{N}$ since $M=zN$. Therefore the 
first order term in~(\ref{landau}) will be $\pm \, \mbox{const.}\times
N^{1/2}$. Here we've let the individual contact variance
$\overline{\dei^2} = b^2$. Similar arguments of the effects
of heterogeneity on the barrier were considered in~\cite{Wolynes97cap}.

On the other hand, the second order term in~(\ref{landau})
is proportional to $\left< \delta V^2 \right>$ and scales like
$N$ and is always negative. By the reasoning in eq.~(\ref{number}) the
average, over realizations of native disorder, of the thermal
fluctuation is  
\be
\overline{\left< V^2 \right> - \left< V\right>^2} = 
\sum_{i,j=1}^M \, \overline{\dei \dej} \;\; \overline{\left[ \left< \di \dj
\right> - \left< \di \right> \left< \dj \right> \right]} \: .
\ee
Since the perturbations are independent of each other the cross terms
in the sum vanish: $\overline{\dei \dej} = \overline{\dei^2} \delta_{ij}
= b^2 \delta_{ij}$, and
\be
\overline{\left< V^2 \right> - \left< V\right>^2} = \sum_{i=1}^M
b^2 \; \overline{\Qi \left( 1- \Qi\right)} \: ,
\label{nocross}
\ee
where the last equality follows from the fact that the fluctuations of
particles obeying Fermi-Dirac statistics (c.f. eq.~\ref{Qstar}) 
obey the property 
$\left<\di^2\right> - \left<\di\right>^2 = \left<\di\right>\left(
1-\left<\di\right> \right)$.
The sum in~(\ref{nocross}) has the form of $M$ positive terms and thus
scales extensively ($\sim M$) with the size of the system as opposed
to the first order term.
Thus the free energy change due to random perturbations in the
native energies is negative in the thermodynamic limit.
Since native contacts are less formed in the unfolded state than in the
transition state, the change in barrier height is also negative in
the thermodynamic limit. 

That higher order terms do not reverse the trend in barrier height can be
ensured by the Peierls-Bogoliubov inequality $F \leq F_o + \left< V
\right>_o$ where $F_o$ is the free energy in absence of the random
component and $V$ is the random part of the Hamiltonian averaged over
the unperturbed states, which is just the first order term in
eq.~(\ref{landau}). Thus the transition state free energy (per volume)
$F(\Qdag)/N$ is always less than the
unperturbed free energy $F_o(\Qdag)/N$ in the thermodynamic limit, and
since $F(0) \cong F_o (0)$ in the unfolded state, the barrier is
always lowered. 

\subsection{Distributing native energies uniformly in a well-designed
protein maximizes the thermal entropy}
\label{sect:ebar}

If we consider for illustration that native energies are chosen from a
distribution which is the independent product of Gaussians
\be
P(\setei) \approx \prod_{i=1}^M \, \exp\left(-\frac{\left(\ei -
\ebar\right)}{2 b^2} \right) \: ,
\ee
then the native energies will typically fluctuate around their
average $\ebar$ on a scale $b$, with the set of energies $\ei =
\ebar$ being the most likely distribution.
We are ignoring for now any constraints on the native energies that
might be present in real proteins, e.g. for functional requirements.
We now show that this distribution of native energies maximizes the
thermal entropy at any degrees of nativeness, for well-designed
proteins.

The entropy at $Q$ is given by
\be
S(Q) = - \sum_{\alpha}' p_{\alpha} \ln p_{\alpha}
\ee
where the sum is over all states at $Q$, and $p_{\alpha}$ is
the probability that state $\alpha$ is occupied from the ensemble of
states at $Q$.
We approximate a well-designed protein by a G\={o} model. Then
if all the native energies are equal ($\ei = \ebar$), all states at
$Q$ have the same energy $E_{\alpha} = Q \En$. Then the partition
function $Z$ is $\Omega(Q) \exp( -Q \En/T)$, where $\Omega(Q)$ is the
total number of configurational states at $Q$.
Then the probability any state
is occupied is $1/ \Omega(Q)$, and the thermal entropy becomes
\be
S(Q) = \ln \Omega(Q)
\ee
which is the configurational entropy-  the largest the thermal entropy
can possibly be. This result is recovered again using the general
free energy functional theory in section~\ref{sect:totent}, and can be
seen in fig.~\ref{fig:Splot}. 

It should be noted that even though the entropy at the bottleneck is
maximized for this choice of energies, the barrier is not minimized,
and in fact may be lowered further by the addition of energetic
heterogeneity as described above.

\section{Free Energy Functional}
\setcounter{equation}{0}
\label{sect:f}

In this section we derive  the free energy functional to be used in the
main analysis. This should  solidify the  concepts outlined in the
previous section regarding
the effects of heterogeneity on the folding mechanism.
We first show how the functional is related to the Hamiltonian, as in
section~\ref{sect:simple}. Then in section~\ref{sect:ent} the entropic
terms present in the functional are derived. In~\ref{sect:qstar} the
thermal contact probabilities are obtained by minimizing the free
energy functional.

\subsection{Obtaining the functional from a Hamiltonian}

We can motivate the form of the free energy
functional from landscape arguments, i.e. by considering energy
distributions of states with structural similarity to the native. 
Consider a contact Hamiltonian ${\cal H}$ of the form 
\be
{\cal H}\left( \left\{ \Dab \right\} | \left\{ \Dabn \right\} \right)
 = \sum_{\alpha<\beta} \eabn \Dab \Dabn + \eab\Dab
\left(1-\Dabn\right) 
\label{Hamilt}
\ee
which gives the energy of a particular configuration defined by the set of
contact interactions $\{\Dab \}$. This Hamiltonian
gives energy $\eabn$ to the contacts which are native contacts, and
energy $\eab$ to non-native contacts. 
We embody the principle of minimum frustration~\cite{BryngelsonJD87}
by making the mean of the distributions from which native contact
energies are chosen be lower than the mean of the distribution
for non-native contact energies.
Native contacts may also have a smaller variance, depending on the
effective number of effective letters in the sequence.
The energies in~(\ref{Hamilt}) are  internal free
energies of spatially short-ranged interaction between effective monomeric
units, after averaging over side chain and solvent degrees of freedom.
The double sum is over residue indices, and 
$\Dab=1$ if residues $\alpha$ and $\beta$ are in
contact in a configuration, $\Dab = 0$ otherwise. $\Dabn =1$ if
these residues are also in contact in the {\it native}
configuration, and $\Dabn = 0$ otherwise.
$\eabn$ and $\eab$ are again the energies of native and non-native contacts
respectively.

We obtain the thermodynamics for this system by considering
statistical properties of an ensemble of partially native
states.
Once the density of states $n\left( E|\setQi\right)$ is known the
thermodynamics at temperature $T$ can be obtained. 
We obtain a statistical average of $n\left( E|\setQi\right)$ from a
knowledge of the overall number of partially native states, and the
probability each of these states has a given energy.
A similar derivation for a homogeneous order parameter $Q$ was
calculated in~\cite{PlotkinSS97}.
The probability a configuration with a particular set of native contacts
$\{ \Dab\Dabn \}$ has energy $E$, given the 
native state has energy $\En$, is given by
\be
P\left(E|\En, \left\{ \Dab\Dabn \right\} \right) = 
\left< \delta \left[ E- {\cal H} \setDab \right] \delta\left[ \En -
{\cal H} \setDabn \right] \right>_{non-nat}
\ee
where the averaging is over the non-native contact coupling energies,
$$
\left< \cdots \right>_{non-nat} = \int \prod_{non-nat} \, P\left(\eab \right) \,
d\epsilon_{\alpha\beta} \: .
$$
Residual features in the folding mechanism
may be present due to non-self-averaging effects of non-native
interactions, resulting in phenomena such as ``off-pathway''
intermediates. We smooth over such 
phenomena with the above averaging, leaving only an average non-native
background field, while native interactions are explicitly retained. 
Thus ``on-pathway'' intermediates, or fluctuations in the free energy
landscape as in fig.~\ref{chey} due
to native structural or energetic heterogeneity 
are retained in this procedure. 
Note ${\cal H} \setDabn $ is just the sum of the
native interaction energies.
Averaging the Fourier-transformed delta functions over 
non-native interactions chosen from a Gaussian
distribution, 
$$
P\left(\eab \right) = \frac{1}{\left(2\pi b^2\right)^{1/2}}\exp
\left(-\frac{\eab^2}{ 2 b^2} \right)
$$
results in
\be
P\left(E|\En, \left\{ \Dab\Dabn \right\} \right) =
\frac{1}{\left( 2 \pi Mb^2 \left(1-Q\right)\right)^{1/2}} \exp \left(-
\frac{ \left( E-\sum_i \ei \Qi \right)^2}{2 M b^2 \left(1-Q\right) }
\right)
\ee
where the sum over native contacts $\sum_{\alpha\beta}  \eabn \Dab
\Dabn$ is written in the shorthand single index notation $\sum_i \ei
\Qi$,  i.e. $\Qi \equiv \Dab\Dabn$. Here $\Qi=0,1$ but in 
the free energy functional, fractional values are allowed as in the
derivation of eq.~(\ref{Fpara}) (the entropy per spin would be
zero if only integer values of the spin degree of freedom were
allowed). We will see that the thermal values of the contact probabilities 
$\Qi^{\star} = \left<\right.\Dab\Dabn\left. \right>_{\mbox{\tiny T}}$
are the fractional values that minimize the functional
(c.f. eq.~(\ref{Qstar}) in sect.~\ref{sect:qstar}).

When the  log density of states $\log n\left( E|\setQi\right)$ is
large, it can be replaced by the disorder-averaged number
$\Omega(\setQi) P(E|\En,\setQi) $, since the relative 
fluctuations in the number die 
off as $M^{-1/2}$ for uncorrelated disorder:
\be
\log n\left( E|\setQi\right) \approx S\left(\setQi\right) - 
\frac{ \left( E-\sum_i \ei \Qi \right)^2}{2 M b^2 \left(1-Q\right) }
\: .
\label{nofE}
\ee
The term $S\left(\setQi\right)$ is the configurational entropy,
discussed  below. 
The thermal energy $E(T|\setQi)$ is obtained from the density of
states above through $\D \log n(E)/\D E = T^{-1}$:
\be
E\left(T|\setQi\right) = \sum_i \ei \Qi - \frac{M
b^2}{T}\left(1-Q\right) \: .
\label{ET}
\ee
This procedure is applicable in the high
temperature regime when the number of states occupied at such
temperatures is large. The energy consists of an integration over an
energy density i.e. by an energy per contact times the
probability that contact is made, $\ei \Qi$, summed over all
contacts, minus a term corresponding to a lowering of the thermal
energy due to the net effect of non-native traps.
Substituting~(\ref{ET}) into~(\ref{nofE}) gives
the thermal entropy
\be
S\left(T|\setQi\right) = S\left(\setQi\right) - \frac{M b^2}{2 T^2}
\left(1-Q\right) \: ,
\label{SthermQ}
\ee
which consists of the entropy of the 
polymer chain subject to the geometric constraints $\setQi$ of contact
formation, $S\left(\setQi\right)$, and a lowering term due to the
presence of non-native traps (fluctuations in
Boltzmann weights due to the fluctuations in state energies 
reduces the effective total number of states occupied).
The temperature dependence of $S(\setQi)$ appears through the
implicit temperature dependence of the contact probabilities $\Qi$
(see eq.~(\ref{Qstar})).

At this point, since no exact solution for the entropy of a
three-dimensional polymer containing topological constraints is known,
we must either resort to an exact solution of an approximate, idealized
model system, or 
an approximate phenomenological treatment of the
real, exact system.
We choose the latter approach for the theory, and the former approach
in the lattice simulations. While still an approximation, the
entropy we derive 
captures the same quantitative effects we see in the simulations,
which of course contains an exact computation of the entropy for the
idealized lattice model.
When computing the entropy in the contact representation,
we must calculate how much entropy the unconstrained polymer has, $N s_o$, 
how much polymer entropy is lost to form a set of contacts consistent
with an overall fraction $Q$ of native structure, ${\cal S}_{\mbox{\tiny BOND}} ( \{ \Qi \} | \setli)$, and how much
``mixing'' entropy is contained in the diversity of contact
patterns consistent with that overall fraction $Q$ of native
structure, ${\cal S}_{\mbox{\tiny ROUTE}}(\{ \Qi \})$:
\be
S\left(\setQi\right) = N s_o + 
{\cal S}_{\mbox{\tiny ROUTE}} \left( \left\{ \Qi \right\}\right)
+ {\cal S}_{\mbox{\tiny BOND}} \left( \left\{ \Qi \right\} | \setli
\right) \, .
\label{eq:Stotal1}
\ee
These contributions are discussed in detail  in section~\ref{sect:ent}.

The free energy functional 
at temperature $T$ and nativeness $Q$ is written as $E-T S$ in terms
of the field $\setQi$, using
eq.s~(\ref{ET}), (\ref{SthermQ}) and~(\ref{eq:Stotal1}) :
\bea
F\left(T| \left\{ \Qi\left(Q\right)\right\} | \setei ,\setli \right)
&=& \sum_i \ei \Qi 
- T {\cal S}_{\mbox{\tiny ROUTE}} \left( \left\{ \Qi \right\}\right)
- T {\cal S}_{\mbox{\tiny BOND}} \left( \left\{ \Qi \right\} | \setli
\right)  \nonumber \\
&& + \overline{E}(Q,\eta) - N T s_o - \frac{M b^2}{2
T}\left(1-Q\right)  \; .
\label{f}
\eea
The terms depending on $\setQi$ in eq.~(\ref{f}) involve integrations over
the native density field $\setQi$, while the remaining terms depend
only on the  uniform ``background field'' $Q$.  
We have included a mean energy  $\overline{E}(Q,\eta)$ dependent on
$Q$ and total packing fraction $\eta$, the 
total configurational entropy $N
s_o = N \ln \nu$, where $\nu$ is the number of configurations per
residue in the unconstrained polymer, and the correction to the
free energy due to non-native ruggedness $-\Delta E^2(Q)/2 T$.
These uniform terms are not central to our main analysis, which
considers specifically the effects of native heterogeneity in
structure and contact energy.

We note in passing  that for the ensemble of sequences with only
overall stability $\sum_i \ei = \En$ specified, rather than the whole
distribution $\setei$ as in eq.~(\ref{f}) above, 
the non-native ruggedness decreases with $Q$ as $\sim
(1-Q^2)$~\cite{PlotkinSS97}, rather than as $(1-Q)$ as above.
This results from averaging over the native coupling energies under the
constraint $\sum_i \ei = \En$.

The native stability gap is composed of  a sum 
of 2-body interaction energies between $M$ pairs of native residues.
Cooperative contributions to the energy
function~\cite{KolinskiA93:jcp,PlotkinSS97} necessary for {\it de
novo} prediction~\cite{KolinskiA93:jcp,VendruscoloM99} 
and accurately representing barriers~\cite{PlotkinSS97,EastwoodMP00}
are not  studied here, since native  stability is present {\it a
priori} in the free energy of our model, and thus 
we focus specifically on the properties of already
well designed sequences to a given structure, for which cooperative effects
should induce quantitative but not qualitative changes in the results
presented here. 

For a G\={o}-like set of interactions, collapse and folding occur
simultaneously. $\Ebar$ in eq.~(\ref{f}) can then be set
to zero since all energetic contributions $\sum_i \Qi \ei$ are from
native contacts. At 
the other extreme, if there is no change in density with folding and the
total number of contacts of any kind is a constant,
the term $\sum_i \Qi \ei$ can be interpreted as the {\it extra} energy native
contacts get. Then  $\Ebar$ is a constant, which again has no effect on
the free energy. In an intermediate regime there is some non-native
density coupling to progress along the folding reaction
coordinate~\cite{PlotkinSS97}. Since this subtle effect is secondary to
and unnecessary 
for the analysis below, we ignore it and treat $\Ebar$ as a constant.

\subsection{Entropic Terms}
\label{sect:ent}

If we imagine the ensemble of configurations that have a given amount
of order, say a given number $MQ$ of native contacts, then within
this ensemble there are a multiplicity of sub-ensembles of states
having different sets of $MQ$ contacts, which we identify as a measure
of the number of distinct routes in folding to the native state. 
Each sub-ensemble contains many
states corresponding to the entropy of the disordered polymer around
the particular native core (e.g. see fig.~\ref{core}). We define the
entropy that corresponds to the 
degeneracy of contact patterns having functional  order $\{ \Qi(Q) \}$ as 
${\cal S}_{\mbox{\tiny ROUTE}} \left( \left\{ \Qi (Q)\right\}\right)$
(${\cal S}_{\mbox{\tiny ROUTE}} >0$), and 
the configurational entropy lost from the coil state to induce the
ordering specified by $\left\{ \Qi \right\}$ as
${\cal S}_{\mbox{\tiny BOND}} \left( \left\{ \Qi \right\} | \setli
\right)$ (${\cal S}_{\mbox{\tiny BOND}} <0$).

\subsubsection{Route Entropy}
\label{subsect:rout}
In capillarity models of nucleation~\cite{Becker35}, ${\cal S}_{\mbox{\tiny
ROUTE}}$ corresponds to the log of the translational partition
function~\cite{Lothe62,Reiss68,Lothe69} which scales logarithmically
with system size, 
plus the entropy of surface fluctuations of droplets of a given
size~\cite{Langer67,Fisher67} which correspond to logarithmic terms in the
expansion of 
the free energy density. This entropy is small compared to the total
conformational entropy, however at the spinodal
where $F(Q)$ becomes downhill (e.g. long-dashed curves in
fig.~\ref{fig:Fplot} ), the 
nucleus is of small amplitude and highly
ramified~\cite{Gunton83,UngerC84}. 
In this regime the droplet structure is 
percolative as in spinodal decomposition of binary
fluids, and the capillarity approximation is poor.
Field-theoretic descriptions for the structure of the droplet are
typically used in this regime~\cite{Cahn58,Langer69}.
Binary fluid approximations to
the route entropy in proteins which scale extensively with system
size have been used in this
limit~\cite{BohrHG92,PlotkinSS96,PlotkinSS97,PandeVS97:fd,ShoemakerBA97,ShoemakerBA99,ShoemakerWang99}.
The amount of route diversity in folding has also been analyzed in
terms of the Shannon entropy~\cite{FernandezA00:jcp}, which is similar
in spirit to the following treatment~\cite{PlotkinSS00:pnas}.
We make no capillarity or spinodal assumptions, and treat the 
route entropy ${\cal S}_{\mbox{\tiny ROUTE}} \left( \left\{ \Qi
\right\}\right)$ as a fairly simple modification of the entropy of 
a binary fluid mixture~\cite{Landau80}:
\bea
\exp {\cal S}_{\mbox{\tiny ROUTE}}^o \left( Q\right) &=& 
\frac{M!}{MQ! \left( M-MQ\right)! } \cong
\left(\Omega_i^o\right)^M 
\label{mix1} \\
\Omega_i^o\ &=& Q^{-Q} \left(1-Q\right)^{-\left(1-Q\right)}
\label{mix2}
\eea
which we interpret here as the product of the complexities per contact
$\Omega_i^o$ and is readily generalized to the case where the
complexities are not all equal: $\exp {\cal S}_{\mbox{\tiny ROUTE}}^o
\left(\left\{\Qi\right\}\right)\Rightarrow\prod_{i=1}^{M}\Qi^{-\Qi}
\left(1-\Qi\right)^{-\left(1-\Qi\right)}$, as in eq.~(\ref{Fpara}).
The principle modification introduced here for proteins is that, due to chain
connectivity, as contact density increases, there is less sterically
allowed space for a monomer to move around when one of its constraining
contacts is broken. Thus not all $M!/MQ! \left( M-MQ\right)!$ contact
patterns have an entropy $\approx N s_o + {\cal S}_{\mbox{\tiny
BOND}}$. In other words making some native contacts forces spatially
nearby contacts to be made because the corresponding monomers are
forced to be in each other's proximity. So there is a reduction from
the putative complexity $\left(\Omega_i^o\right)^M$ since not all $M$
contacts are independently contributing to mixing, with several
contact patterns corresponding to the same constrained state. Here we
remove this degeneracy by dividing out the $\left(\Omega_i^o\right)^{M
a\left(\setQi\right)}$ states that have been overcounted. Making a mean-field
approximation for the local field around contact $i$ which reduces its
complexity, $\sum_{\alpha \neq \beta} Q_{\alpha\beta}/\sum_{\alpha \neq
\beta} \left( 1\right) \simeq Q$, the new total complexity is 
$\prod_{i=1}^M \Omega_i^{\left[1-a\left(Q\right)\right]}$. Here
$a(Q)$ is a monotonically increasing function of $Q$, from
$a(Q\rightarrow 0) = 0$ to $a(Q\rightarrow 1) = 1$, since a nearly
fully constrained polymer has all its entropy on the surface, making
the mixing entropy per monomer negligible in the thermodynamic limit.
We introduce the form $a(Q)=Q^{\alpha}$ with $\alpha$ a 
parameter determined by the best fit to the lattice data (see
fig.s~\ref{fig:mix} and~\ref{fig:sroutehet} and table~\ref{table1}
). The route entropy appearing in eq.~(\ref{f}) then
becomes~\cite{PlotkinSS00:pnas}: 
\bea
{\cal S}_{\mbox{\tiny ROUTE}} \left( \left\{ \Qi \right\}\right) &=& \log
\prod_{i=1}^{M} \Omega_i^{\lambda\left(Q\right)} = 
\lambda\left(Q\right) \sum_{i=1}^{M} \left[ -Q_i \ln
Q_i - (1-Q_i)\ln\left(1-Q_i\right) \right] 
\label{smix} \\
\lambda(Q) &\equiv& 1-Q^{\alpha}
\label{eq:lambdaQ}
\eea
The factor $\lambda(Q)$ measures the entropy
reduction due to the coupling of chain connectivity with the native
topology under study.
The power $\alpha$ in $\lambda\left(Q\right)$ should be a decreasing
function of the persistence length, and also 
of system size $N$, since for 
larger systems more polymer is buried and thus more strongly
constrained by surrounding contacts. 
Fluctuations in contact probabilities $\Qi$ will lower the route
entropy (see eq.~(\ref{flucR}) and also
figure~\ref{fig:sroutehet}). 
An alternative derivation for the route entropy in a protein
is given in Appendix~B.

\subsubsection{Bond Entropy}
\label{sect:bondS}

The calculation of the total entropy lost due to contact formation is
rendered difficult because the entropy loss of a given contact depends
not only on the contact's sequence-length or bare loop-length $\li$, but
also on the configuration of contacts $\setQi$ already present when
the contact is formed. In spite of this difficulty some general
statements can still be made, as follows.

If we make the assumption that the entropy loss to form contact $i$
depends explicitly only on the sequence length of contact $i$, as well
as the full contact pattern present $\setQi$, then the most
general form for the change in entropy due to
contact formation, to go from
configurations having one set of $\Qi$'s parameterized in terms of a
variable $t$, $\left\{ \Qi\left(t_o\right) \right\}$, to another state
having $\left\{ \Qi\left(t_f\right) \right\}$, is
\be
{\cal S}_{\mbox{\tiny BOND}} \left( \left\{ \Qi\left(t_f\right) \right\} | 
\left\{ \Qi\left(t_o\right) \right\} \right) = \sum_i \int_{t_o}^{t_f}
\, {\cal D} \Qi\left(t\right) \,\,s_i\left(\li,\left\{ \Qj\left(t\right)
\right\} \right) \: .
\label{S}
\ee
Here $s_i\left(\li,\left\{ \Qj\left(t\right) \right\} \right)$ is the
entropy loss to form contact $i$ having sequence separation $\li$, in
the presence of the contact pattern $\left\{ \Qj\left(t\right)
\right\}$, which is itself parameterized through $t$~\cite{note:resum}.
Each $s_i\left(\li,\left\{ \Qj\left(t\right) \right\} \right)$ in
eq.~(\ref{S}) is functionally integrated along the $M$-dimensional
path specified by $\left\{ \Qi\left(t\right)\right\}$.
However the entropy as a function of the set $\setQi$ must be a state
function, meaning that the value of the integral depends only on the
end points and not on the path taken. The condition for path
independence is obtained as follows.
We can envision a small subsection of the $M$-dimensional path as
traversing  a hypercube of volume $\prod_{i=1}^{M} \dQi$. Then path
independence means the entropy increment ${\cal S}_{\mbox{\tiny BOND}}
\left( \left\{ \Qi \right\} | \left\{ \Qi + \dQi \right\} \right)$ is
independent of the order the edges are traversed in going from 
$\left\{ \Qi \right\}$ to $\left\{ \Qi + \dQi \right\}$. Consider two
possible paths labeled $(1)$ and $(2)$ 
along two of these coordinates $\{ \Qi, \Qj \}$, 
as shown in fig.~\ref{fig:paths}.
Along path~$(1)$, the entropy change to second order in $\delta Q$
is 
\bea
{\cal S}_{\mbox{\tiny BOND}}^{(1)} &=& \int_{\Qi}^{\Qi+\dQi} \! \dQi' \, 
s_i \left(\li,\Qi' , \Qj\right) + \int_{\Qj}^{\Qj+\dQj} \! \dQj'\, 
s_j \left(\lj,\Qi+\dQi , \Qj'\right) \nonumber \\
&\cong& s_i \left(\li,\Qi, \Qj\right) \dQi + s_j\left(\lj\Qi,
\Qj\right) \dQj + \frac{\dQi^2}{2} \,\frac{\D \si}{\D \Qi}\left(\li,\Qi,
\Qj\right)  + \frac{\dQj^2}{2} \,\frac{\D \sj}{\D
\Qj}\left(\lj,\Qi,\Qj\right)  \nonumber \\
&+& \dQi \, \dQj\, \frac{\D \sj}{\D
\Qi}\left( \lj , \Qi,\Qj\right) 
\label{eq:p1}
\eea
while along path~$(2)$ the entropy change is the same as
expression~(\ref{eq:p1}) except that the last term is replaced by
$\dQi \, \dQj\, \frac{\D \si}{\D\Qj}\left( \li , \Qi,\Qj\right)$. For
these two expressions to be equal
\bea
\frac{\D s_j}{\D \Qi} \left( \lj, \left\{ \Qk
\right\} \right) = \frac{\D s_i}{\D \Qj} \left( \li, 
\left\{ \Qk \right\} \right) && \;\;\; \mbox{for}\;\;i \neq j \: .
\label{state}
\eea
For $M$ dimensions, it follows that eq.~(\ref{state}) holds for all
pairs $(i, j)$, yielding $M (M-1)/2$ nontrivial 
constraints on the form of the
configurational entropy loss at each value of $Q$. 

When the entropy loss satisfies eq.~(\ref{state}), the total entropy
difference only depends on the initial and final 
states and can be rewritten as
\be
{\cal S}_{\mbox{\tiny BOND}} \left( \{ \Qi^f \} | 
\{ \Qi^o \} \right) = \sum_i \int_{\Qi^o}^{\Qi^f} \, 
d\Qi \,\, s_i\left(\li,\left\{ \Qj\right\} \right) \: .
\label{Sb}
\ee

Now we seek an approximate formula for $\si$ that satisfies
eq.~(\ref{state}). In forming 
a contact $i$ from the unconstrained molten globule or coil state, 
the segment of 
polymer loses the entropy of a free chain with the length of that segment, 
$s_i \left(\li, \setQj \cong \left\{ 0 \right\}\right) = \ln
\left(a/\li\right)^{3/2}$ where $a$ is a $Q$ independent constant
related through a sum rule  
to polymeric properties (see eq.~(\ref{a})). However ``zippering up'' contacts
formed in a nearly fully 
constrained polymer cost almost no entropy: $s_i\left(\li, \setQj
 \approx \left\{ 1 \right\}\right) \cong 0$. 
To account for this we introduce an  
effective loop length $\leff (\li, \setQj  )$ 
into $s_i\left(\li, \setQj\right) = \ln (a/\leff)^{3/2}$.
We ignore here possibly important changes in the power of the ideal chain
exponent $3/2$, since it becomes cumbersome to incorporate an
exponent dependent on $\setQi$ and to 
simultaneously satisfy eq.~(\ref{state}).

Because of the path independence of the configurational entropy loss
${\cal S}_{\mbox{\tiny BOND}} ( \{ \Qi^f \} | \{ \Qi^o \} )$, the
change in entropy for a small 
change in one of the contacts $\Qi^f\rightarrow\Qi^f+\delta\Qi$ 
is simply the integrand evaluated at the upper limit:
\be
\frac{\D {\cal S}_{\mbox{\tiny BOND}} }{\D \Qi} \left( \{ \Qj^f \} | 
\left\{ \Qj^o \right\} \right)  = s_i\left(\li,\{
\Qk^f\} \right)
\ee
which can be shown from eqs.~(\ref{state}) and~(\ref{Sb}) by using the
definition of the derivative.

In this paper we satisfy eq.~(\ref{state}) with the following 
ansatz for the functional form of $\leff$:
\be
\leff \left( \li , \setQk \right) = f\left( \li\right)\,
g\left(\setQk\right) = f\left( \li\right)\, g\left(\frac{1}{M}\sum_k
\Qk \right)
\ee
so that the loop length is decreased by a function of the mean of the
contact density field, $g(Q)$. This is in the spirit of the Hartree
ansatz in the one-electron theory of metals, where electrons interact
only through an averaged field.
The condition $\leff\left(\li ,Q=0
\right) = \li$ gives $f\left( \li\right)=\li$ and $g(0)=1$. The
condition that $\leff\left(\li ,Q=1\right) \approx 1$ gives $g(1)
\approx 1/\lbar$ (since $g(Q)$ cannot depend on $\li$), where $\lbar =
(1/M)\sum_i \li$. 
To approximate the $Q$ dependence of $\leff$ we note
that the probability of a monomer being constrained at $Q$ is roughly
$Q$ under the assumption of a uniform contact probability.
Then given a chain of unbonded monomers, the probability of it being
length $L$ is then $p_{L} = Q (1-Q)^{L-1}$. So the average
length of strings of unbonded monomers at $Q$ is then $\overline{L}
= \sum_{L} L p_{L}/\sum_{L} p_{L} \cong 1/Q$, which can
be interpreted roughly as the total length of polymer $N$ over the total
number of bonds $\approx N Q$~\cite{FloryPJ56:jacs}, or the total
length over the total number of constrained monomers. We approximate the
effective loop length at $Q$, $\leff\left(\li,Q \right)$,  in the same
way by dividing the total loop length $\li$ by the number of bonded
residues in the loop (or the approximate 
number of bonds in the loop) $\cong \lbar Q$, so that finally 
\bea
s_i\left(\li,\left\{ \Qk\right\} \right) &\approx& 
\frac{3}{2} \ln \left(\frac{a}{\leff\left(\li, Q\right)}\right) 
\label{sbond}  \\
\leff\left(\li, Q\right) &\approx& \frac{\li}{\left(\lbar-1\right) Q +1} 
\label{leff} 
\eea
Note $\leff$ has the mean-field behavior for large $\li$ and also has
the right limiting behavior as $Q\rightarrow 0$ and $Q\rightarrow 1$.
Eq.s~(\ref{leff}) and~(\ref{sbond}) are accurate for weak dispersion in
loop lengths; for larger values of $\delta\li$ they must be modified
(see comments after eq.~(\ref{eq:smfflory}) ).

Expressions~(\ref{Sb}) and~(\ref{sbond}) reduce to the Flory form for the
configurational entropy loss in the mean field 
limit~\cite{FloryPJ56:jacs,GutinA94,PlotkinSS96} when $\li=\lbar$
and $\Qi=Q$. Then eq.~(\ref{Sb}) becomes
\be
{\cal S}_{\mbox{\tiny BOND}}^{\mbox{\tiny (MF)}} \left( Q | 0 \right) = 
\int_{0}^{Q} \, dQ \,\, M \, \ln \left( \frac{a \left[ 1+ \left( \lbar
-1 \right) Q \right] }{\lbar} \right)^{3/2} \: ,
\label{Sbmf}
\ee
which can be interpreted as a summation of entropy losses from $0$ to
$Q$: 
\bea
{\cal S}_{\mbox{\tiny BOND}}^{\mbox{\tiny (MF)}} \left( Q | 0 \right) 
&=& \sum_{Q'=\Delta Q}^Q \, \Delta S(Q') = 
\sum_{Q'=\Delta Q}^Q \, \ln  \frac{\Omega \left(Q' \right)}{\Omega
\left(Q'-\Delta Q\right)} \nonumber \\
&=& \ln \left( \frac{\Omega\left(\Delta
Q\right)}{\Omega\left(0\right)} \, \frac{\Omega\left(2 \Delta
Q\right)}{\Omega\left(\Delta Q\right)} \, \cdots \, \frac{\Omega\left(
Q \right)}{\Omega\left(Q-\Delta Q\right)} \right) = \ln
\frac{\Omega\left( Q \right)}{\Omega\left( 0 \right)} \nonumber \\
&=& S^{\mbox{\tiny (MF)}}\left( Q\right) 
- S^{\mbox{\tiny (MF)}}\left( 0\right) \nonumber \: .
\eea
When $\lbar Q \gg 1$, eq.~(\ref{Sbmf}) gives
\be
\frac{{\cal S}_{\mbox{\tiny BOND}}^{\mbox{\tiny (MF)}}}{M} \left( Q |
0 \right) = \frac{3 Q}{2} \left( \ln a -1 + \ln Q \right)
\label{sflory}
\ee
which is essentially the Flory result derived earlier in the
mean-field limit.

In the presence of heterogeneity, 
equations~(\ref{Sb}) and~(\ref{sbond}) give
\bea
{\cal S}_{\mbox{\tiny BOND}} \left( \left\{ \Qi\right\} | 0 \right)
&=& \frac{3}{2} M Q  \ln a - \sum_{i=1}^{M} \Qi\ln \li + \sum_{i=1}^{M}
\int_0^{\Qi} \, 
d\Qi' \, \ln \left[ 1 + \frac{\lbar-1}{M}\sum_k \Qk \right] \nonumber
\\
&=& \frac{3}{2} M \left( Q \left( \ln a\right) - \frac{1}{M} \sum_i \Qi
\ln \li  - Q + \frac{\left[ 1+ \left(\lbar -1\right)Q\right]}{\lbar-1}
\ln \left[1+\left(\lbar-1\right)Q\right] \right) 
\label{sinit1}
\eea
where the last integral can be done by charging up each $\Qi$ one at a
time (in any order) to its value at $Q$, i.e. the integral is 
$\sum_i\int_0^{\Qi} \, d\Qi' \, \ln \left[ 1 + \frac{\lbar-1}{M}\left(
\sum_{j<i} \Qj + \Qi' \right) \right] $. This gives an expression
identical to the
mean-field result for this term, since the integrand only depends on
$Q$ and is integrated up to each $\Qi(Q)$.

Because the free energy of the native state $F\left( \left\{ 1 \right\} |
\setei ,\setli \right)$ is $\En$ (c.f. eq.~(\ref{f})), 
all the polymer entropy is lost
upon folding in the model. Therefore there is a sum rule
for the entropy loss, 
\be
{\cal S}_{\mbox{\tiny BOND}} \left( \left\{ 1 \right\} | 0 \right)
= \sum_{i=1}^{M} \int_0^1 d\Qi \, s_i\left(\li,\left\{ \Qj\right\}
\right) = - N \ln \nu = -N s_o \: ,
\ee
which, using eq.~(\ref{sinit1}), determines the coefficient $a$ in the
entropy of bond formation:
\be
\ln a \left( \nu, \left\{ \li \right\} \right) =
-\frac{2}{3} \frac{s_o}{z} + 1 + \overline{\ln \ell} -
\frac{\lbar}{\lbar-1} \ln \lbar \: .
\label{a} 
\ee
The coefficient $a$ depends on the distribution of $\li$ as well as
the entropy per monomer $s_o$.
Using~(\ref{a}) and~(\ref{sinit1}),  the final expression for the
entropy loss arising from contact formation is
\be
{\cal S}_{\mbox{\tiny BOND}} \left( \left\{ \Qi\right\} | 0 \right)
= {\cal S}_{\mbox{\tiny MF}} \left(Q,\lbar \right) - 
\frac{3}{2} M \left< \delta Q \delta \ln \ell \right>
\label{Sfinal} 
\ee
where the first term in~(\ref{Sfinal}) is the mean-field entropy loss
\be
{\cal S}_{\mbox{\tiny MF}} \left(Q,\lbar \right) =
-Q N s_o - \frac{3}{2} M Q \frac{\lbar \ln \lbar}{\lbar -1}
+ \frac{3}{2} M \frac{1}{\lbar -1} \left[1
+ \left(\lbar-1\right)Q\right] \ln\left[1+
\left(\lbar-1\right)Q\right]
\label{eq:smf}
\ee
and the second term in~(\ref{Sfinal}) is the change in entropy loss
due to fluctuations (again the notation $\left< X_i \right> \equiv
\overline{X} \equiv \frac{1}{M} \sum_i X_i$ is used):
\be
M \left< \delta Q \delta \ln \ell \right> = 
\sum_i \left(\Qi-Q\right) \left( \ln \li - \overline{\ln \ell} \right)
\: .
\label{fdefn}
\ee
From inspection of eq.s~(\ref{Sfinal}-\ref{fdefn}) we can confirm that ${\cal
S}_{\mbox{\tiny BOND}} (Q=0) = 0$ and ${\cal S}_{\mbox{\tiny BOND}}
(Q=1) = -N s_o$.
When $\lbar Q  \gg 1$,~(\ref{eq:smf}) reduces  to
eq.s~(\ref{sflory}),~(\ref{a}):
\be
{\cal S}_{\mbox{\tiny MF}} \left(Q,\lbar \gg 1 \right) \approx
-Q N s_o + \frac{3}{2} z N  Q \ln Q \: ,
\label{eq:smfflory}
\ee
which has lost the information about the mean loop length and only
retained information about the total chain length $N$, as in the Flory
mean-field theory.
The first term in~(\ref{eq:smf}) or~(\ref{eq:smfflory}) is the loss in
entropy to constrain a given fraction of the protein and is linear in
$Q$. The remainder in~(\ref{eq:smf}) or~(\ref{eq:smfflory}) is the extra
entropy loss this constraint induces on the remaining free parts by
pinning down regions of the polymer chain. The analogous quantity in
the capillarity theory is the surface entropy cost in forming a
nucleus of folded
structure~\cite{FinkelsteinAV97:molbiol,Wolynes97cap}.
In capillarity theories, the surface entropy cost scales like $N^{2/3}$,
whereas in mean-field theories it scales like $N$.
Eq.~(\ref{eq:smf}) can be thought of as a generalization of
eq.~(\ref{eq:smfflory}) to finite mean return length $\lbar$ for a
finite-sized system, and
eq.~(\ref{Sfinal}) can be thought of as generalizing~(\ref{eq:smf}) to
include fluctuations in the return length.

The effect of fluctuations in~(\ref{Sfinal}) is typically to increase
the bond entropy of partially native states. The trend in the folding
barrier with heterogeneity results from the interplay of this effect
with the effects of fluctuations on the route entropy and native
energetic fluctuations. The 
magnitude of the effect scales extensively with the size of the
system. To illustrate, recall that for a G\={o} model the total
entropy at $Q$ is $N s_o + 
{\cal S}_{\mbox{\tiny ROUTE}} ( \{ \Qi \}) +
{\cal S}_{\mbox{\tiny BOND}} (\{\Qi(Q)|0\} )$
(c.f. eq.s~(\ref{SthermQ}) and ~(\ref{eq:Stotal1})). Thus  
if we look at loops longer than the average ($\li > \lbar$,
and since the log function is concave down, $\ln \li > \overline{\ln
\ell}$), then they are less likely to be formed
(c.f. eq.~\ref{Qstar}), so that $\Qi < Q$ and the second term 
in~(\ref{Sfinal}) is negative, thus raising the bond entropy. If $\li <
\lbar$, $\Qi > Q$ and the effect is the same. The halo entropy of the
system $N s_o + {\cal S}_{\mbox{\tiny BOND}} (\{\Qi(Q)|0\} )$, or
Flory entropy as we refer to it later, increases when we relax the
condition that all contacts must be 
equally probable, and allow differences in contact probability based
on their entropic likelihood (see fig.~\ref{fig:stotflory}).

From~(\ref{eq:smfflory}) it can be seen that there is an entropy
crisis (${\cal S}_{\mbox{\tiny MF}} < 0$) at values of 
$Q < 1$ when $2 s_o/3 z
\lesssim 1$. This is essentially because in the mean field
approximation contacts 
are shared equally between residues; only one contact is needed to
constrain a residue, however there may be more than one contact per
residue. The increase in entropy from heterogeneity alleviates (but
not necessarily eliminates) this problem. The route entropy described
above in section~(\ref{subsect:rout}) further increases the total
entropy. However if there is a crisis, then at $Q$ values higher than 
that where the entropy crisis occurs, the mean field description is no
longer valid. Typical values of the parameters from
off-lattice simulations of Chymotrypsin inhibitor or the $\alpha$-spectrin
SH3 domain~\cite{ClementiC00:jmb} give $s_o \cong 3.4$, $z \cong 2.4$, 
$2 s_o/3 z \cong 0.94$; here the entropy crisis occurs rather late in
folding, if at all, because of entropy increase by the above-mentioned
effects. At $Q$ values above the crisis, fluctuations from the
mean-field contacts per residue must be accounted for. One way to
achieve this is to switch to a residue representation for the entropy:
the number of states is counted by considering the combinatorics of
strings of residues which are frozen or melted out~\cite{PlotkinSS97}.

On the other hand, eq.~(\ref{sbond}) breaks down for sufficiently large 
structural heterogeneity. Inspection of~(\ref{sbond}) shows that the
entropy loss has the same derivative as a function of $Q$ for all
contacts, but the initial values are different. This 
leads to some problems with the shorter loops for high $Q$ values,
which it is worth noting as a word of caution here. The crude way 
in which the entropy loss for a loop is coupled to the degree of
nativeness of the rest of the protein leads to a non-negative entropy
loss to 
close some of the shorter loops near $Q \approx 1$. We resolved this
problem  by actually truncating the entropy loss formula
for the shorter loops when they reached a value of zero.
Putting eq.~(\ref{a}) into (\ref{sbond}), letting $\lbar Q \gg 1$, and
expanding to first order in $\dl/\lbar$ (weak dispersion limit) we
obtain the approximate value of $Q$ where the entropy loss crosses
zero, namely 
$Q_v \approx 2 s_o/ 3 z + \dli/\lbar$. When $\dli = 0$ this is
consistent with the Flory analysis above, however when $\dli < 0$
(shorter loops) $Q_v$ is decreased. We  truncate the
entropy formula at zero for $Q > Q_v$. 

As a simple illustration, consider a structure whose loop
distribution is given by first returns of a self-avoiding walk, 
$P(\ell) \approx (3/2) \ell^{-5/2}$, and thus from eq.~(\ref{sbond}) $P(s) =
\exp (s - s(\li=1))$, where $s(\li=1) = 5/2 - (9/4) \ln 3 - s_o/z +
(3/2) \ln (1 + 2 Q)$ is the entropy of closure for the smallest loops
(where the problem is the worst). Using the above values for $s_o$ and $z$, 
all $s_i$ are negative until $Q_v \gtrsim 0.76$. Since the barrier peak
typically occurs at $Q$ values smaller than this, errors due to
truncation would be  small for these structures. 
On the other hand protein structures tend to have distributions with a
wider dispersion than the random globule, and in these cases
the problem would be worse.
Applying the theory
to the lattice structure of fig.~\ref{fig1}, we must truncate the entropy
loss for  loops with $\li = 3$ at $Q_{v 3} \approx 0.4$ and for loops
with $\li = 5$ at $Q_{v 5} \approx 0.75$; for all other loops there is
no entropy crisis. Numerically there is some
quantitative error
introduced by this truncation, since in the theory these loops  no
longer contribute to the total entropy loss above $Q_v$, whereas in
the actual simulation they do. 
Of course, implementing a cutoff in loop entropy causes the total
entropy to deviate from a state function by
eq.~(\ref{state}). Theories of polymer entropy which take more complete
account of correlations should remedy this and are a topic of future
work. For now we content ourselves with the Hartree style entropy
formulation in eq.~(\ref{sbond}), implementing a cutoff if needed.
In general however truncating doesn't
qualitatively change trends in the  barrier except
possibly in pathological cases of limited relevance.

Equations~(\ref{f}),~(\ref{smix}) and~(\ref{Sfinal})
together give an analytic  
expression for the free energy for a fast-folding protein which includes
heterogeneity in the folding mechanism:
\be
F\left( \left\{ \Qi\left(Q\right)\right\} | \setei ,\setli
\right) = F_{\mbox{\tiny MF}} ( Q, \ebar , \lbar ) 
+ \delta F \left( \{ \dQi \} | \{\dei\} , \{ \dli \} \right)
\label{FTOT}
\ee
where we've written the total free energy in terms of a mean-field
term plus a fluctuation due to variations in energy, loop length,
and contact probability.
In~(\ref{FTOT}), 
$F_{\mbox{\tiny MF}}/M$ is the mean-field free energy per
monomer~\cite{PlotkinSS97}:
\be
\frac{F_{\mbox{\tiny MF}}}{M} = \ebar Q -T \frac{s_o}{z} -T 
\frac{{\cal S}_{\mbox{\tiny MF}}}{M} \left(Q,\lbar \right) 
-T \frac{{\cal S}_{\mbox{\tiny ROUTE}}}{M} (Q) - \frac{b^2}{2 T} \left(
1-Q\right) + \frac{\overline{E}}{M}
\label{FMF}
\ee
with ${\cal S}_{\mbox{\tiny MF}}$  given by eq.~(\ref{eq:smf}),
and ${\cal S}_{\mbox{\tiny ROUTE}}$  given by eq.~(\ref{smix}) with
all $\Qi =Q$. 
The fluctuation in~(\ref{FTOT}) is given by
\be
\frac{\delta F}{M} \left( \{ \dQi \} | \{\dei\} , \{ \dli \} \right) = 
 \left< \d Q \, \d \epsilon
\right> + T\lambda(Q) \left< \Qi\ln\frac{\Qi}{Q} +
(1-\Qi)\ln \frac{1-\Qi}{1-Q}\right> +\frac{3}{2} T \left<
\d Q \, \d \ln \ell 
\right> 
\label{FF}
\ee
Equation~(\ref{FMF}) contains $5$ adjustable parameters which
characterize the system: $N,  s_o, z, b$ and $\Ebar$, 
and eq.~(\ref{FF}) contains $1$ adjustable parameter: $\alpha$ in
$\lambda(Q)$ of eq.~(\ref{eq:lambdaQ}). Once chosen, these
parameters are fixed for the rest of the analysis.
We've chosen some values for the parameters in table~\ref{table1} 
to compare with the  lattice simulations.
All other quantities such as $\ebar$, $\lbar$, $\delta \ell^2$, etc.
arise from the structural and energetic distribution of a given
protein at overall nativeness $Q$ and temperature $T$. In our analysis
we study trends
in the thermodynamics by varying these distributions.

As noted above, the free energy functional consists of an 
integration over a free energy density whose only information about
the surrounding medium is through the average field present ($Q$):
$F = \sum_i f_i(\Qi,Q)$. Explicitly accounting for cooperative 
entropic effects due to correlations between
contacts~\cite{ChanHS90,ShoemakerBA99,ShoemakerWang99,DillKA93:pnas} 
would be an important extension of the model, and terms that lead to such
effects have been introduced into the functional 
in similar models~\cite{ShoemakerBA99,ShoemakerWang99}.

We can make connection with the intuitive arguments discussed
previously by investigating the effects of heterogeneity on each of
the three terms in eq.~(\ref{FF}). As mentioned above, for longer
loops the contact probability is expected to be less than average, and
for shorter loops $\Qi$ is expected to be above average. So relaxing
the $\Qi$ values to accommodate this makes the third term in~(\ref{FF})
negative, lowering the free energy. Also, since the fluctuation $\dQi$
is expected to be positive when a contact is stronger ($\dei$ is
negative), the first term in~(\ref{FF}) is negative and the free
energy is lowered. Lastly, the second term in eq.~(\ref{FF}) consists
of two terms inside the average which are both concave up, i.e. have a
positive second derivative w.r.t. $\Qi$. Thus the average of the
terms is greater than the term evaluated at the average, i.e.
\bea
\left< \Qi \ln \frac{\Qi}{Q} \right> &>& \left< \Qi \right> \ln 
\frac{\left< \Qi\right>}{Q} = 0 \nonumber \\
\left< (1-\Qi)\ln \frac{1-\Qi}{1-Q}\right> &>& \left( 1 - \left< \Qi
\right> \right) \ln \frac{ 1 -\left<\Qi \right>}{1-Q} = 0  \: ,
\label{flucR}
\eea
and so the second term in~(\ref{FF}) is positive. Fluctuations away
from uniform 
ordering raise the terms in the free energy due to route entropy. This
effect competes with the two lowering effects above. To find which
terms dominate, we find the functional dependence of the
contact probabilities $\Qi$ on the energies $\ei$ and entropies $\si$
in the next subsection,
and then investigate the trend on barrier height in
section~\ref{section:opt}. 

\subsection{The most likely distribution of contact probabilities} 
\label{sect:qstar}

Equations~(\ref{FTOT}),~(\ref{FMF}), and~(\ref{FF}) describe the free
energy for an arbitrary distribution of contact probabilities $\{ \Qi
(Q) \}$, subject only to the constraint that the average probability
$\left< \Qi \right>$ is $Q$. 
The most likely distribution $\setQistarQ$ of the contact
probabilities $\Qi(Q)$, i.e. the thermal distribution, 
is obtained by minimizing the free energy $F\left( 
\left\{ \Qi\left( Q  | \setei ,\setli \right)\right\}\right)$
subject to the constraint $\sum_i \Qi\left(Q\right) = M Q$,
i.e. $\delta( F + \mu\sum_j \Qj ) = 0$, or
\be
\sum_i  \left[ \frac{\D}{\D \Qi} F\left( \setQi | \setei ,\setli
\right) + \mu \right] \delta \Qi = 0
\label{varQi}
\ee
for arbitrary and independent variations $\delta \Qi$ (c.f. the
analogous expression eq.~(\ref{eq:dfdspin}).
Substituting eqs.~(\ref{FTOT}),~(\ref{FMF}), and~(\ref{FF})
into eq.~(\ref{varQi}) yields a Fermi-Dirac
distribution for the most probable thermodynamic occupation
probabilities $\setQistar$ for a given $\setei$ and $\setli$: 
\be
\Qi^{\star}\left( Q,\setei,\setli\right) = \frac{1}{1+\exp \left[
\frac{1}{\lambda T} \left( \mu  + \ei - 
T s_i\left(\li,Q\right) - T \lambda' \left<
s_{\mbox{{\tiny ROUTE}}}^o \right> + \frac{b^2}{2T} 
\right) \right] } \; ,
\label{Qstar}
\ee
where $\lambda' = d\lambda\left(Q\right)/dQ$ (c.f. eq.~\ref{eq:lambdaQ}) and
$\left< s_{\mbox{{\tiny ROUTE}}}^o \right>(Q) = \left< -\Qj\ln\Qj -
\left( 1-\Qj\right)\ln\left( 1-\Qj\right)\right>$. Thus each
probability $\Qi^{\star}$,
referred to below simply as $\Qi$, is a function of all the $\setQj$,
and must be solved for self-consistently.
Non-native ruggedness introduces a term with
anomalous $1/T^2$ temperature dependence in the distribution. By the
structure of eq.~(\ref{Qstar}), all
contact probabilities $\Qi$ are between zero and one.

The Lagrange multiplier $\mu$ is determined by the constraint
$\sum_i \Qi^{\star} = M Q$, and so is a function of $Q$ and the distributions
of $\setei$ and $\setli$. It can be interpreted as proportional to an
effective force along the  $Q$ coordinate since 
\be
\mu = - \frac{1}{M} \frac{\D F}{\D Q} 
\label{eqmu1}
\ee
by the properties of the Legendre transformation (see
Appendix A). Thus again since the free energy $F$ is of course a
function of $Q$ and the 
distributions $\setei$ and $\setli$, $\mu = - (1/M) \D F/\D Q$ is also.

The second variation of $F\left( \setQi | \setei ,\setli
\right)$ (neglecting terms of order ${\cal O} \left( 1/M \right)$)
is indeed positive
\be
\frac{\D^2 F}{\D Q_j \, \D Q_i} = \lambda T \frac{\delta_{ij}
}{\Qi \left( 1 - \Qi \right)} > 0 \: ,
\ee
verifying that the extremal values of $\Qi$ are the ones which minimize
$F\left( \setQi | \setei ,\setli \right)$.

\section{Changing folding mechanisms by tailoring native interaction
energies and altering native structural motifs} 
\label{section:opt}

Most single domain proteins most fold over a free energy barrier of a
few $\kB T$ at the transition temperature. This barrier is small
compared to the total thermal energy in the system, reflecting the
exchange of energy for entropy as a protein
folds~\cite{Hao94,PlotkinSS97}. However the barrier height can vary
significantly depending on which parts of a protein are most stable in
the native structure, i.e. how the native energy is distributed
throughout the native structure.  In section~\ref{sect:egiven} we look at the
effects on the thermodynamics when native interactions are changed in
a controlled manner.  We find that a distribution of native energy
which induces a uniform folding mechanism will maximize the
barrier. For model systems of small proteins this barrier is about
twice as large as the barrier when native energies are uniformly
distributed.  Increasing heterogeneity in the folding mechanism
systematically decreases the folding barrier and may eliminate it
entirely, at least in the absence of cooperative interactions. The
corresponding folding rate increases, as long as the protein remains
well-designed. In section~\ref{sect:structandE} we develop a
perturbation expansion of the free energy to incorporate structural as
well as energetic heterogeneity, and the effect on the free energy of
correlations between them.  In~\ref{sect:ill} we further apply the
functional theory to calculate various thermodynamic quantities and
compare the results with simulations of the lattice $27$-mer, and
in~\ref{sect:mech} we continue the comparison by applying the
functional theory to properties of the folding mechanism.  In
section~\ref{sec:routmeas} we show how the folding barrier decreases
with the degree of route-like folding in the system, so long as the
protein remains well-designed.  In~\ref{sect:kinetics} we explicitly
investigate the kinetics of folding times in the system and find that
folding kinetics is well-characterized by the thermodynamic folding
barrier.  In~\ref{sect:mean} we illustrate the effect of contact order
or mean contact length on the folding barrier in the model.
In~\ref{sect:structvar} we investigate the effects of structural
variance on a hypothetical ensemble of well-designed protein fold
motifs. We find that for fixed average loop length $\lbar$, native
structures that have larger dispersion $\delta \ell$ in the
distribution of return lengths tend to have smaller folding barriers.
Finally in section~\ref{sect:match} we show how native energies can be
tuned or native structures can be sought to match a desired free
energy potential, which we illustrate for a simple two-state potential
as well as a potential with an on-pathway intermediate.

\subsection{Energetic heterogeneity for a given structure}
\label{sect:egiven}
\setcounter{equation}{0}

First we consider the free energy as a function of $Q$ and the field of
energies $\setei$, given the field of loop lengths $\setli$.
Each contact probability $\Qi$ in the free energy (eq.~\ref{FF}) is 
considered through eq.~(\ref{Qstar}) to be a function of $Q$, its energy,
its loop length, and the Lagrange multiplier $\mu (Q,\setej,\setlj )$,
which is itself a function of $Q$ and the distributions $\setej$ and
$\setlj$. Thus the free energy depends both implicitly and explicitly
on $\setei$.

We now seek to relax the values of $\setei$, at fixed stability (fixed
total native energy)
\bea
\sum_j \ej  &=& \En \: , 
\label{constraintE} \\
\sum_j \dej &=& 0
\label{eq:dezero}
\eea
to the distribution $\left\{
\ei^{\star}\left(\setlj\right)\right\}$ that extremizes the free energy
barrier.
Under variations of the energies $\left\{ \delta \ei \right\}$ for a
given structure $\setli$, the free energy becomes
\be
F\left\{ \eio+\delta \ei\right\} = F\left\{ \eio\right\} + \sum_i
 \left(\frac{\delta F}{\delta \ei}\right)_{\eio} \delta\ei +
\frac{1}{2!}\sum_{i,j} \left(\frac{\delta^2
F}{\delta\ei\delta\ej}\right)_{\eio,\ejo} \delta\ei \, \delta\ej +
\ldots \: ,
\label{taylor}
\ee
where $\delta/\delta \ei$ is the total derivative with respect to $\ei$.
So the distribution $\left\{
\ei^{\star}\left(\setlj\right)\right\}$ that extremizes the free
energy barrier subject to the constraint eq.~(\ref{constraintE})
is the solution of $\delta (\DF^{\ddag}-p\sum_j\ej ) =
0$, or 
\be
\sum_i \left[ \frac{\delta \DF^{\ddag}}{\delta\ei} -p\right]\dei =0
\label{eq:mm}
\ee
for arbitrary and independent variations $\dei$ in the energies.
The Lagrange multiplier $p$ imposes the constraint that the total
native energy $\En$ is constant.
Changes in the barrier height are roughly equal to changes in the free
energy at the barrier peak, since the free energy in the unfolded
state $Q_o \approx 0$ is more weakly dependent on $\setei$, i.e.
$\delta \DFdag/\dei \cong \delta F(\Qddag)/\dei$, because 
$\delta F(Q_o\approx 0 )/\dei \cong 0$; less native structure is
present in the unfolded state. The effect on the free energy of
perturbations in $\setei$ is largest at intermediate $Q$; there is no
effect at the end points because at $Q=0$ there are no native
interactions, and at $Q=1$ all native interactions are present and
must add up to the total native stability $\En$, which is fixed. In
fact in the equations for the free energy perturbation this effect is
manifested by the factor of $Q (1-Q)$ which multiplies every term, see
e.g. eq.s~(\ref{df22}), (\ref{pertF}), and~(\ref{eq:dqdeqq}).

Because of the implicit functions mentioned above
\bea
\frac{\delta F}{\delta\ei} &=& \frac{\D F}{\D\ei} +
\sum_j \frac{\D F}{\D\Qj} \frac{\D\Qj}{\D\ei} + \sum_j \frac{\D
F}{\D\Qj} \frac{\D\Qj}{\D\mu} \frac{\D\mu}{\D\ei} 
\label{totder} \\
&=& \frac{\D F}{\D\ei} + \mu \sum_j \left[ \frac{\D\Qj}{\D\ei} +
\frac{\D\Qj}{\D\mu} \frac{\D\mu}{\D\ei}  \right] \: .
\label{tot2}
\eea
However the term in square brackets is just the
total derivative $\dQj/\dei$, so the sum vanishes because $Q$ is a
fixed parameter independent of $\ei$~\cite{note:lag}:
\be
\sum_j \frac{\dQj}{\dei} = \frac{\delta}{\dei} \sum_j \Qj = 
\frac{\delta}{\dei} \left( MQ\right) = 0 .
\ee
Differentiating eq.~(\ref{f}) immediately yields:
\be
\frac{\D \Delta F^{\ddag}}{\D \ei} \cong \Qi (\Qddag)\: ,
\label{eqphi}
\ee
so the perturbative change in the free energy barrier by varying a
contact's energy is equal to the probability that contact was formed
at $\Qddag$ (c.f. eq.s~(\ref{dqde}) and~(\ref{number}) ).

This is strongly related to experimental
$\phi_i$ values, which measure
the change in the log rate after mutation over the change
in difference in equilibrium populations of the folded and unfolded
states~\cite{FershtAR92,Itzhaki95}. When the prefactor to the rate is unaffected
by the mutation this is equivalent to the change with mutation in the
barrier height over the change in the difference of the free energy
minima~\cite{FershtAR92,Onuchic96}, which we refer to as $\phi'$:
\be
\phi_i' \equiv \frac{\left(\D F^{\ddag}/\D \ei\right) 
- \left(\D F_u/\D \ei\right)}{ \left(\D F_f/\D
\ei\right) - \left(\D F_u/\D \ei\right) } = 
\frac{ \Qi \left(\Qddag\right) - \Qi \left( \Qu
\right)}{\Qi\left(\Qf\right) - \Qi \left( \Qu \right)} \: .
\label{eqphi1}
\ee
When the nativeness in the unfolded state can be neglected, $\Qi(\Qu)
\approx 0$, and when the native contacts in the folded state are
essentially  fully formed,  $\Qi (\Qf) \approx 1$. Then
eq.~(\ref{eqphi1}) becomes
\be
\phi_i' \equiv \frac{\delta \Delta F^{\ddag}}{\dei} = 
\frac{\D \Delta F^{\ddag}}{\D \ei} = \Qi (\Qddag)\: .
\ee
Equating $\phi$ values with contact probabilities assumes that contact
probability 
be used as a kinetic reaction coordinate. In fact it has been observed in
simulations that $\phi$ values correlate with $\Qi$ values as well as
any other reaction coordinate currently proposed~\cite{NymeyerH00:pnas}.

Continuing now to find the energies $\eistar$ which extremize the free
energy, eq.~(\ref{eq:mm}) gives finally: $\Qi 
(\Qddag,\mu^{\ddag}=0,\eistar,\li) = p$:
the free energy is extremized when all the
$\Qi$ values are tuned to the same number at the barrier peak. This
folding scenario is that of a symmetric funnel: the
protein is equally likely to order from any place within it.
Thus since $\sum_i \Qi = M Q$,
\be
\Qi\left(\Qddag, \mu^{\ddag}=0,\eistar,\li \right) = \Qddag \: .
\label{qh}
\ee
Solving eq.~(\ref{qh}) for the energies using eq.~(\ref{Qstar}) gives
\be
\eistar = T \si + T \frac{d}{dQ}\left( \frac{}{}\lambda \left[-Q\ln Q -
(1-Q)\ln(1-Q) \right]\right)_{\Qddag} - \frac{b^2}{2T} \: .
\label{eim}
\ee
Subtracting $\ebar$ from $\ei$ by averaging  eq.~(\ref{eim})
yields: 
\bea
\eistar - \ebar &=&  T\left( \si - \overline{s} \right)
\nonumber \\
&=& -\frac{3}{2} T \left( \ln \li - \overline{\ln \ell}\right)
\label{eq:el}
\eea
where eq.~(\ref{leff}) was used. The free energy fluctuations $\dfi
=0$ in a uniform folding mechanism.
Thus contacts pinching off longer
loops ($\li \gtrsim \overline{\li}$) have lower (stronger) energies ($\ei <
\overline{\ei}$) to make all the contact probabilities equal at the
barrier peak~\cite{note:note}.
If correlations
between contacts are fully accounted for, the $\Qi$ values deviate
slightly from $Q$ away from the barrier peak, but the fluctuations
away from uniform ordering are still strongly suppressed (see
fig.~\ref{fig:QivsQroute}C ).

Evaluating the second derivative stability matrix in
eq.~(\ref{taylor}) shows 
$\Qi = \Qddag$ in eq.~(\ref{qh}) to be an unstable
maximum, as follows. From eq.~(\ref{eqphi})
\bea
\frac{\delta^2
F}{\dej\dei} = \frac{\delta \Qi}{\dej} &=& \frac{\D\Qi}{\D\ej}
+\frac{\D\Qi}{\D\mu} \frac{\D\mu}{\D\ej} \nonumber \\
&=& -\frac{\Qi\left(1-\Qi\right)}{\lambda T} \left( \dij +
\frac{\D\mu}{\D\ej} \right)
\label{d2fde}
\eea
by eq.~(\ref{Qstar}).
Thus the second order change in the free energy at the extremum is
\be
\sum_{i,j} \left( \frac{\delta^2 \DFdag}{\dej \dei} \right)_{\eistar,
\ejstar} \dei \dej = - \frac{\Qdag \left( 1-\Qdag\right)}{\lT} \left[
M \,\overline{\de^2} + \sum_{i,j} \left( \frac{\D\mu}{\D\ej}\right)
\dei\dej \right] .
\label{dblesum}
\ee
Since the perturbations $\dei$ are independent, cross terms in the
double sum of eq.~(\ref{dblesum}) vanish, making the sum equal to 
\be
\sum_{i=1}^{M} \left(\frac{\D\mu}{\D\ei}\right) \dei^2 \: .
\label{negterm}
\ee
This term is negligible for the following reasons.
First, note that $\D F/\D\ei = \Qi$ is $ \sim {\cal O}(1)$. Then since
$\D \mu/\D\ei = -(1/M)(\D/\D Q)(\D F/\D\ei)$ by
eq.~(\ref{eqmu1}), the terms $\D \mu/\D\ei$ in eq.~(\ref{negterm}) are
$\sim {\cal O}\left(1/M\right)$. So the sum of $M$ terms
in~(\ref{negterm}) is 
$\sim {\cal O} \left(1\right) \overline{\de^2}$, whereas the first
term in eq.~(\ref{dblesum}) is $\sim {\cal O} \left(M\right)
\overline{\de^2}$ and dominates in the thermodynamic limit.
Thus to order ${\cal O}(1/M)$:
\be
\left(\frac{\delta^2
\DFdag}{\delta\ej\delta\ei}\right)_{\ei^{\star},\ej^{\star}} = - \dij
\frac{\Qddag (1-\Qddag)}{\lambda^{\ddag} T}  
\label{2nd}
\ee
which is clearly negative, meaning that tuning the energies so that
$\Qi = \Qddag$ maximizes the free energy at the barrier peak.
This is consistent with the intuitive arguments in
section~\ref{sec:add}. Recall for example in section~\ref{sec:landau} we showed
using thermodynamic perturbation theory that random perturbations to
the contact energies always lowered the free energy barrier.

Substituting eq.s~(\ref{eqphi}), (\ref{qh}), and~(\ref{2nd})
into~(\ref{taylor}) gives
\be
\Delta F^{\ddag}\left\{ \eistar+\delta \ei\right\} \cong
\Delta F_{\mbox{\tiny MF}}^\ddag - M \frac{\Qddag (1-\Qddag)}{2
\lambda^{\ddag} T} \: \overline{\de^2} \: .
\label{df22}
\ee
It should be noted this expression is very similar to eq.~(\ref{DFrem})
obtained using a simple REM argument, and also to eq.s~(\ref{landau})
and~(\ref{nocross}) using thermodynamic perturbation theory. The only
major difference here 
is the factor of $\lambda^{\ddag}$ which accounts for the reduction in
route entropy due to chain connectivity.

For an energetic standard deviation of about a $\kB T$ from the
optimal distribution,  
the barrier goes down by about $\sim N \kB T/2$ (we've let $M\approx 2
N$, $\lambda^{\ddag} \approx 1- \Qddag$ since the exponent $\alpha$
in~(\ref{eq:lambdaQ}) is about $1$, and $\Qddag \approx 1/2$).
The barrier governed rate
increases with native energetic heterogeneity as
\be
k = k_o \exp\left(-\frac{\Delta F^\ddag}{T}\right) = \kH
\exp\left(\Qddag (1-\Qddag) \frac{M\overline{\de^2}}{2 \lambda^{\ddag}
T^2}\right) 
\label{rateFunc}
\ee
which should be compared with eq.~(\ref{rate1}).

\subsection{Including structural heterogeneity}
\label{sect:structandE}

The theory also allows us to investigate the effects of native
structural variance on the barrier, as well as the correlations between
structure and energetics.
A perturbation analysis shows that structural variance lowers the
barrier as discussed in section~\ref{sec:of}, and that entropically
likely contacts  should be made 
stronger to lower the barrier, as discussed in
section~\ref{sec:likely}. 
In the model, entropically likely contacts are short-ranged. However
they may occasionally be long-ranged when entropy is more precisely
accounted for by accurately accounting for  correlations between
contacts (see figure~\ref{fig:simthrytest}).

Consider perturbing the  free energy
of a homogeneous system with $\li =\lbar$, $\ei=\ebar$,
$\Qi =\Qddag$, by letting $\li = \lbar+\dli$ and
$\ei=\ebar+\dei$. Then,
\bea
&&\DF^{\ddag}\left\{ \ebar+\dei, \lbar+\dli \right\} =
 \DF^{\ddag}_{\mbox{\tiny MF}}\left\{ \ebar, \lbar \right\}
+ \sum_i  
\left(\frac{\delta \DF^{\ddag}}{\dei}\right)_{\ebar,\lbar} \dei
+ \sum_i  
\left(\frac{\delta \DF^{\ddag}}{\dli}\right)_{\ebar,\lbar} \dli 
\nonumber \\
&&+ \frac{1}{2!}\sum_{i,j} 
\left(\frac{\delta^2 \DF^{\ddag}}{\dei\dej}
  \right)_{\ebar,\lbar} \dei \dej 
+ \frac{1}{2!}\sum_{i,j} \left(\frac{\delta^2 \DF^{\ddag}}{\dli\dlj}
  \right)_{\ebar,\lbar} \dli \dlj 
+ \frac{1}{2!} \sum_{i,j} \left(\frac{\delta^2 \DF^{\ddag}}{\dli\dej} 
  \right)_{\ebar,\lbar}\dli \dej  \ldots \: .
\label{tel}
\eea
The first term in the expansion 
$\DF^{\ddag}_{\mbox{\tiny MF}}\left\{ \ebar, \lbar \right\}$ is
the mean-field free energy eq.~(\ref{FMF}). The second term 
is zero at the extremum where $\Qi=\Qddag$ by eq.s~(\ref{eqphi}),
(\ref{qh}) and~(\ref{eq:dezero}), and the fourth term is 
given in eq.~(\ref{df22}). The calculation of the third term proceeds
along the same lines as the derivation of eq.~(\ref{eqphi}). Like
eq.~(\ref{tot2}), $\delta \DF/\delta \li$ contains a term involving 
an explicit 
derivative of $\li$, and implicit derivatives
which are identically zero.
The explicit term itself vanishes when evaluated for homogeneous
fields. From eq.~(\ref{FF}):
\be
\left(\frac{\delta \DF^{\ddag}}{\dli}\right)_{\ebar,\lbar} = \frac{3}{2} T
\left(\frac{\Qi - \Qddag}{\li}\right)_{\ebar,\lbar} = 0\: .
\label{dfdl}
\ee
Calculation of the fifth term involves calculating $\delta \Qj/\dli$,
which proceeds analogously to the derivation of eq.~(\ref{2nd}) 
via~(\ref{d2fde}):
\be
\frac{\delta \Qj}{\delta \li} \cong -\dij \, \frac{3}{2}
\frac{\Qj\left(1-\Qj\right)}{\lambda \lj}
\label{dqdl}
\ee
which is again diagonal and negative as is eq.~(\ref{2nd}); raising
the energy of a contact or 
increasing its loop length decreases that contact's probability of
formation.
From eq.s~(\ref{dfdl}) and~(\ref{dqdl}), the fifth and sixth terms  in
eq.~(\ref{tel}) can be calculated, yielding
\bea
\DF^{\ddag}\left\{ \ebar+\dei, \lbar+\dli \right\} &=&
\DF^{\ddag\, 0}\left\{ \ebar, \lbar \right\} 
- M \frac{\Qddag \left( 1 - \Qddag\right)}{2 \ldag T} \:
\overline{\de^2} \nonumber \\
&-& M T \frac{9}{8} \frac{\Qddag\left( 1 - \Qddag\right)}{\ldag}
\frac{\overline{\dl^2} }{\lbar^2}
- M \frac{3}{4} \frac{\Qddag\left( 1 - \Qddag\right)}{\ldag}
\frac{\overline{\dl \de}}{\lbar} 
\label{pertF}
\eea
The third term in eq.~(\ref{pertF}) indicates that structural
dispersion also lowers the barrier, in agreement with the arguments
given in sect.~\ref{sec:of}. The fourth term indicates that the
free energy barrier is additionally lowered in the model when shorter range 
contacts become stronger energetically ($\dli <0$ and $\dei <0$) or
longer range contacts become weaker energetically ($\dli >0$ and
$\dei >0$). This means in general that the free energy is additionally lowered
when fluctuations are correlated so as to further increase the variance in
contact participations, in agreement with the intuitive arguments
given in section~\ref{sec:likely}. Note again that all reductions in
free energy due to structural and/or energetic heterogeneity scale
extensively with system size.

\subsection{Illustrations using a lattice model protein and functional theory}
\label{sect:ill}

We illustrate in 
figures~\ref{fig:Eplot}--~{\hspace{-0.1cm}\ref{fig:stotflory}}
the general effects of introducing heterogeneity to a model system.
Results are shown for a G\={o} protein, modeled with the free energy
functional theory using reasonable parameters and loop-length
distribution given in table I, and also
simulated on a lattice (see also reference~\cite{PlotkinSS00:pnas}
for a concise treatment of some of these phenomena).
The lattice protein is a chain of $27$ monomers constrained to reside
on vertices of a three-dimensional cubic lattice (see
fig.~\ref{fig1}). Details of the model 
and its behavior can be found in
ref.s~\cite{LeopoldPE92,SaliA94:nat,SocciND95:jcp,AbkevichVI95,SocciND96:jcp}.
Monomers have non-bonded contact interactions with a G\={o} potential
(native interactions only). Corner, crankshaft, and end moves are
allowed. Free energies and contact probabilities are obtained by
equilibrium Monte Carlo sampling using the histogram
method~\cite{SocciND95:jcp}. Sampling error is $< 5 \%$.
Coupling energies were chosen by first running a
simulated annealing algorithm to find the set of native energies $\{
\ei^\star \}$ that makes all the contact probabilities equal at the
barrier peak: $\Qi(\{ \ei^\star \}) = \Qddag$ (c.f. table I for the
energies which induce uniform folding for the native structure in
fig.~\ref{fig1}). Energies 
are always constrained to sum to a fixed total native energy: $\sum_i
\ei = M \ebar$, i.e. overall stability is fixed.
Energies are relaxed from $\{ \ei^\star \}$  by letting  
\be
\ei = \eistar + \alpha ( \ebar - \eistar ) \: . 
\label{Etune}
\ee
For the simulation results, the values $\alpha =
0$, $1$, $1.35$, $2.05$ were used in the figures, and for the
theoretical results, the values $\alpha = 0, 0.5, 0.65, 1.9$ are
used. When $\alpha = 0$,
$\setei = \seteistar$ and all the contact probabilities are the same
at the barrier peak; the folding
barrier is maximized. When $\alpha
= 1$, all the energies are equal ($\setei = \{ \ebar \}$); the entropy
at the barrier peak is maximized. When
$\alpha \approx 1.35$ ($\alpha \approx 0.65$ in the theory) the free
energy barrier vanishes, and when 
$\alpha \approx 2.05$ ($\alpha \approx 1.9$ in the theory), folding
occurs by only a few routes- 
essentially one route through the transition state.
The discrepancy between the theory and the simulation for the values
of $\alpha$ required to produce a given 
folding mechanism is most likely because the theory for loop
closure,  eq.s~(\ref{sbond}) and~(\ref{leff}) overestimates the
dispersion in entropy losses for contact formation. 

In the figures to follow we compare
thermodynamic quantities and folding mechanisms for various energy
functions, consistently using the following convention; thin solid:
$\asim = 0$, $\athry = 0$; thick solid: $\asim = 1.0$, $\athry = 0.5$;
long dashed: $\asim = 1.35$, $\athry = 0.65$; short dashed: $\asim =
2.05$, $\athry = 1.9$.

\subsubsection{Total Energy}
\label{sect:Esect}

Figure~\ref{fig:Eplot} shows a plot of the total energy as a function
of $Q$ obtained from lattice simulations of a G\={o} model to the structure
shown in fig.~\ref{fig1}.  When native energies are
uniform, the total energy decreases linearly with $Q$, since the total
energy (eq.~(\ref{ET}) with $b=0$) is $\sum_i \Qi \ei = \ebar \sum_i
\Qi = \En Q$. When energies are tuned so that contact probabilities at
the barrier peak are all equal ($\Qi (\eistar, \Qdag) = \Qdag$, see
eq.~(\ref{qh}) ), the total energy at the barrier peak position is
equal to the total energy when $\ei = \ebar$, since $\sum_i \Qi(\Qdag)
\ei = \Qdag \sum_i \ei$ which again equals $\En \Qdag$. So as the
coupling energies are relaxed from $\seteistar$ to $\setebar$ in
eq.~(\ref{Etune}), the 
energy at the barrier peak hardly changes, as can be seen from the
figure (shifting of the peak position is a small effect).  Further
perturbing the coupling energies results in an overall decrease in
total energy. Temperatures are adjusted to $\Tf$ for all curves in
fig.~\ref{fig:Eplot}, however $\Tf$ is roughly constant for the upper
$3$ curves (see fig.s~{\ref{fig:sdplot} and~\ref{fig:TTFF}); for the case when the
energies are tuned to induce route-like folding: $\setei =
\seteir$, the folding temperature is about a factor of $6$ lower.
While heterogeneity effects on the total energy in
fig.~\ref{fig:Eplot} and entropy in
fig.~\ref{fig:Splot} below may appear fairly small, the effect on the
barrier and corresponding rate may be large, since the barrier arises
from the cancellation of these two large terms, and the rate is
proportional to the exponential of this difference.

\subsubsection{Total Entropy}
\label{sect:totent}

Figure~\ref{fig:Splot} shows a plot of the total entropy as a function
of $Q$ from simulations of the G\={o} model mentioned above.  When the
native energies are all the same, $\setei = \setebar$ (maximal solid
curve in fig.~\ref{fig:Splot} ), the entropy is larger than for any
other distribution  as shown in section~\ref{sect:ebar}.
We derive this result now using the free energy functional theory.

First, the total entropy at $Q$ depends on the relative occupation
probabilities and so does not depend explicitly on the total
energy. 
One can see this by taking the derivative:
\be
\frac{\D S}{\D \En} = \frac{\D S}{\D \left(\sum_i \ei\right)} = 
\frac{1}{M} \sum_i \frac{\D S}{\D \ei} = 0
\ee
since the entropy does not depend explicitly on the coupling energies
($\D S /\D \ei = 0$).
The last equation follows from the properties of the directional
derivative (c.f. Appendix~A).
This independence of canonical entropy on total energy is analogous to
the property that the Lagrangian for a system of 
particles not under any external forces is independent of the center
of mass coordinate. One can solve Newton's equations without the
constraint and obtain a zero frequency normal mode corresponding to
the whole system in uniform motion. Keeping the constraint introduces
a new equation and unknown (the Lagrange multiplier), and solving the
eigenvalue problem yields that the multiplier is zero. 
Another example is a thermal bath of photons, which has no constraint
on the total number of photons. Thus the Lagrange multiplier for this
constraint- the chemical potential, is zero.
The Lagrange constraint here 
appears in the form $p \sum_i \ei$ as in eq.~(\ref{eq:mm}), which has
the form of a Legendre transform with the Lagrange multiplier $p$
being proportional to $\D S/\D \En$, just as the Lagrange multiplier
$\mu$ in eq.~(\ref{eqmu1}) is proportional to $\D F/ \D (MQ)$. So the
multiplier $p$ is identically zero.

We take the extremum of the entropy, now explicitly introducing the
constraint that $Q$ is fixed:
\be
\delta \left[ S + \nu \sum_j \Qj \right] = \sum_i \left[ \frac{\d
S}{\d \ei} + \nu \frac{\d \Qj}{\d \ei}\right] \d \ei = 0
\label{eq:Sext}
\ee
for arbitrary and independent variations $\dei$. Since the entropy
does not depend explicitly on $\ei$, 
\be
\frac{\d S}{\d \ei} = \sum_j \frac{\D S}{\D \Qj}\frac{\d \Qj}{\d \ei} \: .
\ee
Using $S = (E-F)/T$, and eq.s~(\ref{ET}) and~(\ref{varQi}),
\be
\frac{\D S}{\D \Qj} = \frac{\ej}{T} + \frac{\mu}{T} \: .
\label{dsdqj}
\ee
Now by the properties of the Legendre transform, 
\be
\nu = - \frac{1}{M} \frac{\D S}{\D Q}
\ee
see the analogous eq.~(\ref{force}). Again using $S=(E-F)/T$,
eq.~(\ref{force}), and the directional derivative on $E$
(c.f. eq.(\ref{direct}) ), we obtain
\be
\nu = - \frac{\ebar}{T} - \frac{\mu}{T} \: .
\label{eqnu}
\ee
Putting eq.s~(\ref{dsdqj}) and~(\ref{eqnu}) into eq.~(\ref{eq:Sext})
gives 
\be
\sum_j \left[\frac{\ej - \ebar}{T} \right] \frac{\d \Qj}{\d \ei} = 0
\label{Sext2}
\ee
and using eq.(\ref{d2fde}), i.e. 
\be
\frac{\d \Qj}{\d \ei} = - \dij \frac{\Qj \left( 1 -\Qj\right)}{\lT}
\label{eq:dqdeqq}
\ee
eq.~(\ref{Sext2}) becomes 
\be
\left(\ei - \ebar\right) \frac{\Qi \left( 1 -\Qi\right)}{\lT} = 0 \: ,
\label{mineq}
\ee
whose only solution is $\ei = \ebar$, i.e. all the coupling energies
are equal. Taking the second derivative yields
\be
\left(\frac{\delta^2
S}{\delta\ej\delta\ei}\right)_{\ei=\ebar,\ej=\ebar} = - \dij 
\frac{\Qi (1-\Qi)}{\lT}  
\label{2ndS}
\ee
which is negative indicating the extremum is a maximum. Thus the
thermal entropy is maximized for uniform coupling energies, for
proteins well-designed enough to be modeled by G\={o}-like models. 

Keeping the Lagrange constraint $\sum_i \ei = \En$ modifies
eq.~(\ref{mineq}) to
\bea
\dei \, \ai + p &=& 0 \;\;\;\;\;\;\; 1 < i < M  \nonumber \\
\sum_i \dei &=& 0
\label{L1}
\eea
where $\ai \equiv \Qi ( 1 -\Qi)/\lT$ and $p$ is the undetermined
multiplier. Equations~(\ref{L1}) constitute $M+1$ equations for $M+1$
unknowns (the $\dei$ and $p$). 
The solution then amounts to finding the determinant
of the matrix
\be
\mbox{Det} \left[
\begin{array}{cccccc}
\alpha_1  & 0           & \ldots & 0 & 1 \\
0       & \alpha_2 & \ldots & 0 & 1 \\
\vdots  & \vdots   & \ddots &  \vdots & \vdots \\
0       &   0          & \ldots &  \alpha_{\mbox{\tiny M}}& 1 \\
1 	& 1	    & \ldots & 1 &  0
\end{array} \right] = 0
\ee
which is readily evaluated, and consists of $M$ negative terms, which
are all possible ways to pick $M-1$ of the $M$ $\ai$'s, e.g. if $M=4$,
the determinant is $-\alpha_1\alpha_2\alpha_3 -
\alpha_1\alpha_2\alpha_4 - \alpha_1\alpha_3\alpha_4 -
\alpha_2\alpha_3\alpha_4 $. Thus the determinant is never zero, so
there is no solution other than the trivial solution, where all $\dei
= 0$, which is again the condition that $\ei = \ebar$.
Putting this condition in eq.s~(\ref{L1}), we again recover that that
the Lagrange constraint $p=0$.

\subsubsection{Free Energy}
\label{sect:F}

Figure~\ref{fig:Fplot} shows the total free energy $F(Q)$ in units of
$\ebar$ for the simulations and functional theory, for various energy
functions.  For the simulations $F(Q)$ is obtained from
$E(Q)-\Tf(\setei) S(Q)$ where $E(Q)$ is shown in
fig.~(\ref{fig:Eplot}) and $S(Q)$ shown in fig.~(\ref{fig:Splot}).
The transition temperature $\Tf(\setei)$ is defined as the temperature
giving a native ($Q=1$) occupation probability of $50\% $ (see
fig.~\ref{fig:sdplot}).  For the theory $F(Q)$ is obtained from
eq.s~(\ref{FTOT}), (\ref{FMF}), and~(\ref{FF}), and the transition
temperature $\Tf$ is defined as the temperature where the probability
for $Q>0.9$ is $50\%$ (the transition temperature is not strongly
dependent on this cutoff).  For $\eistar$ (thin solid curve) the
folding barrier is maximized, see eq.~(\ref{2nd}). Here the profile is
largely a function only of the mean loop length $\lbar$ and system
size $N$ (see eq.~(\ref{FMF})), since most structural features are
tuned away by adjusting the native energies. The only other residual
structural features which remain are the effects of native structure
on the exponent $\alpha$ in the route entropy reduction,
eq.~(\ref{smix}}), which is probably a small effect compared to the
effects we consider here.  For uniform native energies, the transition
state energy hardly changes as explained in section~\ref{sect:Esect},
but the entropy increases to its maximal value as explained in
section~\ref{sect:totent}. So the barrier height initially decreases
for entropic reasons (see also fig.s~\ref{fig:sdplot}
and~\ref{fig:FvsR}).  For the lattice model the barrier is reduced by
about a factor of $2$, (thick solid in upper plot) and for the theory
the barrier disappears entirely at the transition temperature, the
curve residing between the long-dashed and short-dashed curves in the
lower plot.  The thick solid curve in the theory plot is for about
half the energetic dispersion required to tune the system to
homogeneous folding; this results in the same reduction in folding
temperature as the simulations. Further perturbing the energies to
$\seteio$ eliminates the barrier entirely at the transition
temperature making the transition second order (long dashed curves).
For the simulations, between $\setebar$ and $\seteio$ the barrier
decreases because the energy at the transition state lowers while the
entropy doesn't change to first order. In the theory there is
sufficient entropic variance to kill the barrier before relaxing all
energies to a uniform distribution $\setebar$, however $\setei$ for
the long dashed line in the theory has a relative variance $\de/\ebar$
of only about $0.15$ compared to about $0.47$ for the maximum barrier.
Cooperative
effects~\cite{KolinskiA96:prot,PlotkinSS97,ShoemakerWang99,EastwoodMP00}
or entropic reduction arising from the coupling of chain stiffness
with folding~\cite{PlotkinSS00:un} restores a barrier for uniform
native energies.  As energies are further perturbed to a distribution
$\seteir$ causing folding to occur by a single dominant route (short
dashed), folding becomes strongly downhill at the the transition
temperature, which drops sharply by about a factor of $6$.  For a
system with non-native as well as native interactions, the free energy
would be less downhill and much more rugged at these
temperatures. Folding would be exceedingly slow because the protein
would spend a long time in individual traps.

Figure~\ref{fig:sdplot} shows the total barrier height
$\Delta F^{\ddag}$, in units
of the mean native 
contact strength $\ebar$, vs. the RMS native energetic variance in the same
units. Plotted are results from simulation of a lattice $27$-mer
G\={o} model; the analytic theory gives
qualitatively the same results (see fig.~\ref{fig:FvsR}). Native
interactions $\setei$
are tuned to a distribution $\{\ei^{\star}\}$ that symmetrizes the
funnel to uniform ordering  
at the barrier peak, giving the largest
barrier at the upper right of the figure (see eq.~\ref{2nd}). 
The energies here are
anticorrelated with their loop lengths in that more
negative energies are required for the longer loops to have equal free
energies (see eq.~\ref{eq:el}). A large dispersion in interaction
strengths is required to achieve this scenario, which may be
impossible to achieve in practice (see table~I).
Interactions are then uniformly relaxed via eq.~(\ref{Etune}) to
$\{ \ei \} = \{ \ebar \}$ (i.e. $\dei = 0$) and the barrier is reduced
by a factor of $2$, due to entropic gains in the transition state
ensemble (TSE) (see eq.~(\ref{2ndS})). The $\ei$ are then
continued to evolve - now $\dei$ begins to 
correlate with loop length and anti-correlate with contact probability
(c.f. eq.~\ref{Etune} ), which further increases the heterogeneity. 
The barrier decreases now because energetic gains win over entropic
losses in the TSE, and eventually the barrier vanishes at coupling
energies  $\{ \ei^o \}$. We should emphasize here that $\{ \ei^o \}$
is not unique; many sufficiently heterogeneous distributions of native
stability may kill the barrier.
All the while the folding temperature $\Tf$ (dashed curve) defined here as
the temperature where the native state at $Q=1$ is $50\%$ occupied,
decreases by only $10\%$ and remains well above the putative estimate
of the glass temperature $\Tg \cong \Tf/1.6$ (thin horizontal
dot-dashed line at lower right). 
The fact that $\Tf/\Tg$ was roughly constant as native interactions
were varied indicates that the prefactor to the folding rate, $k_o$ in
eq.~(\ref{eq:k1}) is roughly constant (see sect.~\ref{sect:kinetics}). Folding
rates are governed largely by 
the free energy barrier in this regime (see fig.~\ref{fig:tau} below) and can
be treated with thermodynamic analysis.
Because the entropy of the bottleneck is not constant as $\de/\ebar$
is varied, another valid measure of $\Tf/\Tg$ for this 
system is $\Tf$ times the root entropy per residue at the bottleneck
(thin solid line with squares). This measure actually shows a slightly
increasing 
trend as $\ei \rightarrow \ebar$ and only slight decrease as $\{ \ebar
\} \rightarrow \{ \ei^o \}$.

\subsection{Folding Mechanisms: Route Entropy, Halo Entropy, and
Contact Probability} 
\label{sect:mech}

We continue by investigating different possible folding mechanisms to
the same structure, induced by different distributions of native
stability.So far we have seen that in G\={o}-like models of
well-designed proteins, governed by an energy function with pair
interactions, the folding mechanism to a given structure may involve a
barrier of various heights (fig.~\ref{fig:Fplot}) while the total
entropy and energy are relatively unaffected (figs.~\ref{fig:Eplot}
and~\ref{fig:Splot}), depending on how native interactions are
distributed throughout the protein. The barrier is sensitive to
perturbations because it arises from the cancellation of large terms:
the total entropy and energy~\cite{Hao94,PlotkinSS97}.
We have seen so far that it is quite difficult to tune away all the effects
of native topology, or induce folding through one or a few routes.
Now we investigate some of the properties of partially native states
when folding occurs by various mechanisms. We again vary the
distribution of native energies by varying values of $\alpha$ in
eq.~(\ref{Etune}).  

Figure~\ref{fig:sroutehet} shows the route entropy ${\cal
S}_{\mbox{\tiny ROUTE}}$ over the number of contacts $M$ as a function
of $Q$, for the $27$-mer lattice model (Top) and for the functional
theory (Bottom). In both cases eq.~(\ref{smix}) is used to calculate
the mixing entropy from the contact probabilities $\setQi$.  For
native contact distribution $\seteistar$ which induces homogeneous
folding through the barrier peak (thin solid), the curve essentially
reproduces that in fig.~\ref{fig:mix}. This distribution maximizes the
route entropy. For a uniform distribution of native energies (Top
figure, thick solid) there is a reduction from the homogeneously
ordering case solely due to the topology of the native
structure. Different native structures will have different
characteristic curves.  For the theory curves, the same native energy
distributions used above in fig.~\ref{fig:Fplot} are used here.  For
interactions $\seteio$ which kill the barrier, the route entropy is
further reduced (long dashed).  The upper $3$ curves in both plots are
funneled folding mechanisms with barrier heights varying from their
maximum to zero. The bottom curves (short dashed) are route entropies
for a folding mechanism involving a small number of routes.  If ${\cal
S}_{\mbox{\tiny ROUTE}} =0$, only one route to the native state is
allowed, because then all contact probabilities $\Qi$ are only zero or
one at any degree of nativeness, and if this is the case each
successive bond made in folding must always be the same one at each
and every degree of nativeness.  It is interesting that in both theory
and simulation there are fluctuations in the route entropy. These
correspond to the necessary fractional values of contact formation
probability present when a particular contact begins to be formed, as
long as contact probabilities $\Qi$ go from zero to one over a finite
width $\Delta Q$ (see fig.~\ref{fig:QivsQroute}).

Figure~\ref{fig:QivsQroute} shows the contact formation probability
$\Qi(Q)$ as a function of $Q$ for all contacts. The top row is the
lattice simulation result, the bottom row from the analytical theory,
eq.~(\ref{Qstar}). The left column is for a route-like folding
scenario, the right column for a uniform folding scenario at the
barrier peak. Energies are chosen from eq.~(\ref{Etune}) with values
for $\alpha$ shown above each figure. There are less curves in the
analytic theory because contacts with the same loop length (which have
the same energy) have the same $\Qi(Q)$.  In the route-like folding
scenario, contacts remain unformed until their free energy of contact
formation $f_i = \ei - T \si(Q)$ equals $-\mu (Q) =(1/M) \D F /\D Q$,
for larger $Q$ they are then essentially formed (see eq.~(\ref{Qstar})
with $s_{\mbox{{\tiny ROUTE}}}^o = 0$ and $b=0$). Here $\mu(Q)$ in
eq.~(\ref{Qstar}) serves as a chemical potential for each contact.
Fluctuations in route entropy for a route-like folding scenario can be
understood through the behavior of contact formation probabilities.
For this folding mechanism contact probabilities are approximately
either zero or one. But as $Q$ increases, new contacts must be formed
and so their formation probabilities must go from zero to one, over
some finite interval $\Delta Q$ for any finite sized system with a
reasonable distribution of 
native energies. In this range $\Delta Q$, the contact(s) increasing
from zero to one have fractional occupation probabilities and so
contribute to the route entropy, causing the bumps seen in
figure~\ref{fig:sroutehet}.  In the uniform folding scenario, all 
$\Qi(\Qdag) \approx \Qdag$ at the barrier peak position. They deviate
away from $Q$ away from the barrier peak in the lattice model because
of the nontrivial dependence of the loop entropy on
nativeness. Deviations away from $Q$ in the theory are due to the
cut-off introduced into the entropy function mentioned above, so that
eq.~(\ref{eq:el}) is not strictly valid. When contact probability is
governed by native topology, as in figure~\ref{fig:QiQ}, the variance in
$\Qi$ values is in between the two extreme cases shown here in
figure~\ref{fig:QivsQroute}.

We can make a connection with earlier polymer theory by investigating
the total entropy minus the route entropy ${\cal S}_{\mbox{\tiny
FLORY}} \equiv N s_o + {\cal S}_{\mbox{\tiny BOND}}$ (see
fig.~\ref{fig:stotflory}).  This can be thought of as the residual
entropy of the effective halo dressing the partially native
structures, and is analogous to the polymer entropy in a cross-linked
system studied by Flory and
others~\cite{FloryPJ56:jacs,WallFT51,DeamRT76}.  As mentioned in the
comments after eq.~(\ref{eq:smfflory}), fluctuations away from uniform
folding raise the bond entropy and thus the Flory entropy, as can be
seen in figure~\ref{fig:stotflory} for both simulations and
theory. For the uniform folding scenario, there is a value of $Q_v<1$
where the Flory entropy runs out (see again comments after
eq.~(\ref{eq:smfflory})). In a uniform folding scenario above $Q_v$,
all residual entropy is combinatoric- if $M Q_v$ crosslinks were
permanently made with equal probability anywhere in the protein, there
would be no entropy left in the system.  The mean-field prediction
gives $Q_v \approx 2 s_o/3 z \approx 0.87$. The lattice value of $Q_v
\approx 0.65$ is considerably lower probably because ideal chain
statistics used in the theory overestimates the entropy actually
present in a partially constrained lattice polymer of length
$N=27$. The folding mechanism having few routes has essentially the
largest Flory entropy here because following the recipe of
eq.~(\ref{Etune}) weights a native core with naturally large halo
entropy. Weighting other cores would in general not give the same
result.

We continue the comparison between theory and simulation by plotting
several quantities in figure~\ref{fig:simthrytest}, for a folding
mechanism with intermediate barrier height ($\alpha_{\mbox{\tiny
SIM}}=1.0$, $\alpha_{\mbox{\tiny THRY}} = 0.5$ in eq.~(\ref{Etune})),
and for energies $\eistar$ which induce a uniform folding mechanism in
fig.~\ref{fig:simthrytest}C.
Plot~\ref{fig:simthrytest}A shows the correlation between theory and
simulation of contact probability at the barrier peak $\Qi(\Qdag)$
given in the theory by eq.~(\ref{Qstar}).
Contacts with the same loop length have the
same $\Qi$ value in the theory but not in the simulation (due to
entropic differences even for contacts having the same loop
length). However the average of the simulational contact probabilities
(squares in fig.~\ref{fig:simthrytest}A) correlates well with the
theory (correlation coefficient $= 0.93$), indicating that the theory
captures the average trend, but there are potentially important
many-body effects in the calculation of the entropy loss upon contact
formation.
Figure~\ref{fig:simthrytest}B shows the trend in contact probability
with increasing loop length, by replotting the theory and data against
log loop length. The solid line is the theoretical result
of eq.~(\ref{Qstar}) with eq.s~(\ref{sbond}) and~(\ref{leff}). Again
the data fluctuate significantly about 
this curve, but the theory captures the trend in the average values
(squares) (as before, correlation coefficient $0.93$).
Figure~\ref{fig:simthrytest}C shows the results of tuning the energies
to induce uniform folding, for theory and simulation. The solid line
is the result of 
equation~(\ref{eq:el}): $\dei = \Tf \dsi$. The data points with error
bars are the values extracted from lattice simulations. Again the
theory captures the overall trend (correlation coefficient $0.71$),
but there are still significant fluctuations about the average.

\subsection{Measures of routing}
\label{sec:routmeas}

Since the free energy barrier is maximized for  a uniform funnel
folding mechanism (eq.~\ref{2nd}), we expect the barrier height to
be decreasing 
function of the dispersion in $\Qi$ values at
the barrier peak $\overline{\dQ^2}(\Qddag) = \left< (\Qi-\Qddag)^2
\right>$. 
Let us introduce a measure of ``routing'' ${\cal R}(\Qddag)$ through
the bottleneck by the function
\be
{\cal R}\left(Q\right) = \frac{\left< \delta Q^2\right>}{\left< 
\delta Q^2\right>_{{\mbox{\tiny MAX}}}} = 
\frac{\left< \delta Q^2\right>}{Q \left( 1-Q\right)} .
\label{eq:route}
\ee
The denominator is the most route-like the system can get at $Q$,
i.e. if  
$MQ$ contacts were made with probability $1$ and $M-MQ$ contacts were
made with probability $0$, then $\left< (\Qi-Q)^2 \right> = (1/M) (MQ
(1-Q)^2 + (M-MQ) Q^2) = Q(1-Q)$. Thus ${\cal R}\left(Q\right)$ is
between $0$ and $1$. 
${\cal R}\left(Q\right)$ is proportional to the lowest order correction to the
route entropy (\ref{smix}) when fluctuations $\dQ$ are present: 
\be
{\cal S}_{\mbox{\tiny ROUTE}} \left( \left\{ Q+\dQi \right\}\right) \cong
{\cal S}_{\mbox{\tiny ROUTE}}^o - \frac{M \lambda}{2}{\cal
R}\left(Q\right) \: . 
\ee
In the (non-perturbative) limit ${\cal R}\left(Q\right)=1$, ${\cal
S}_{\mbox{\tiny ROUTE}}=0$ and only one route to the native state is
allowed, i.e.
since all $\Qi$ are only zero or one at any degree of nativeness, 
each successive bond added at that degree of nativeness must always be
the same one. 

Using eq.~(\ref{eq:dqdeqq}) we can relate
the fluctuations in optimal energies 
$\dei$ in terms of fluctuations from the uniform contact probabilities
$\dQi$ as $\dei = -\lT \, \dQi /\Qddag (1-\Qddag)$, and then substitute
this along with~(\ref{2nd}) into eq.~(\ref{taylor}) to obtain the
decrease in barrier height 
with route measure:
\be
\delta \DFdag \cong -  M \frac{\lambda^{\ddag} T}{2}
\frac{\overline{\dQ^2}}{\Qddag\left(1-\Qddag\right) } 
= - M \frac{\lambda^{\ddag} T}{2} \, {\cal R}(\Qddag) \: .
\label{dfhet}
\ee
As noted above (see e.g. comments after eq.~\ref{pertF}), the
reduction in the barrier height due to ordering 
heterogeneity scales extensively with system size.
A dispersion in contact participations $\overline{\dQ^2} = 0.05$ which
is about $20\%$ of the maximal dispersion 
($\approx 1/4$, taking $\Qddag\approx 1/2$) lowers the
barrier by about $0.1 N\kB T$ or about $5 \kB T$ for a chain of
length $N \approx 50$, believed to model a protein with $\approx 100
aa$~\cite{OnuchicJN95:pnas}. 
We should note here that renormalizing real amino acids into
coarse-grained monomers may underestimate the heterogeneity effect,
because small-scale free energy fluctuations do not average out upon
coarse-graining, but will still add up extensively.
Plots of the route measure as a function
of $Q$ for the various possible folding scenarios were given
in reference~\cite{PlotkinSS00:pnas}; essentially the same information
is contained in the route entropy plotted here in figure~\ref{fig:sroutehet}.

Figure~\ref{fig:FvsR} shows the barrier height at the transition
temperature $\Tf$ in units of $\ebar$, vs. the route measure at the
barrier peak ${\cal R}(\Qddag)$ ($\Tf$ is itself weakly dependent on
${\cal R}(\Qddag)$; see e.g. fig.~\ref{fig:sdplot}). There is a
monotonically decreasing trend for both theory and simulation. The
solid line in figure~\ref{fig:FvsR} is the theoretical result for a
model with parameters in table~\ref{table1}.  The barrier vanishes for
roughly the same magnitude of route measure in both theory and
simulation, even though different dispersions in contact energies are
required to produce these route measures.  We can examine the effects
of heterogeneity on a uniform, idealized funnel, where all loops have
length $\li = \lbar \approx 9$, the average of the lattice model
native structure of figure~\ref{fig1}; and initially all energies $\ei
= \ebar$ (long dashed line in fig.~\ref{fig:FvsR}).  Here too breaking
the ordering symmetry $Q\rightarrow\setQi$ by random perturbations on
$\ei$ lowers the thermodynamic barrier, as explained in
sections~\ref{sect:simple} and~\ref{sec:add}; the barrier goes down
because the energy decreases more than the route-entropy decreases.
Perturbing the structure by including entropic dispersion via $\dli$
yields a similar result (c.f. eq.~(\ref{pertF})). The short dashed
line in figure~\ref{fig:FvsR} is the perturbation result for this
model from eq.~(\ref{dfhet}), which agrees well with the full
non-perturbative result for small ${\cal R}$.

\subsection{Kinetics of folding times}
\label{sect:kinetics}

We can verify that the heterogeneity effect on the barrier translates
directly to an increase in folding rate by simulating long Monte Carlo
runs and measuring the distribution of folding times at the transition
temperature $\Tf$ for
various energy functions $\setei$ (see fig.~\ref{fig:tau}).

To estimate  the increase in rate assuming the prefactor is not
important, take the barrier heights from fig.~\ref{fig:FvsR} and the
neglect the change in transition temperature $\Tf$ since it only
depends weakly on ${\cal R}$ (see fig.~\ref{fig:sdplot}).
Then the ratio of folding times, say at  ${\cal R} = 0$ and ${\cal R}
= 0.2$, is 
\be
\frac{\tauF \left({\cal R} = 0\right)}{\tauF\left({\cal R} =
0.2\right)} = \frac{\exp \left(\DFdag\left({\cal R} = 0\right)/\Tf
\right)}{\exp \left(\DFdag\left({\cal R} = 0.2\right)/\Tf\right)}
\approx 
\frac{\exp \left( 2.1/0.7\right)}{\exp \left(1.2/0.7\right)} \approx
3.6
\ee
which is in reasonable agreement with the measured ratio of about $3$
(see fig.~\ref{fig:tau}). For the heterogeneous G\={o} models
considered here, the folding time is well approximated by an Arrhenius
law with constant prefactor, indicating that the time scale in the
reconfigurational kinetics is not strongly influenced by the degree of
routing, at least up until the barrier vanishes.

We can expand
equation~(\ref{eq:k1}) to first order in changes in the barrier height
$\delta \Fdag$, folding temperature $\delta \Tf$, and prefactor $\delta k_o$ to
obtain a condition that must be satisfied for the rate to increase under
perturbations of native energy:
\bea
\frac{\d \Fdag}{\Tf} &\lesssim& \frac{\d k_o}{k_o} + \frac{\Fdag \d
\Tf}{\Tf^2} \nonumber \\
&\lesssim& \left( \frac{\D  \ln k_o}{\D T} + \frac{\Fdag}{\Tf^2}
\right) \d \Tf \: ,
\label{perteq1}
\eea
where we have made the approximation in the second inequality that the
prefactor is most strongly 
affected by changes in the folding temperature, since the closer $\Tf$
is to $\Tg$ the slower reconfigurational dynamics is within the
protein, and we have seen in fig.~\ref{fig:tau} that the prefactor is nearly
independent of the native energy distribution;
then $\d k_o \approx (\D k_o /\D T) \d T$.
For well-designed folders, the folding transition temperature $\Tf$ is above
the dynamic glass temperature of the system, and reconfigurational
dynamics  is largely 
unactivated~\cite{WangPlot97,TakadaS97:pnas}, i.e. escape from
individual traps does not dominate the kinetics. In this regime, the
prefactor, which is related to the hopping time, 
is nearly temperature independent: $\D  \ln k_o/\D T \ll \Fdag/T^2$.
Then the condition given in eq.~(\ref{perteq1}) becomes
\be
\frac{ \d \Fdag}{\Fdag} \lesssim \frac{ \d \Tf}{\Tf}
\label{eq:FFTT}
\ee
For slower folders, effects on the prefactor may be as important as
effects on the barrier, however for the well-designed folders
considered here the rate increases so long as the relative barrier
height decreases more strongly than the relative change in folding
temperature. This is seen to be the case in figure~\ref{fig:TTFF}.

\subsection{Mean loop length dependence of the barrier height}
\label{sect:mean}

Experimental evidence has shown a strong correlation of folding rate
with a quantity in our model equal  to the mean loop
length divided by 
the total chain length~\cite{Plaxco98}. Since no strong correlation
with $N$ is observed at least for typical protein sizes, we are
interested in testing if the barrier
height in our model  correlates with $\lbar$, at fixed $N$.

We seek the change in free energy $\delta F$ upon a change in the quantity
$(1/M) \sum_i \li$. This can be found by utilizing the directional
derivative (see Appendix A and eq.~\ref{direct}):
\be
\frac{\d F}{\d \lbar} = M \frac{\d F}{\d \left(\sum_i \li\right)} =
\left(\sum_i \hat{i}\right) \cdot \left( \sum_j \frac{\d F}{\d \lj}
\hat{j} \right) = \sum_i \frac{\d F}{\d \li} \: .
\ee
Using again the analog of eq.~(\ref{tot2}) that we used already to
obtain eq.~(\ref{dfdl}), the total derivative of $F$ with respect to
$\li$ is equivalent to the partial derivative. The free energy 
$F$ depends on $\li$ explicitly only through the bond entropy
eq.~(\ref{Sfinal}), which is composed of a mean-field term depending
on the sum $\lbar$ plus a fluctuation term, eq.s~(\ref{eq:smf})
and~(\ref{fdefn}). Noting that 
$$
\frac{\D {\cal S}_{\mbox{\tiny MF}} (Q,\lbar)}{\D \li} = \frac{1}{M}
\frac{\D {\cal S}_{\mbox{\tiny MF}} (Q,\lbar)}{\D \lbar}
$$
we obtain 
\be
\frac{\D F}{\D\lbar} = -T \frac{\D {\cal S}_{\mbox{\tiny BOND}}}{\D\lbar}
= \frac{3}{2} \frac{M T}{\left(\lbar-1\right)^2} \left[ \ln \left(1+
\left(\lbar-1\right)Q \right) - Q\ln\lbar\right] + \frac{3}{2}M T \left<
\frac{\dQ}{\ell} \right> \: .
\label{sder}
\ee
The first term in expression~(\ref{sder}) is always positive for $Q>0$. 
The second term weights loops with smaller $\li$ more heavily, and for
these loops $\dQ > 0$, so the second term is always positive when
entropic effects are considered alone. The native energies would have
to be specially  tuned to change the sign of this term. Moreover the
whole expression is zero when $Q=0$, so we conclude that the effect of
increasing the mean loop length is to increase the barrier height
$\DFdag$. This effect is illustrated in fig.~\ref{fig:LL} for the
simple case where $\li =\lbar$, where the second term in~(\ref{sder}) is
zero; this is a lower limit to the actual increase in barrier.

As eq.~(\ref{sder}) implies, the change in barrier height with mean
loop length is an entropic effect; proteins with native structures
having larger mean loop length have lower entropy near the transition
state. Another perhaps simpler way to see this is to note that the
entropy of loop closure must become larger (more negative) as the loop
length for that contact is increased. From eq.s~(\ref{sbond})
and~(\ref{leff}) and setting  $\li =\lbar$ for purposes of illustration,
\be
\frac{\D \si}{\D \lbar} \approx - \frac{1-Q}{\lbar \left( 1 + \left(
\lbar -1 \right) Q\right)} < 0 \: .
\label{eq:dsdlbarCO}
\ee
Therefore more entropy is lost in contact formation for structures
with larger mean loop length. Furthermore since
\be
\frac{\D^2 \si}{\D Q \D \lbar} \approx \frac{1}{\left(
1+\left(\lbar-1\right) Q\right)^2} > 0 \: ,
\ee
this effect is largest at low degrees of nativeness (e.g. from
eq.~(\ref{eq:dsdlbarCO}) at $Q=0$, $\D
\si /\D\lbar \approx -1/\lbar$ while at $Q=1$ $\D \si /\D\lbar \approx
0$): the entropy becomes more of a convex down function as $\lbar$ is
increased, see fig.~\ref{fig:Sco}. Since the free energy barrier
arises from the incomplete cancellation of entropy and energy (which
is independent of $\lbar$) as $Q$
increases, a more convex down entropy indicates a larger barrier
height.

\subsection{Dependence of barrier heights and rates on structural
variance} 
\label{sect:structvar}

By eq.~(\ref{pertF}), if we let $\ei=\ebar$ and fix $\lbar$, the
folding barrier is lower for structures with larger variance in loop
energies $\overline{\d \ell^2}$. 
For proteins sufficiently well-designed that the folding rate $\kF$
near the transition temperature is governed by the free energy barrier
as in eq.~(\ref{eq:k1}), then
\be
\ln  \frac{\kF ( \overline{\dl^2} )}{\kF (0)} 
\approx M \Qddag \frac{\overline{\dl^2}}{\lbar^2}
\label{svpert}
\ee
where we have also neglected changes in the folding transition
temperature, since accounting for this is a higher order effect. We
have also let $\ldag \approx 1- \Qddag$ since $\alpha$ in
eq.~(\ref{eq:lambdaQ}) is approximately one (see table I).
Most importantly the perturbation result neglects changes in the
unfolded free energy on structural variance, as well as changes in the
amount of native structure in the unfolded state. These reduce the
trend on the rate due to structural variance. In general we should use
\be
\ln  \frac{\kF ( \overline{\dl^2} )}{\kF
(0)}  \cong \frac{\DF^{\ddag}
(0)}{\Tf(0)} -
\frac{\DF^{\ddag}(\overline{\dl^2}
)}{\Tf(\overline{\dl^2})}
\label{svfull}
\ee
for the log ratio of rates. The barrier height is then obtained from
eq.s~(\ref{FTOT}) and~(\ref{Qstar}).
It is seen in figure~\ref{fig:sv} that there is a significant
increase in folding rate for structures having larger variance in
loop lengths. Structural variance is generated here for a system with
parameters in table I, but the loop lengths are given by
\be
\li = \lbar + \alpha \left( \li^o - \lbar \right)
\label{lrec}
\ee
where $\li^o$ is taken from the loop length distribution in table I.
As $\alpha$ varies from zero to one, the mean loop length $\lbar$ remains
unchanged ($\lbar \cong 9.14$), but the structural variance
$\overline{\dl^2}$ increases (see fig.~(\ref{fig:sv})).

\subsection{Tuning energy functions to match desired potentials}
\label{sect:match}

We have so far focused on how folding thermodynamics and mechanism are
affected by properties of the native structure and distribution of
native stability. We can also turn the problem around and seek the
native structural properties or stability distribution that would give
a specified free energy profile or folding mechanism. To illustrate, 
fig.~\ref{fig:F} shows the free energy potential $F(Q)$ for the 
$27$-mer chain. A fit to the lattice data of ref.~\cite{SocciND96:jcp}
for example can be made by annealing the stability distribution
$\setei$, contact length distribution $\setli$, and/or other parameters
with a cost function  
\be
\mbox{Cost} = \int_{0}^1 \, dQ \,\left(F_{target}(Q) -
F\left(T,\ebar,\{ \ei \},\lbar, \{ \li \}, s_o,b,\alpha | Q\right)\right)^2 \: .
\label{eq:anneal}
\ee
Fits may be made to a fairly arbitrary potential by adjusting the
native coupling energies or loop length distribution (the structure),
e.g. potentials with and without intermediates, potentials having
sharply peaked or flat barriers, etc. See fig.~\ref{fig:F}B for an example of
annealing the energies to fit a target potential with an on-pathway
intermediate. 
Using a sufficiently accurate numerical theory of entropy loss, this
method should be able to distinguish between intermediates governed by
native structural properties, native energetic stability, or
misfolding.

\section{Summary and conclusions}

In this paper we have introduced refinement and insight into the
funnel picture by considering heterogeneity in the folding of
well-designed proteins. 
We have explored in minimally frustrated sequences how folding is
effected by 
heterogeneity in native contact energies, as well as the entropic
heterogeneity inherent in folding to a specific 
three-dimensional native structure.
The general method we utilized here should be amenable to systematic
refinement, and should be sufficiently accurate to compare with
experimental results.

Specifically we found that heterogeneity, whether energetic or
entropic in origin, will always lower the folding free energy barrier;
see sections~\ref{sect:simple}, \ref{sec:add}, and \ref{sect:egiven},
equations~(\ref{2ndising}), (\ref{DFrem}), (\ref{eq:hetcore}),
(\ref{df22}), (\ref{pertF}), and (\ref{dfhet}), and figures~\ref{fig:Fplot}, 
\ref{fig:sdplot}, and \ref{fig:FvsR}. 
This was shown using arguments from the random energy model
(section~\ref{sect:remarg}), transition state theory
(section~\ref{sect:tst}), the optimum fluctuation method
(section~\ref{sec:of}), thermodynamic perturbation theory
(section~\ref{sec:landau}), and free energy functional theory
(sections~\ref{sect:simple} and~\ref{sect:egiven}).
For sufficiently well-designed
proteins the corresponding rate also increases; see
equations~(\ref{rate1}), (\ref{eq:optfrate}), and (\ref{rateFunc}),
section \ref{sect:kinetics}, and figures~\ref{fig:tau},
\ref{fig:TTFF}, and \ref{fig:sv}.
The effects of heterogeneity on barriers and rates are stronger than
the effects on transition temperature; see the discussion in the
introduction on eq.~(\ref{eqtf}), and equation~(\ref{eq:FFTT}), and
figures~\ref{fig:sdplot} and~\ref{fig:TTFF}.
We investigated the effects on the folding barrier due to correlations
between energetics and topology, see sections~\ref{sec:likely}, and
~\ref{sect:egiven}, and eq.s~(\ref{df1st}), 
(\ref{eq:el}) and~(\ref{pertF}), and found that for well-designed
proteins the rate may be increased by making initially likely contacts
stronger while making unlikely contacts weaker. Thus overall stability
is conserved, but the energetic distribution is coupled to the native
structure. 

For the ensemble of well-designed sequences having a given overall
stability, homogeneously ordering sequences have the largest folding
free energy barrier.
For most structures, where
topological factors play an important role, this regime is achieved by
introducing a large dispersion in the distribution of native contact
energies which in practice would be almost impossible to achieve, see
figs~\ref{fig:sdplot} and~\ref{fig:simthrytest}C. As we
reduce the dispersion in the contact energy distribution and the
energies approach a uniform
value $\ebar$, the dispersion of contact participations increases and
thus the number of folding routes decreases, the free
energy barrier decreases and the total configurational entropy
increases to a maximum, see sections~\ref{sect:ebar}
and~\ref{sect:totent}. Again, folding temperature is only mildly
effected; the prefactor 
appearing in the rate is probably only
mildly effected also, since it is largely a function of $\Tf/\Tg$ and
polymer properties~\cite{SocciND96:jcp}.
If many-body forces are not too large the barrier may be reduced to
zero, either by adding random native heterogeneity as in
section~\ref{sec:landau} or by correlating native energy to native structure
so that more probable contacts are stronger, as in
section~\ref{section:opt}. The funnel picture, with different
structural details, is valid for 
the above wide range of native contact energy distributions. 
However, tuning the energies further so that probable contacts have
even lower 
energy (or allowing native energies to have a very large variance)
eventually induces the system to take a single or very few
folding routes at the transition temperature.
A large dispersion of energies is required to achieve this,
and in this regime the folding temperature drops well below
the glass temperature range, where folding rates are extremely slow.

In section~\ref{sect:ent} we derived approximate expressions for the
conformational entropy functional for a well-designed protein, see
eq.s~(\ref{smix}), (\ref{Sfinal}), and~(\ref{eq:smf}).  In
section~\ref{subsect:rout} and Appendix~\ref{sect:appB} we generalized
the entropy of native core placement (the mixing entropy) used
previously in models of folding~\cite{PlotkinSS96,PandeVS97:fd} to
account for the effects of chain connectivity; for a highly 
constrained chain, many contact patterns are degenerate to essentially
the same conformation.  In section~\ref{sect:bondS} we derived a
general condition for the conformational entropy to be a state
function, viz. eq.~(\ref{state}). A Hartree approximation was taken to
account for the entropy loss of loop closure in the presence of other
contacts already formed. This agrees well with the average behavior on
the lattice (see fig.~\ref{fig:simthrytest}), however there are
important fluctuations away from the mean, indicating many-body
correlation effects are present which the theory doesn't account
for. Equation~(\ref{Sfinal}) gives the conformational entropy loss
given a distribution of 
native contact lengths $\setli$. When each $\li \rightarrow \lbar$,
the expression reduces to eq.~(\ref{eq:smf}) which is the entropy loss
for a finite system with mean return length $\lbar$ for all
contacts. When $\lbar \rightarrow \infty$, (\ref{eq:smf}) further
reduces to eq.~(\ref{eq:smfflory}) which is the entropy loss for a polymer
system in the Flory mean-field theory~\cite{FloryPJ56:jacs}.

Residues in proximity are assumed to be in contact energetically; the
reduction in conformational entropy at low $Q$ to the elimination of
conformations which happen to have residues in
proximity~\cite{PlotkinSS96} is not included here because it is a
smaller effect than the other contributions to the entropy.

Several experiments support results from our theory. Enhancement of
folding rates by weighting entropically likely contacts, as found in
sections~\ref{sec:likely} and~\ref{sect:structandE}, has been observed
in Escherichia coli Che~Y~\cite{VigueraAR96}.  Depending on the
variance of native interactions and how native interaction strength
correlates with the entropic likelihood of contact formation,
sequences may be designed to fold both faster or slower to the same
structure as a wild-type sequence.  Enhancements or suppressions of
folding rate to a given structure due to changes in sequence are
modeled in our theory through changes in native interactions, which
induce significant changes in the rate-governing free energy landscape
of a well-designed protein.  A minimally frustrated sequence may fold
to a given native structure by a variety of folding mechanisms (see
fig.s~\ref{fig:Fplot} and~\ref{fig:QivsQroute}), including through
both on and off-pathway 
intermediates (see fig.~\ref{fig:F}).  Thus for example folding in Im7
and Im9 may likely initiate from different places within the native
structure depending on the distribution of
stability~\cite{FergusonN99}. Folding in the IgG binding domain of
protein~L may tend to 
initiate from a specific region of higher stability, indiscernible
from the apparently symmetric native structure~\cite{KimDE00}; contact
formation probability at the transition state depends on both energy
and entropy, as expressed in equation~(\ref{Qstar}). 
However, for a large range
of native energy distributions, barrier heights, and corresponding
rates, a funnel folding mechanism is preserved, in that there are many
routes to the native structure, see fig.s~\ref{fig:mix}
and~\ref{fig:sroutehet}. Folding rates in mutant proteins that exceed
those of the wild type have been receiving much interest in recent
experiments~\cite{VigueraAR96,Munoz96:Rev,HagenSJ96:pnas,KimDE98,BrownBM99};
here we saw how these effects can be understood by applying general
principles of the energy landscape.
Folding rates in the theory were seen to increase with the variance in
contact formation probability, a thermodynamic quantity closely
related to the dispersion in experimental $\phi$ values (see
fig.~\ref{fig:FvsR}). The general trend of reduced rate with larger
contact order~\cite{Plaxco98} is well captured by the theory (see
fig.~\ref{fig:LL}). Additionally, for fixed contact order, folding
rate was shown to increase with larger variance in the contact lengths
which constitute the native structure (see fig.~\ref{fig:sv}).

Fluctuations in rate due to the effect of sequence perturbations on
weakening or strengthening specific non-native kinetic traps or
generally changing non-native interaction strength is an interesting
topic of future research.

It is important to note that the enhancements or reductions in rate we
have explored are mild compared to the enhancement by minimal
frustration (funneling the landscape): the fine tuning of rates may be
a phenomenon manifested by {\it in vitro} or {\it in machina}
evolution, rather than {\it in vivo} evolution. Nevertheless folding
heterogeneity may become an important factor for larger proteins,
where e.g. stabilizing partially native intermediates may increase the
overall rate or prevent aggregation.  Adjusting the backbone rigidity
or the non-additivity of
interactions~\cite{PlotkinSS97,KolinskiA93:jcp} can also modify the
barrier height, possibly as much as the effects we are considering
here.  There may also be functional reasons for non-uniform folding -
malleability or rigidity requirements of the active site may inhibit
or enhance its tendency to order.

The notion expounded here that rates increase with heterogeneity at little expense to
native stability contrasts with the 
view that non-uniform folding in real proteins exists
merely as a residual signature of incomplete evolution to a
uniformly folding protein. 
Moreover, the phenomenon  that fluctuations in native contact energies
contribute extensively to the free energy landscape indicates that the
prediction of numerical values for folding rates and mechanisms from
approximate energy 
functions may be even more difficult than originally suspected,
i.e. even if systematic error in the calculation of potentials is
eliminated, $\cal O (N)$ corrections may still remain.

The amount of route narrowness in folding was introduced as a
thermodynamic measure through
the mean square fluctuations in a local order parameter. 
The route measure may be useful in
quantifying the natural kinetic accessibility of various structures.
While structural heterogeneity is essentially always present,
the flexibility inherent in the number of
letters of the sequence code limits
the amount of native energetic heterogeneity possible.
However some sequence flexibility is in fact required for funnel
topographies~\cite{Wolynes97nsb} and so is probably present
at least to a limited degree. 

We have seen here how a very general theoretical framework can be
introduced to explain and understand the effects of 
heterogeneity in native stability and structural topology on such
quantities as folding rates, transition temperatures, and the degree
of routing in the funnel folding mechanism. Such a theory should be a
useful guide in interpreting and predicting future experimental results on
many fast-folding proteins.

\subsection{Acknowledgments}
We thank Peter Wolynes, Hugh Nymeyer, and Cecilia Clementi
for their generous and insightful discussions. 
This work was supported by NSF Bio-Informatics
fellowship DBI9974199 and NSF Grant MCB0084797.

\renewcommand{\theequation}{\thesection.\arabic{equation}}
\renewcommand{\thesection}{\Alph{section}}
\setcounter{section}{0}

\section{Appendix A}
\setcounter{equation}{0}
Consider the free energy of eq.~(\ref{f}) as the integral over a
semi-local free 
energy density $F(\setQi) = \sum_i f_i (\Qi,Q)=\sum_i f_i (\Qi,\sum_j \Qj)$.
Taking the differential of a new thermodynamic function 
$G = F + \mu \sum_i \Qi $,
\be
\delta G = \sum_i \left[ \left(\frac{\D f_i}{\D
Q_i}\right)_{\mu} + \mu \right] \delta Q_i + \left[\sum_j \Qj\right]
d\mu 
\label{eq30}
\ee
and demanding that $\D G/\D Q_j = 0$ for all $j$ 
Legendre transforms to a new variable $\mu$, with $\D G/\D
\mu = MQ$. This is equivalent to minimizing the free energy subject to
the constraint of fixed $Q$. The equation $\D G/\D Q_j =
0$ means that 
\be
\frac{\D f_i}{\D \Qi} = -\mu
\label{dfdq}
\ee
for all $i$, which enforces equation~(\ref{Qstar}) for each $\Qi$.
The Lagrange multiplier $\mu$ is interpreted as the force
corresponding to the potential $F(Q)$:
\be
\mu = - \frac{1}{M}\frac{\D F(\setQi)}{\D Q}
\label{force}
\ee
by the following arguments. From eq.~(\ref{eq30})
$\left(\D G/\D Q_i\right)_{\mu} = 0$ is equivalent to $\D F/\D Q_i
+\mu=0$ or
\be
\mu = -\frac{\D F}{\D \Qi}\;\;\;\;\mbox{for any $i$}
\label{mudfdq}
\ee
therefore
$\mu = -(\D Q/\D\Qi)(\D F/\D Q) = -(1/M)(\D F/\D Q)$ .
Since the changes w.r.t. to the local
order parameter of  all the local free energy terms are the same
number $\mu$
(eq.~\ref{dfdq}), this number equals the change in the sum ($F$) w.r.t. the
change in the sum of the $\Qi$ ($MQ$) (eq.~\ref{force}).

Another way to see eq.~(\ref{force}) directly is to consider
$\D/\D(MQ) = \D/\D(\sum_{i=1}^M \Qi)$ as the directional derivative
$D_{\mbox{\scriptsize {\bf u}}} F$ of $F$ in
an $M$-dimensional space along the direction $\nabla Q = \sum_i
\hat{i}$ with $\hat{i}$ a unit vector along the $i$th axis, defined by
the $i$th contact. So 
\bea
\frac{\D F}{\D \left( \sum_{i=1}^M \Qi\right)} &=&
\frac{1 }{\left| \nabla Q \right|} D_{\mbox{\small \bf u}} F =
\frac{1}{\sqrt{M}} \frac{\nabla Q}{\left| \nabla Q \right|} \cdot
\nabla F \left(\setQi\right) \nonumber \\
&=& \frac{1}{M}\left(\sum_i \hat{i}
\right) \cdot \sum_j \frac{\D F}{\D\Qj}\hat{j} = -\frac{\mu}{M}
\sum_{ij} \hat{i} \cdot \hat{j} = -\mu \: .
\label{direct}
\eea

For the two-state potentials considered here $\mu$ has two roots
which give the positions of the barrier peak and 
equilibrium unfolded state (where the local force is zero, see inset of
fig.~(\ref{fig:F})).

\section{Appendix B}
\label{sect:appB}
\setcounter{equation}{0}

In this appendix we compare a simple microscopic
model of route entropy ${\cal S}_{\mbox{\tiny ROUTE}} \left( \left\{
\Qi \right\}\right)$ with the semi-empirical model introduced in
section~\ref{subsect:rout}. In the absence of the constraints due to chain
connectivity the route entropy is given by eq.s~(\ref{mix1})
and~(\ref{mix2}). Each native contact pattern $\setQi$ can be
expressed on a native contact map, (see e.g. ref.~\cite{FiebigKM92}). The
idea here is that since several contact patterns correspond to the same
constrained state because of chain-connectivity, the contact-map can
be coarse-grained into cells inside of which one or more contacts
fully constrains the cell. We make a mean-field approximation and
group configurations into cells all of the same $Q$-dependent size,
$\wq > 1$, with $\omega_0 = 1$. To clarify, imagine a system with $6$
possible native contacts, and $2$ made. Eq.~(\ref{mix1}) gives $15$
possible native cores, but if $\omega_{1/3} = 2$ there are
$3!/1!(3-1)! + 3!/2!(3-2)! = 6$ configurations; the first $3$
configurations correspond to $2$ contacts together in any one of the
$3$ renormalized cells, and the second $3$ correspond to one contact
in two separate cells. 

\begin{figure}[htb]
\hspace{.7cm}
\psfig{file=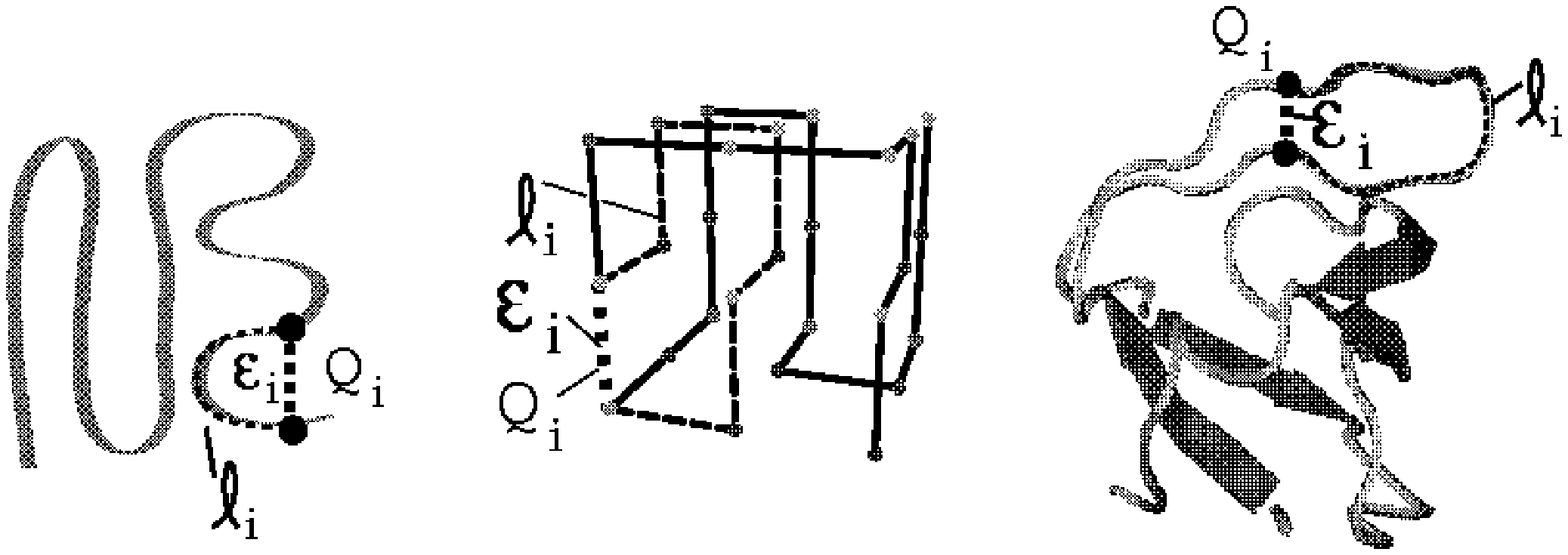,height=17cm,width=6.5cm,angle=-90}
\caption{ } 
\label{fig1}
\end{figure}

\begin{figure}[htb]
\hspace{.7cm}
\psfig{file=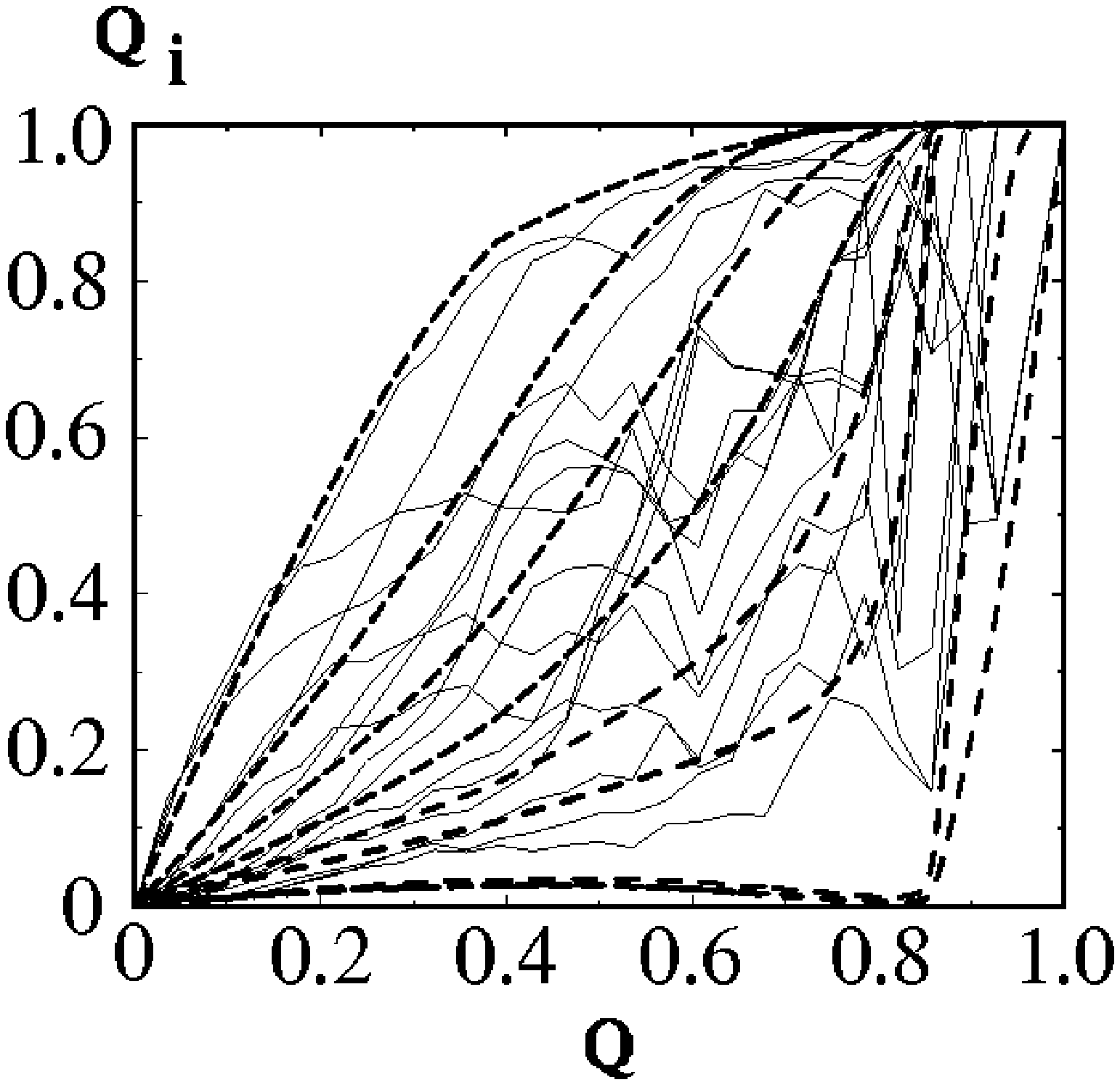,height=8.5cm,width=8.5cm,angle=0}
\caption{ } 
\label{fig:QiQ}
\end{figure}

\begin{figure}[htb]
\hspace{.7cm}
\psfig{file=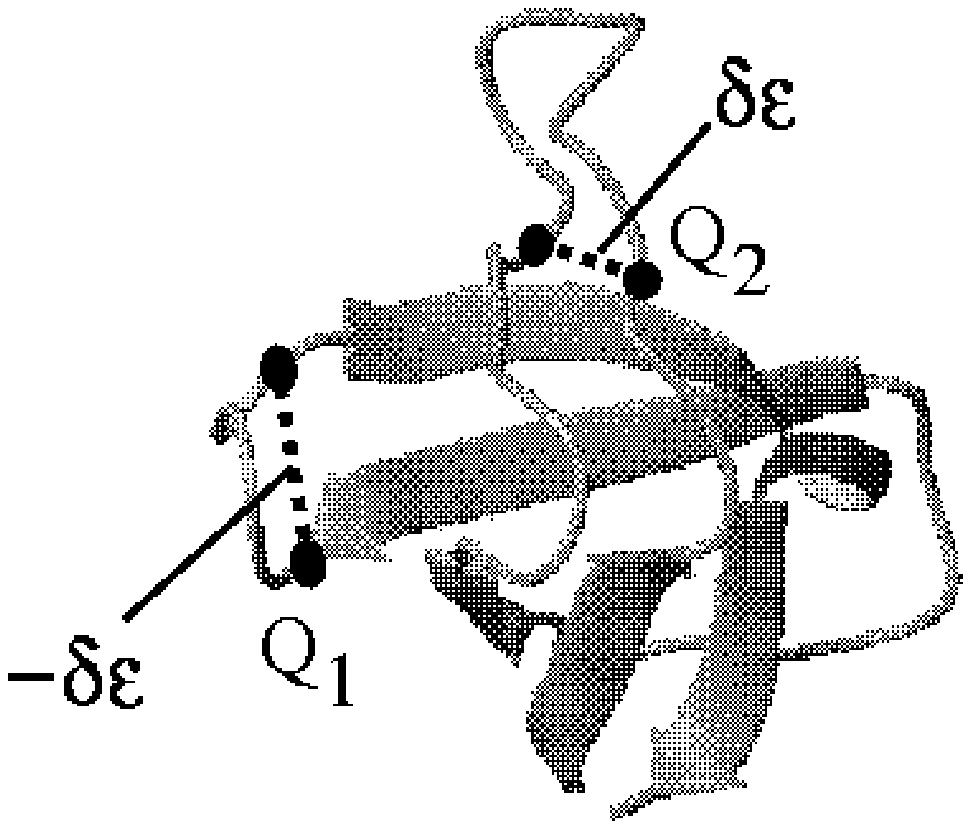,width=8.5cm,angle=0}
\caption{}
\label{fig:eqopp}
\end{figure}

\begin{figure}[htb]
\hspace{.7cm}
\psfig{file=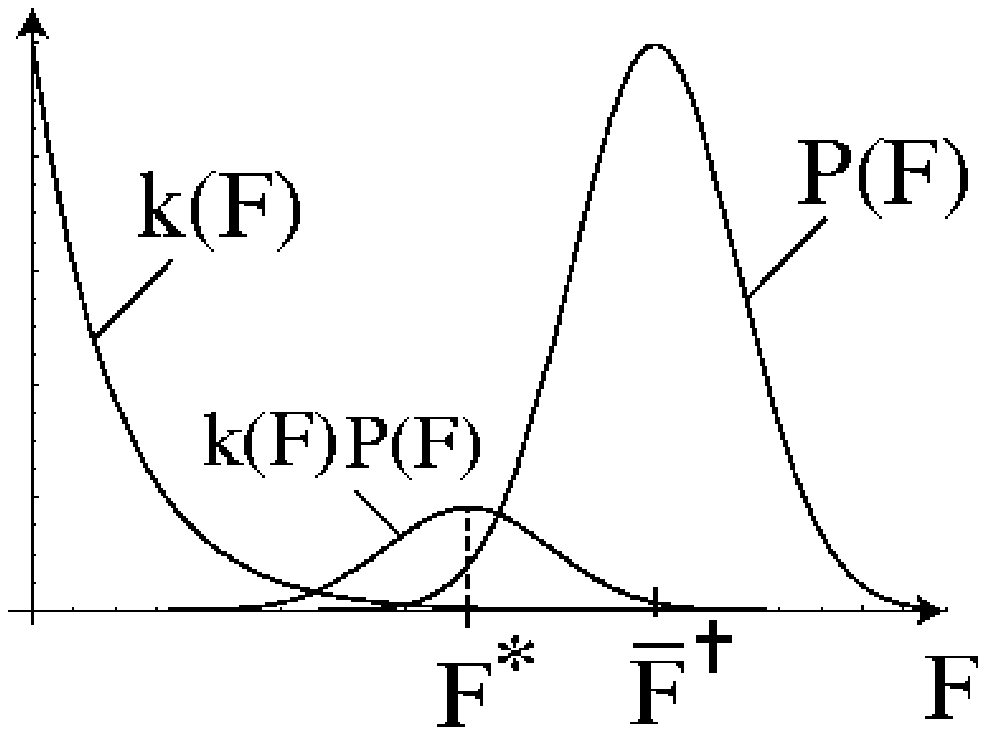,height=6cm,width=8cm,angle=0}
\caption{ }
\label{opt}
\end{figure}

\begin{figure}[htb]
\hspace{.7cm}
\psfig{file=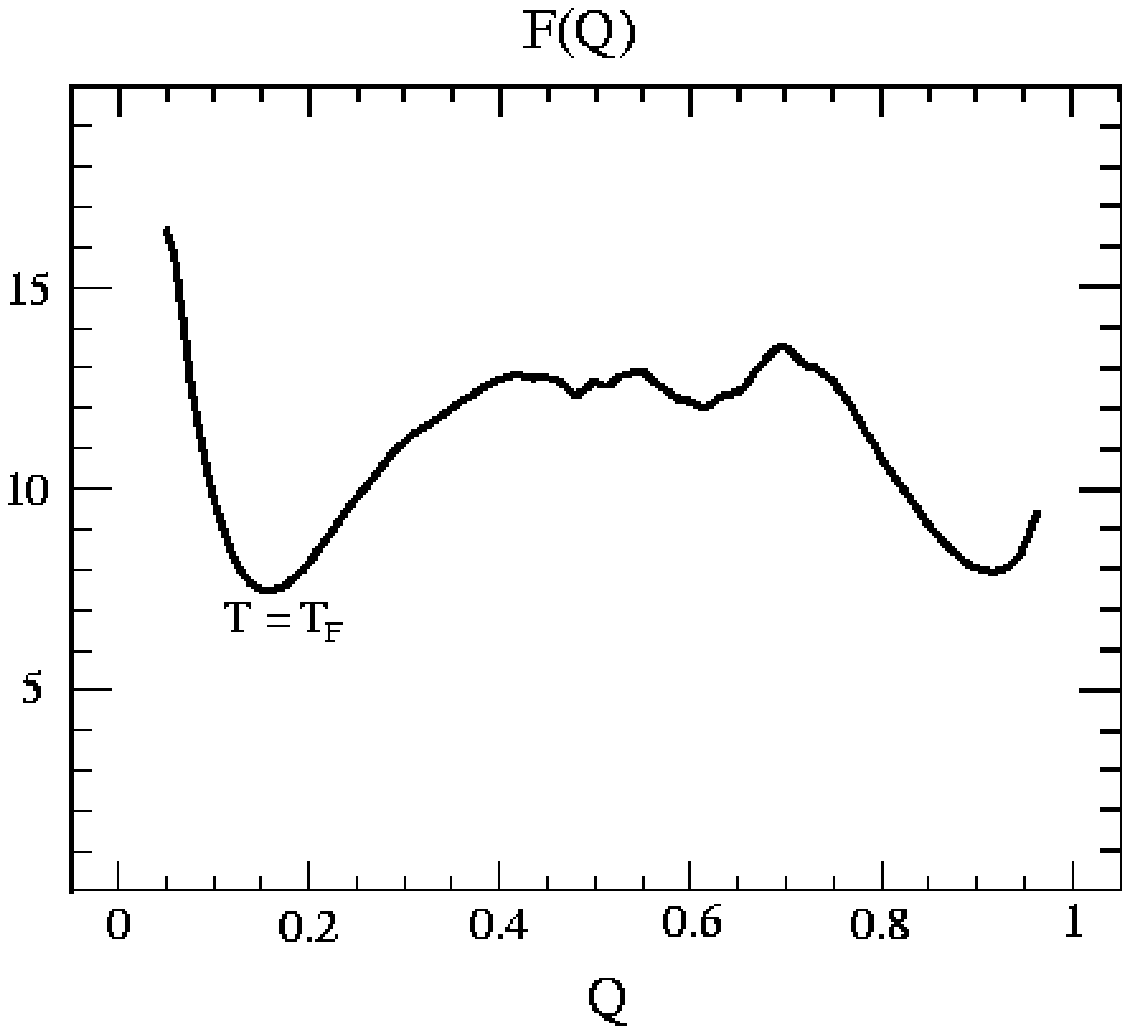,height=8cm,width=8.5cm,angle=0}
\caption{}
\label{chey}
\end{figure}

\begin{figure}[htb]
\hspace{.7cm}
\psfig{file=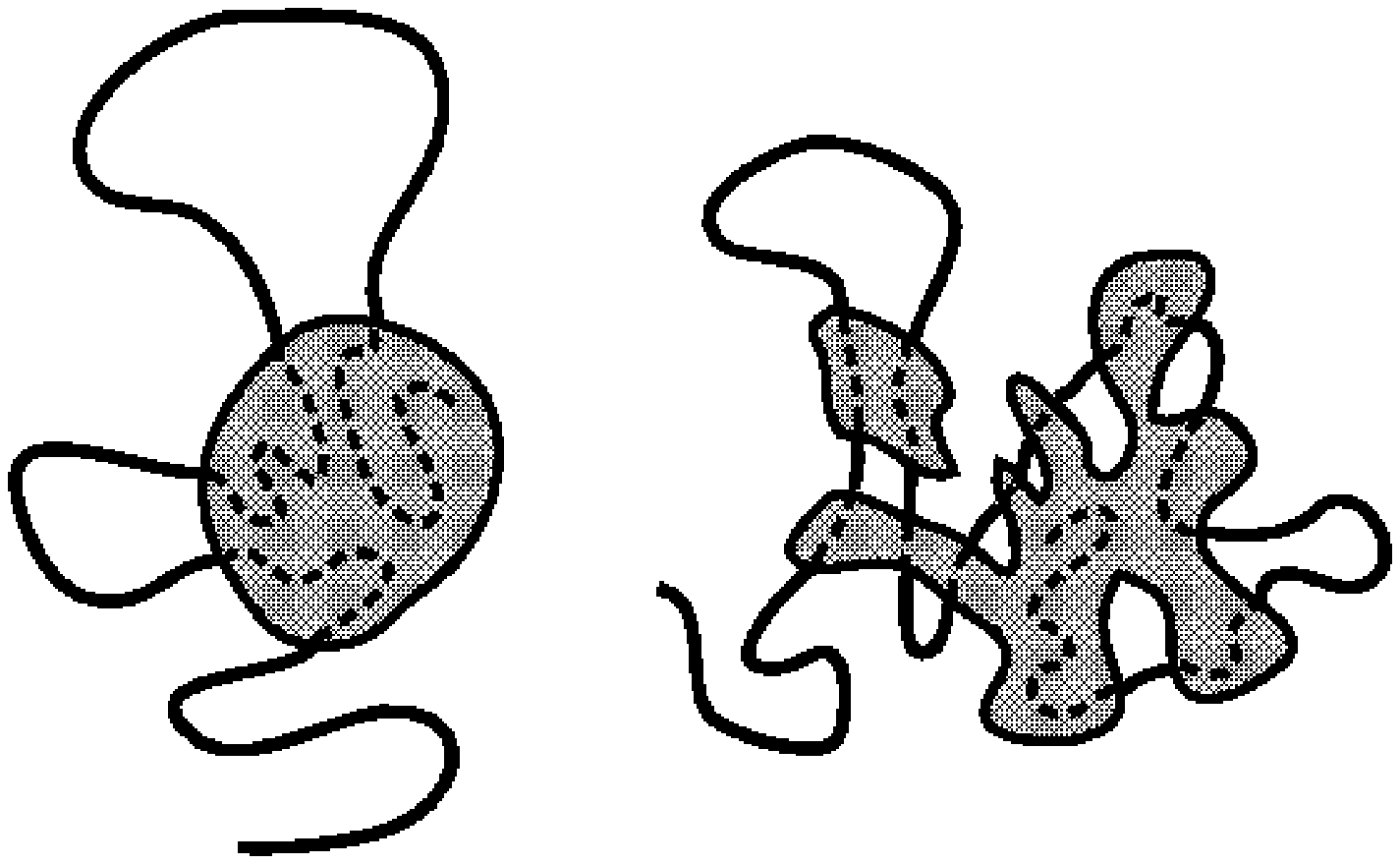,height=8.5cm,width=5cm,angle=-90}
\caption{}
\label{core}
\end{figure}

\begin{figure}[htb]
\hspace{.7cm}
\psfig{file=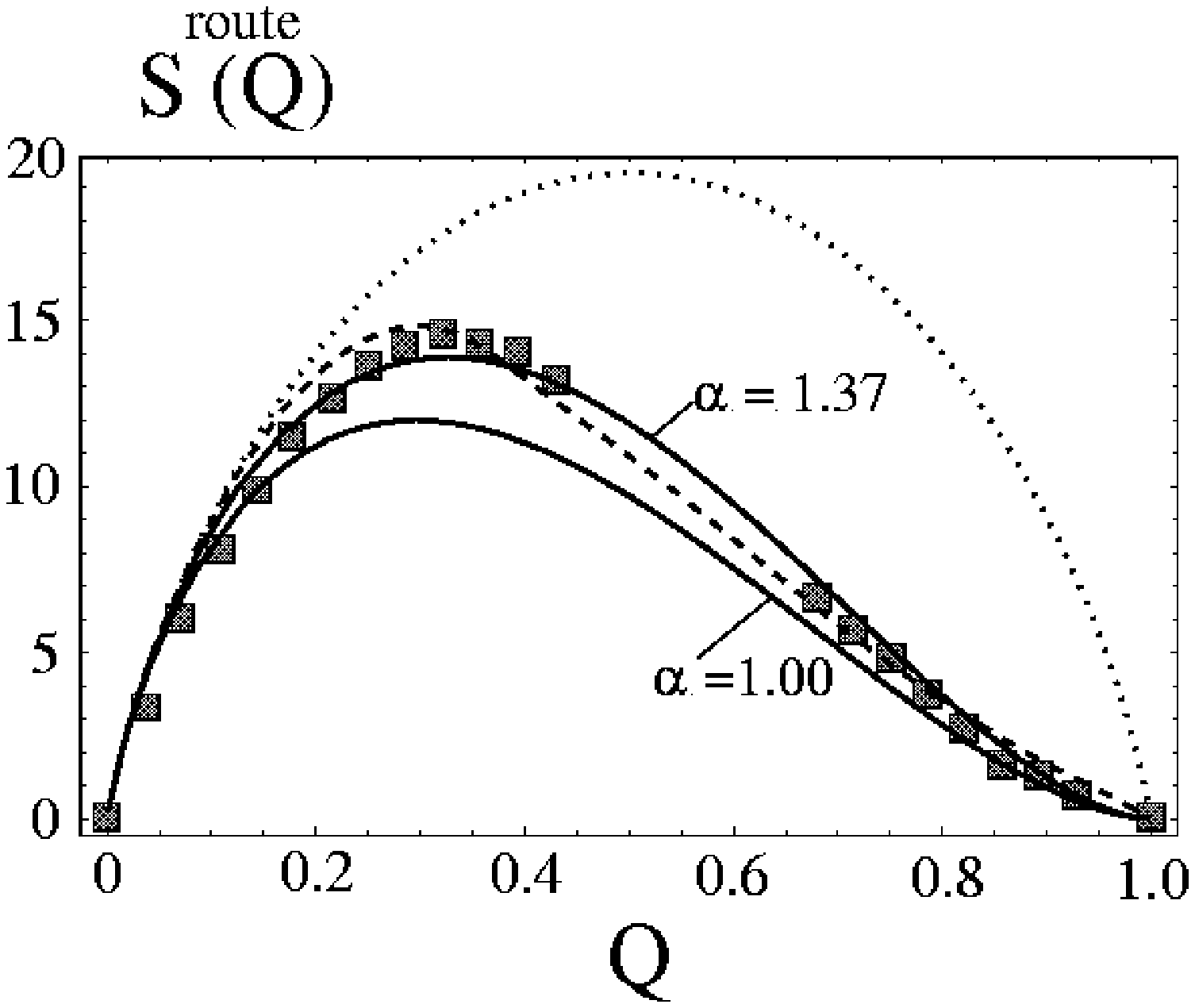,height=8.5cm,width=8.5cm,angle=0}
\caption{}
\label{fig:mix}
\end{figure}

\begin{figure}[htb]
\hspace{.7cm}
\psfig{file=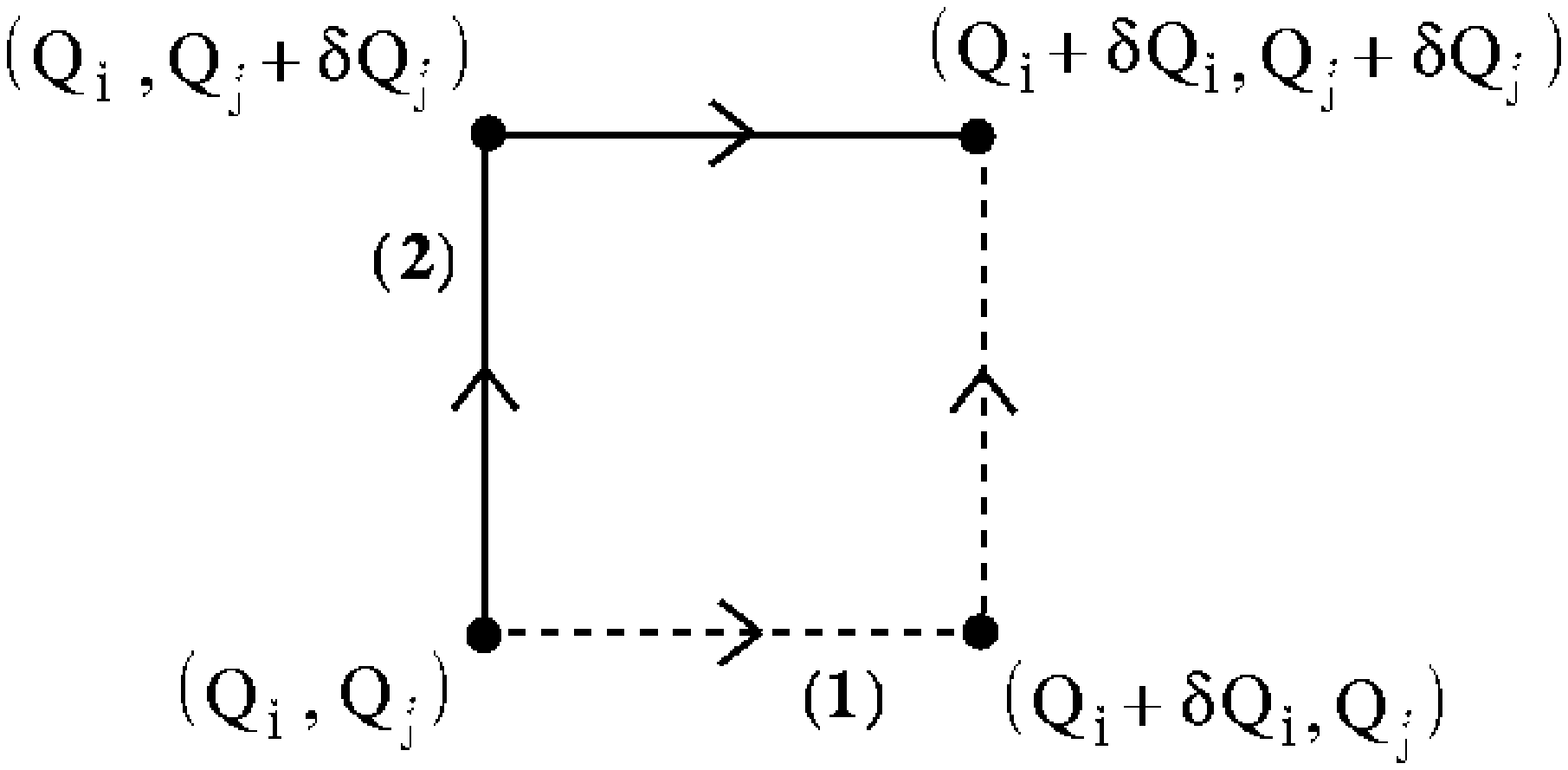,height=8.5cm,width=4.5cm,angle=-90}
\caption{}
\label{fig:paths}
\end{figure}

\begin{figure}[htb]
\hspace{.7cm}
\psfig{file=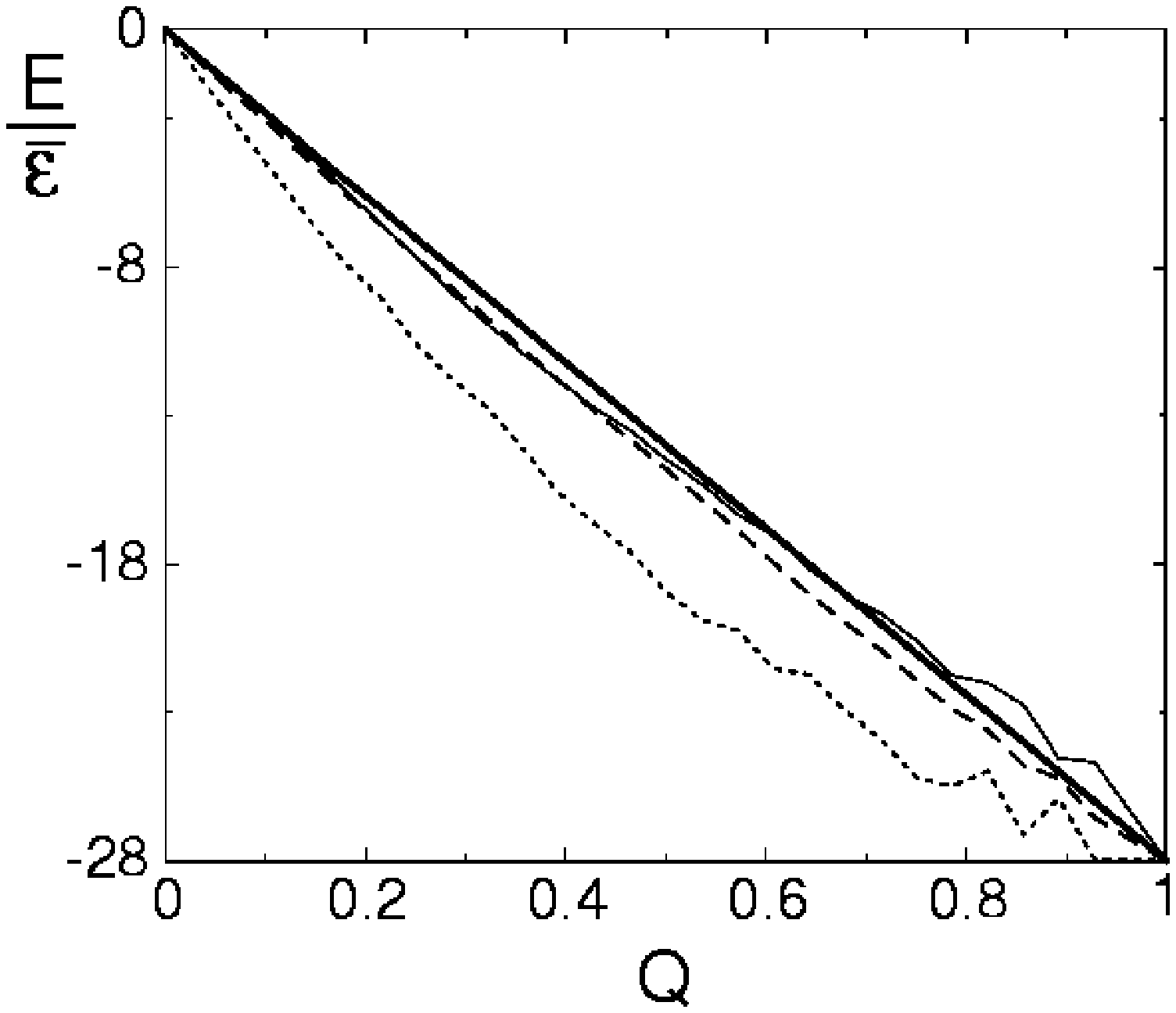,height=7cm,width=8cm,angle=0}
\caption{}
\label{fig:Eplot}
\end{figure}

\begin{figure}[htb]
\hspace{.7cm}
\psfig{file=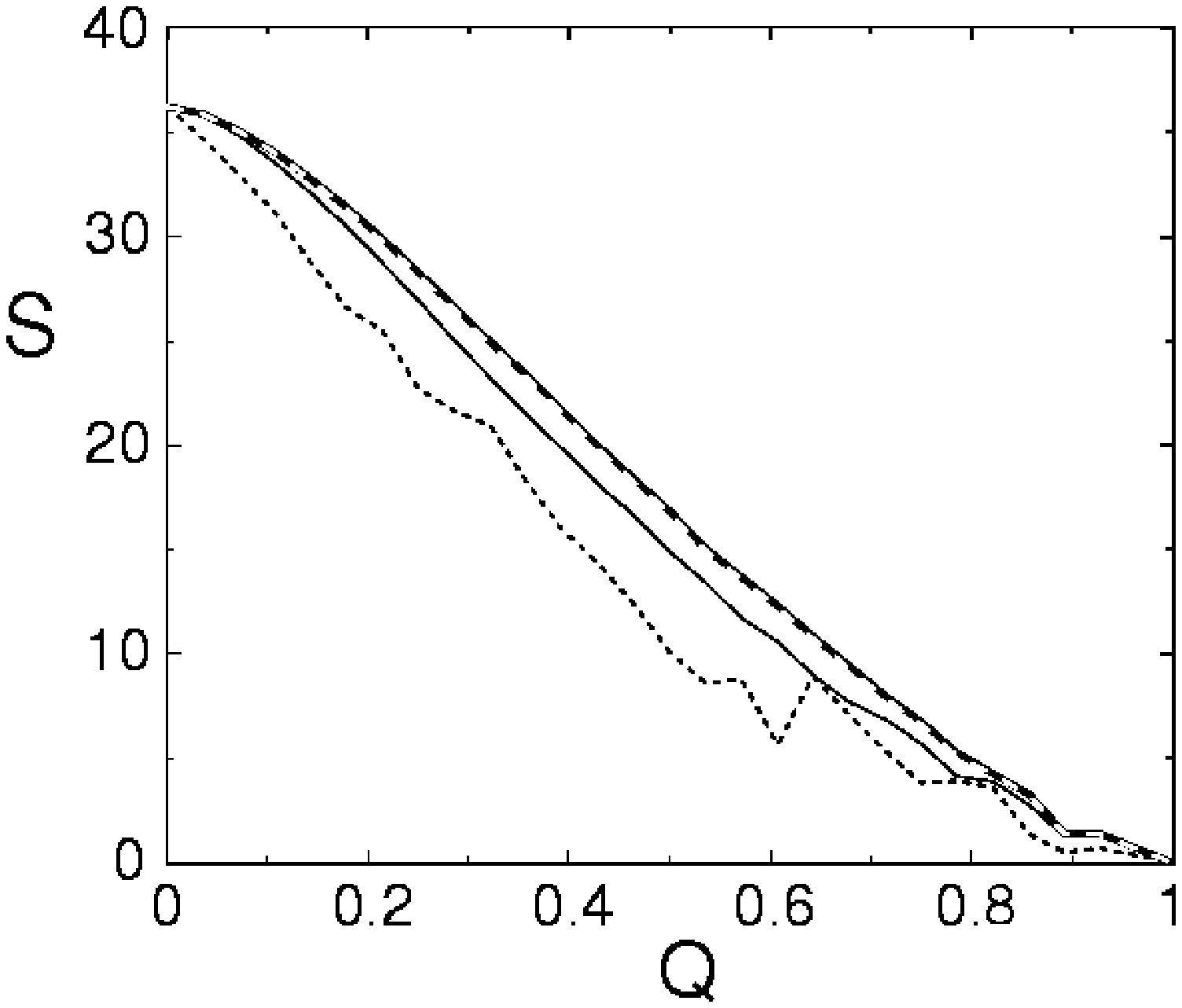,height=7.0cm,width=8.0cm,angle=0}
\caption{}
\label{fig:Splot}
\end{figure}

In general if there are $MQ$ contacts made of $M$ total, with a
renormalization factor $\wq$, there are 
\be
\Omega_{\mbox{\tiny ROUTE}} \left( Q \right)  = 
\sum_{j=0}^{\jmax}  
\frac{ \left(M/\wq\right)!}
{ \left[\mbox{int} \left( MQ/\wq + j + 1/2 \right) \right]!
\left[ M/\wq - \mbox{int} \left( MQ/\wq + j + 1/2\right)\right]! }
\label{omsum}
\ee
total configurations. 
Here $\jmax = \mbox{int} \left[ \mbox{min} \left(MQ,M/\wq\right)-\left(
MQ/\wq + 1/2 \right) \right]$.
For example if $M=8$, $MQ=5$, and $\omega_{5/8}
= 2$, there are $5$ total configurations instead of $56$ assuming
binary mixing.
In the thermodynamic limit the entropy is obtained from the largest
term in the sum of~(\ref{omsum}):
\be
\Omega_{\mbox{\tiny ROUTE}} \left( Q \right)  =  \maxj \;\;
\frac{ \left(M/\wq\right)!}
{ \left[\mbox{int} \left( MQ/\wq + j + 1/2 \right) \right]!
\left[ M/\wq - \mbox{int} \left( MQ/\wq + j + 1/2\right)\right]! } \: ,
\label{eq:maxj}
\ee
so that taking $\D \ln \Omega /\D j = 0$ gives
\bea
j^{\star}\left( Q\right) &\cong& \frac{M}{\wq}\left( 1/2 - Q\right)  \\ 
\mbox{where}&& \; 0 \leq j^{\star}\left( Q\right) \leq 
\mbox{min} \left(MQ,M/\wq\right)- MQ/\wq \: .
\label{bound}
\eea
Letting ${\cal J}^{\star}\left( Q\right) = M j^{\star}\left( Q\right)$
and applying the boundary condition eq.~(\ref{bound}) gives
\be
{\cal J}^{\star}\left( Q\right) = \left\{ \begin{array}{ll} 
	0 			& Q > 1/2 \\
	\left(1/2-Q\right)/\wq  & 1/2 \geq Q > Q^{\ast} \\
	Q\left(1-1/\wq\right)   & Q^{\ast} \geq Q \geq 0 
	\end{array}
\right.
\ee
where $Q^{\ast}$ solves $Q^{\ast} = 1/2\omega_{\Qast}$. The route
entropy is then
\bea
\frac{{\cal S}_{\mbox{\tiny ROUTE}} \left( Q\right)}{M} &=& \frac{1}{M} \ln
\Omega_{\mbox{\tiny ROUTE}} \left( Q, {\cal J}^{\star}\left( Q\right)
\right) \nonumber \\
&=& -\frac{1}{\wq} \ln \wq - \left(\frac{Q}{\wq}+\JQ\right) \ln
\left(\frac{Q}{\wq}+\JQ\right) \nonumber \\
&-& \left(\frac{1}{\wq} -
\frac{Q}{\wq}-\JQ\right) \ln \left(\frac{1}{\wq}
-\frac{Q}{\wq}-\JQ\right) \: .
\label{smixJ}
\eea
In the limit of the cell size $\wq\rightarrow 1$, $\JQ\rightarrow 0$
and eq.~(\ref{smixJ}) reduces to the binary fluid mixing entropy
eq.s~(\ref{mix1}) and~(\ref{mix2}), but for $\wq > 1$ the mixing
entropy is reduced. The dashed line in figure~\ref{fig:mix} shows 
a best fit to the lattice data for $\wq$ of the form $\exp (\alpha Q^{\beta})$.
This gives $\alpha \cong 2.08$ and $\beta\cong 1.85$, whose cell size
$\wq$ is shown in fig.~{\ref{fig:wq}.

\newpage
\begin{table}[t]
\begin{minipage}{1.0\linewidth}
\caption{\scriptsize
PARAMETERS USED IN THE ANALYTIC MODEL TO COMPARE WITH
SIMULATION
}
\begin{tabular}{cccccccc}
{Chain} & {Average } & {Average  } & {Average} & {Configurational} 
& {R.M.S.} & {Reduction} \\
{length} & {bonds} & {(homopolymer)} & {extra} & {entropy} & {energy per} 
& {parameter}  \\
{} & {per residue} & {attraction} & {native} & {per residue} & {non-native} 
& {in route} \\
{} & {} & {per contact} & {stability} & {} & {contact} 
& {entropy} \\
{} & {} & {} & {per contact} & {} & {(roughness)} & {} \\
(N) & ($z$) & ($\overline{E}/zN$) & 
($\overline{\epsilon}$)
&  ($s_o$) & ($b$) & ($\alpha$)  \\
\hline
 $27$ &  $28/27$ &  $-0.0$  &  $-3.0$ &  $1.352$ &  $0.0$ &
 $0.9$ \\
\end{tabular}
\begin{tabular}{p{2.50cm}p{12cm}}
{loop length} & {} \\
{distribution} & 
$\left\{ 3,3,3,3,3,5,5,5,5,7,7,7,7,7,7,9,9,9,9,9,9,11,11,13,21,23,23,23
\right\}$ \\ 
{} & {} \\
{energies inducing} & 
{$-\left\{ 4.9,4.5,2.1,2.6,2.7,5.0,4.1,2.3,6.0,1.0,
2.1,1.6,2.5,4.0, \right. $} \\
{uniform folding} & {
$\left. \;\;\;\;\;1.7,3.0,5.0,3.7,2.0,2.4,1.1,2.1,3.3,3.1,1.0,5.5,2.9,2.0
\right\} $} \\
\end{tabular}
\label{table1}
\end{minipage}
\end{table}

\newpage
\subsection{Figure Captions}

FIG.~\ref{fig1} $\;\;$
Schematic, lattice, and off-lattice representations of the
native structure, 
characterized through the 
distribution of contact energies $\setei$ and contact entropies $\setsi$,
(defined  through the distribution of loop lengths $\setli$). The
probability to form contact $i$ having energy $\ei$ and loop length $\li$
is $Q_i$. 

FIG.~\ref{fig:QiQ} $\;\;$
The fraction of time a contact is made vs. $Q$,
$\Qi^{\star}(Q)$ for folding to the lattice structure shown in
fig.~1. Dashed black curves are the result of the functional theory of
section~\ref{sect:f}, using an energy function given by
eq.~(\ref{Etune}) with $\alpha =0.5$. Thin solid curves are
Monte Carlo simulation results 
for folding to this structure, using eq.~(\ref{Etune}) with $\alpha =
1$. Some contacts are formed relatively
early while others remain unformed until the protein is largely
native; the magnitude of this dispersion in contact formation is
captured fairly well by the theory.

FIG.~\ref{fig:eqopp} $\;\;$
Making contact $1$ stronger in the $\alpha$-spectrin SH3 domain should
speed folding more 
than making contact $2$ stronger by the same amount, since contact $1$
in the distal loop is more likely to be formed in the transition
state than contact $2$ in the RT loop.~\cite{SerranoL98nsb,BakerD98nsb}.

FIG.~\ref{opt} $\;\;$
The reduction in effective thermodynamic barrier due to the
presence of native heterogeneity. The probability of a
nucleation event dies off exponentially with barrier height $k \sim
\exp(-F/T)$, and the probability distribution $P(F)$ of barrier heights is
centered on $\Fbardag$ and has some width in the presence of energetic
and entropic inhomogeneity. The effective nucleation rate $k(F)
\sim P(F) k(F)$ has a maximum at an effective barrier $\Fstar(T) < \Fbardag$
corresponding to the optimum fluctuation. Figure adapted from
ref.~\cite{Oxtoby96}. 

FIG.~\ref{chey}  $\;\;$
Free energy vs. fraction of native contacts $Q$, obtained
from off-lattice simulations using a uniform G\={o}
potential to the native structure of
CheY (data from~\cite{ClementiC00:jmb}). Fluctuations in the free
energy profile 
here depend solely on how the native topology affects the entropy at 
partial degrees of nativeness. The profile observed is the optimal
fluctuation from uniform ordering, given the native structure under
study.

FIG.~\ref{core} $\;\;$
Illustration of partially native configurations consisting of
native cores and the surrounding polymer
halo. The core may be globular or ramified.

FIG.~\ref{fig:mix} $\;\;$
Route entropy for the $27$-mer with native structure shown in
fig.~\ref{fig1}. The 
dotted curve is the putative binary fluid mixing entropy in the
absence of the backbone. The solid curves
includes a prefactor $(1-Q^{\alpha})$ corresponding to the entropic
asymmetry of
applying contacts to an unconstrained polymer and removing contacts
from a fully constrained polymer, described in the text. The data are
for the $27$-mer lattice model shown in
fig.~\ref{fig1}~\cite{Nymeyer:thank}, and are obtained for low
$Q$ values by making  
subsets of $M (1-Q)$ contacts repulsive, $M Q$ contacts attractive,
and then finding the most native-like state in a low temperature
quench.  For high $Q$ values they are obtained by making random sets of 
$M (1-Q)$ contacts repulsive, and counting the remaining native cores
which are distinct. This method finds the reduction in binary fluid
mixing entropy due to chain connectivity and particular native
topology of the protein under study. We assume that $\alpha$ is
essentially constant for a given native structure, independent of the
distribution of native energies.
The computation
treats all contact formation
probabilities on an equal footing (all $\Qi = Q$), and so the route
entropy plotted is an upper
limit to the actual route entropy present. See also
fig.~\ref{fig:sroutehet}. 
Using eq.s~(\ref{smix}), (\ref{eq:lambdaQ}) with $\Qi =Q$ gives
$\alpha=1.37$ for the best fit to the lattice structure in
fig.~\ref{fig1}, while somewhat
smaller values of $\alpha$ fit the free energy function best 
(table~\ref{table1} ). 
The dashed curve is the best fit of a simple microscopic theory to the
route entropy, described in Appendix~B.

FIG.~\ref{fig:paths} $\;\;$
Illustrating constraints on the functional form of the
entropy, given it must be a state function. 
Path (1) dashed, Path (2) solid.

FIG.~\ref{fig:Eplot} $\;\;$
The total energy of the $27$-mer lattice model, in units of $\ebar$, 
as a function of the fraction of native
contacts $Q$, for native energies $\seteistar$ tuned to give a homogeneous
transition state  (thin solid curve), for uniform native energies
$\{ \ebar \}$ (thick solid  line), for native energies $\seteio$ which
induce enough heterogeneity to eliminate the barrier (long dashed),
and for native energies $\seteir$ which induce a folding mechanism by
only a few routes (short dashed). 

FIG.~\ref{fig:Splot} $\;\;$
Total entropy vs. $Q$ for the $27$-mer lattice model. 
The maximal solid curve is for $\setei = \setebar$. For these energies
the entropy has its highest value.
The entropy for the homogeneous funnel with tuned energies
$\seteistar$ (thin solid curve) is significantly lower. 
For native energies $\seteio$ which eliminate the barrier (long
dashed), the entropy is essentially unchanged, since changes in
entropy near the extremum (where $\setei =\setebar$) are second
order. The entropy for a route-like folding mechanism (short dashed)
is lower than the other curves since the route entropy (eq.~\ref{smix}) no
longer contributes significantly to the total entropy at temperatures
where the native state is stable.

FIG.~\ref{fig:Fplot} $\;\;$
Free energy vs. $Q$ in units of $\ebar$ at the folding temperature
$\Tf(\setei)$ for the 
$27$-mer lattice model (TOP) and theory (BOTTOM), for the energy
functions described in the text 
and previous $2$ figures. 

FIG.~\ref{fig:sdplot}  $\;\;$
Barrier height in units of native interaction strength as a function
of RMS native energetic variance in the same units, for simulations of
the G\={o} lattice model. The barrier height is multi-valued here
because of energetic correlations with loop lengths, as described in
the text. Also shown is the folding transition temperature, which
remains fairly constant until the barrier disappears and the RMS
energetic variance becomes large. Another 
measure of how well-designed the protein is weights $\Tf$ by the root
entropy per monomer at the barrier peak, which also remains fairly
constant until the barrier disappears (thin solid line with squares).

FIG.~\ref{fig:sroutehet}  $\;\;$
Route entropy for the $27$-mer lattice model (Top) and free energy
functional theory (Bottom), as given by equation~(\ref{smix}). 
The route entropy is a thermodynamic
measure of how labile or itinerant the native core of the protein is at
intermediate stages of folding. Different line styles represent
different folding mechanisms described in the text (the same code is
used as in fig.s~\ref{fig:Eplot}, \ref{fig:Splot}, and~\ref{fig:Fplot}).

FIG.~\ref{fig:QivsQroute}  $\;\;$
Contact formation probability $\Qi(Q)$ as a function of $Q$. (TOP)
Simulation results, (BOTTOM) Analytical theory. (LEFT) Coupling
energies are tuned to induce a route-like folding mechanism, (RIGHT)
coupling energies are tuned to induce a uniform funnel folding
mechanism through the barrier peak.

FIG.~\ref{fig:stotflory}  $\;\;$
Flory entropy of the system, defined by subtracting the route
entropy from  the total entropy, as a function of the fraction of
native contacts. The top plot is the result from
simulating the lattice $27$-mer. The bottom plot is the result of the
analytical theory (the same line styles as fig.s~\ref{fig:Eplot}, \ref{fig:Splot},
\ref{fig:Fplot}, and~\ref{fig:sroutehet} are used).

FIG.~\ref{fig:simthrytest}   $\;\;$
Comparison of the theory and simulations, for a set of energies
inducing an intermediate folding
barrier height in (A),(B) ($\alpha_{\mbox{\tiny
SIM}}=1.0$, $\alpha_{\mbox{\tiny THRY}} = 0.5$ in eq.~(\ref{Etune})),
and for energies $\eistar$ which induce a uniform folding mechanism (C).
(A) Contact probabilities at the barrier peak. 
(B) Contact probability vs. log loop length. Crosses are the lattice
data, squares are their average, and the solid line is the theory
result (eq.~\ref{Qstar} with eq.s~(\ref{sbond}) and~(\ref{leff})). (C)
plots the energy fluctuation required to tune the 
system to fold uniformly vs. the entropy fluctuation each loop has. The solid line
is the theoretical result eq.~(\ref{eq:el}) and the data points with
error bars are extracted from the simulations.

FIG.~\ref{fig:FvsR}  $\;\;$
Free energy barrier at the transition temperature $\Tf$ in
units of $\ebar$, vs. the route measure at the barrier peak ${\cal
R}(\Qddag)$. (Solid line) theoretical
result for a model with parameters in table~\ref{table1}. (Long
dashed) idealized funnel with $\li = \lbar$, and $\ei = \ebar$
initially. (Short dashed) perturbation result from eq.~(\ref{dfhet}).

FIG.~\ref{fig:tau}  $\;\;$
Simulating a long Monte Carlo run for the $27$-mer lattice model at
$\Tf$ yields an exponential distribution of first passage times to the folded
state from the unfolded state(inset B, plotted for the point with
${\cal R} = 0.11$), indicating a single time-scale
governs the kinetics. The mean of 
the distribution (the folding time $\tauF$) is plotted here for a few
energy functions which 
result in varying amounts of routing through the barrier by
eq.~(\ref{eq:route}). For comparison, the folding time determined from
$\tau = \mbox{const.}\times \exp (-\DFdag/\Tf)$ is shown as a solid
line, with $\DFdag({\cal R})$ taken from the lattice data in
fig.~\ref{fig:FvsR} (there  
are error bars about the points defining this line which are not shown).
The folding time is a decreasing function of the
route measure ${\cal R}(\Qdag)$. (Inset A) The log of the folding time
varies nearly linearly with the barrier height indicating that the
prefactor is roughly constant while the native energy distributions
are varied.

FIG.~\ref{fig:TTFF}  $\;\;$ 
The relative barrier height decreases more strongly than the relative
folding temperature, as a function of route measure at the barrier
peak (eq.~\ref{eq:route}). Thus the condition for the folding rate to
increase, eq.~(\ref{eq:FFTT}), is satisfied.

FIG.~\ref{fig:LL}  $\;\;$ 
Dependence of the free energy profile $F(Q)$ at $\Tf$ on the
mean loop length $\lbar$, for the analytic model with length
$N=50$, $\li =\lbar$, and $\ei = \ebar$ (eq.s~(\ref{FTOT})
and~(\ref{FMF})). $\lbar$ labels each curve. 
The barrier undergoes an increase that is stronger initially.
The inset plots the barrier height as a function of $\lbar$, in units
of $\ebar$.  The trend in barrier height with $\lbar$ shown here is a
lower limit to the full theoretical dependence given in
eq.~(\ref{sder}). 

FIG.~\ref{fig:Sco}  $\;\;$ 
Entropy vs. $Q$ for $\ei = \ebar$ and $\li = \lbar$, for various
$\lbar$. The contact order $\equiv \lbar / N$ ($N=27$ here) labels
each curve. Results are for the theory with parameters in table I (c.f
eq.~\ref{eq:smf}). As $\lbar$ increases, more entropy is lost
initially, leading to a larger free energy barrier and correspondingly
slower folding rate.
(INSET): The model shows a weak increasing dependence of $\theta_m'$
value with contact order, defined here as the relative degree of
partial order at the 
barrier peak:  $\theta_m' \equiv (\Qdag - \Qu)/(1-\Qu)$. The trends
seen here are again lower limits to the full dependence on $\lbar$
given in eq.~(\ref{sder}), we illustrate just the mean-field term
here. 

FIG.~\ref{fig:sv}  $\;\;$ 
Log of the ratio of rates given by eq.~(\ref{svfull}) as a function of
structural variance $\overline{\dl^2}$ at fixed $\lbar$, obtained by
following the recipe of eq.~(\ref{lrec}). (Dashed) Approximate
perturbation result of eq.~(\ref{svpert}). (Solid) Full
non-perturbative result using eq.s~(\ref{FTOT}) and~(\ref{Qstar}),
which accounts for changes in the unfolded free energy with increasing
variance. The barrier is calculated at $\Tf$, which changes only mildly
with $\overline{\dl^2}$ until the barrier height approaches zero at
$\overline{\dl^2} / \lbar^2 \approx 0.25$.

FIG.~\ref{fig:F}  $\;\;$ 
{\bf (A)} $\;$ Squares are the free energy potential obtained in
ref.~\cite{SocciND96:jcp} to the structure shown in fig.~\ref{fig1}B. 
Solid curve is a fit to the data by
annealing the native coupling energies using eq.~(\ref{eq:anneal}}).
({\it INSET}): The effective force $\mu(Q)$. The two zeros of
$\mu$  occur at the positions of the unfolded state $\Qo$ and barrier
peak $\Qddag$. 
{\bf (B)} $\;$ Fitting to a potential containing an on-pathway
intermediate by 
adjusting the coupling energies $\setei$. 
The illustration here is for a system of length $N=57$
with $142$ contacts, having the native loop length distribution of the
$\alpha$-spectrin SH3 domain~\cite{ClementiC00:jmb}. The unfolded
and folded states are at $Q=0,\, 1$ since the bulk limit ($\lbar Q
\gg 1$)  was taken in determining the potentials here
(c.f. eq.~\ref{eq:smfflory}).
A specific part of the
protein is biased energetically with successively larger strength to
achieve an on-pathway intermediate.
This could also have been achieved in the theory by adjusting the
native structure to have contacts which are relatively likely
entropically. Finding the set(s) of loop lengths $\left\{ \li^{\star}
\right \}$ that best fit a given target potential is relevant to the
problem of which native structures would typically have such
a potential, if native energetic features are not as important. 
Progression of the effective force $\mu(Q)$ is also shown in {\bf (C)}; $\mu(Q)$
corresponding to the potential with the intermediate has $3$ roots,
one at the intermediate position, one at the barrier peak, and
one at a smaller barrier peak near the unfolded state. 
{\bf (D)} The barrier height here deviates from its typical dependence on the
route-measure, since the route-measure becomes largest at the value of
order parameter characterized by the intermediate, rather than at the
barrier peak.

FIG.~\ref{fig:wq}  $\;\;$ 
Renormalized mixing cell size $\wq$ as a function of $Q$,
assuming the form $\wq = \exp \left( \alpha Q^\beta \right)$ with
$\alpha = 2.08$ and $\beta = 1.85$, which are the numbers giving the
best fit to the lattice data (dashed curve in fig.~\ref{fig:mix}).

\newpage 
\subsection{Figures}

\begin{figure}[htb]
\hspace{.7cm}
\psfig{file=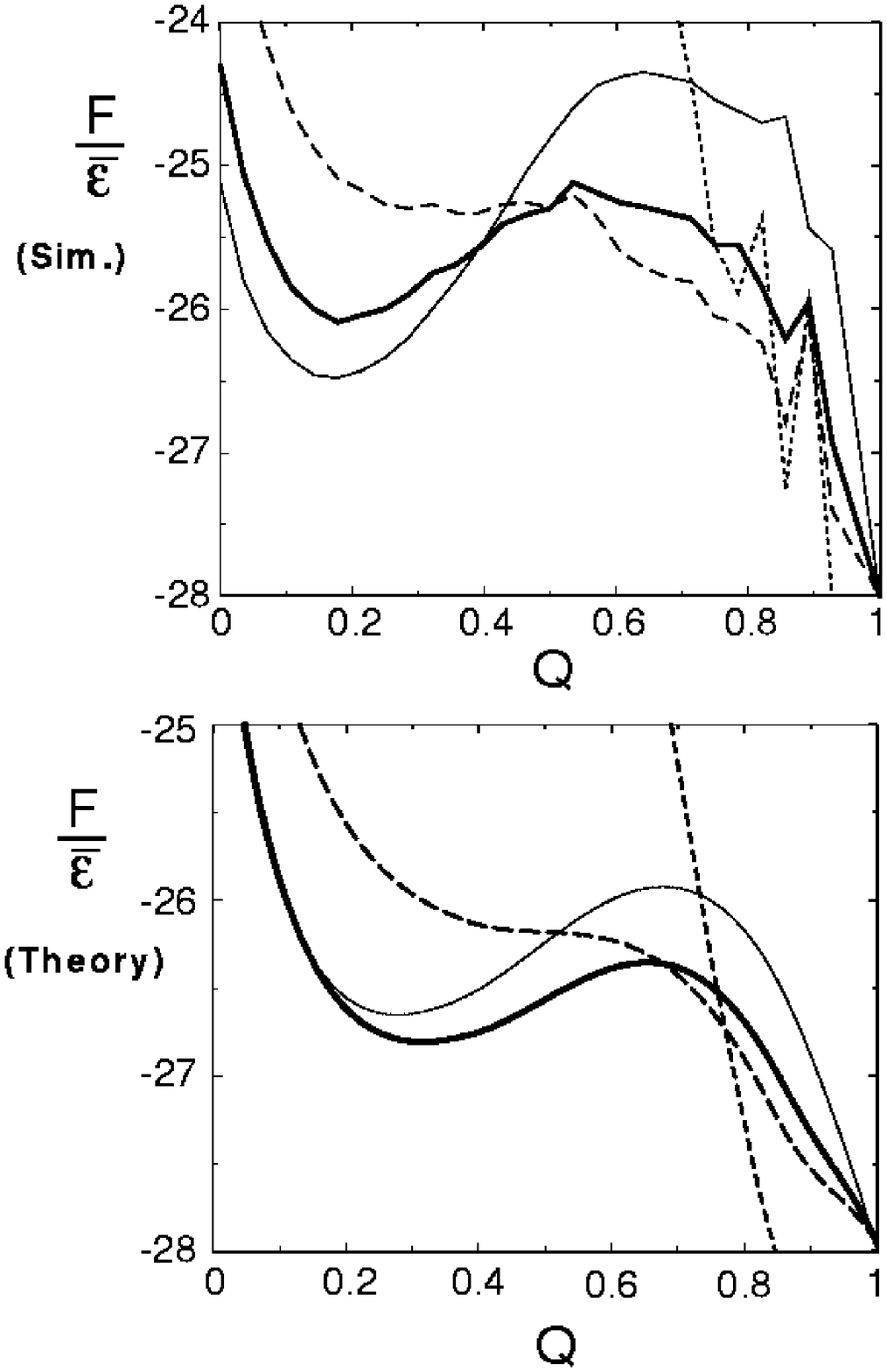,height=13.0cm,width=8.5cm,angle=0}
\caption{}
\label{fig:Fplot}
\end{figure}

\begin{figure}[htb]
\hspace{.7cm}
\psfig{file=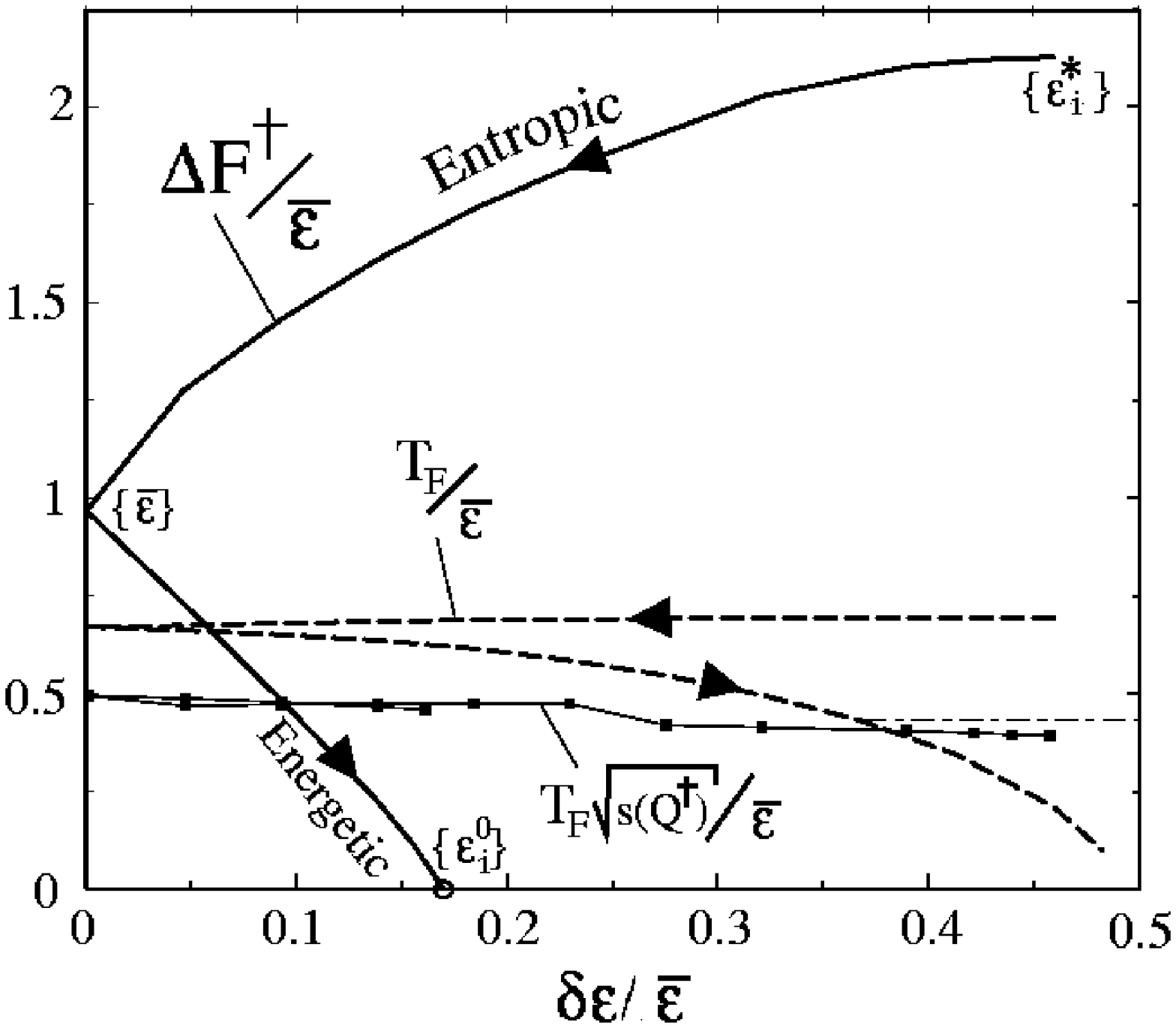,height=8.5cm,width=8cm,angle=-90}
\caption{}
\label{fig:sdplot}
\end{figure}

\begin{figure}[htb]
\hspace{.7cm}
\psfig{file=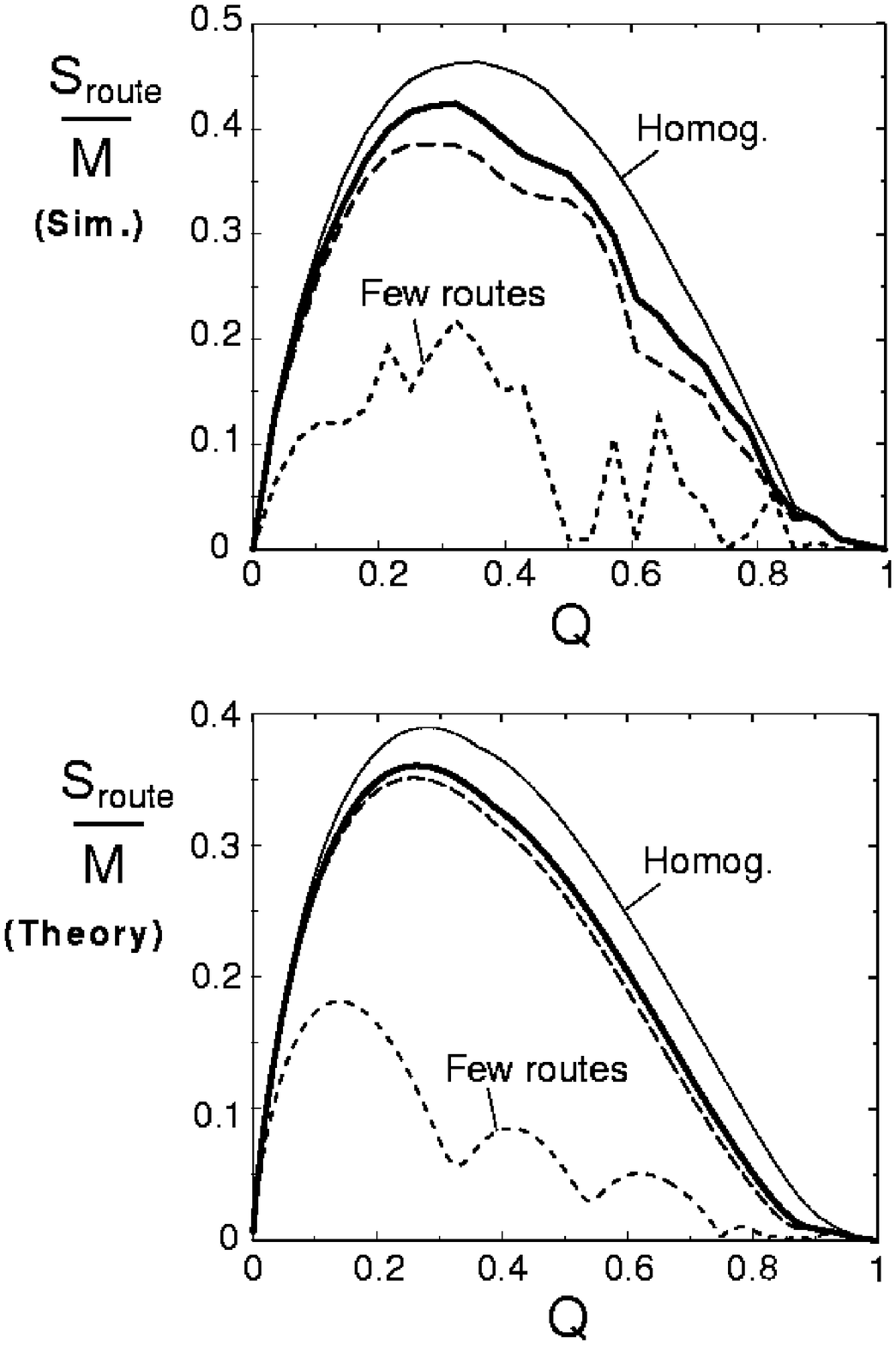,height=12.0cm,width=8.0cm,angle=0}
\caption{}
\label{fig:sroutehet}
\end{figure}

\begin{figure}[htb]
\hspace{.7cm}
\psfig{file=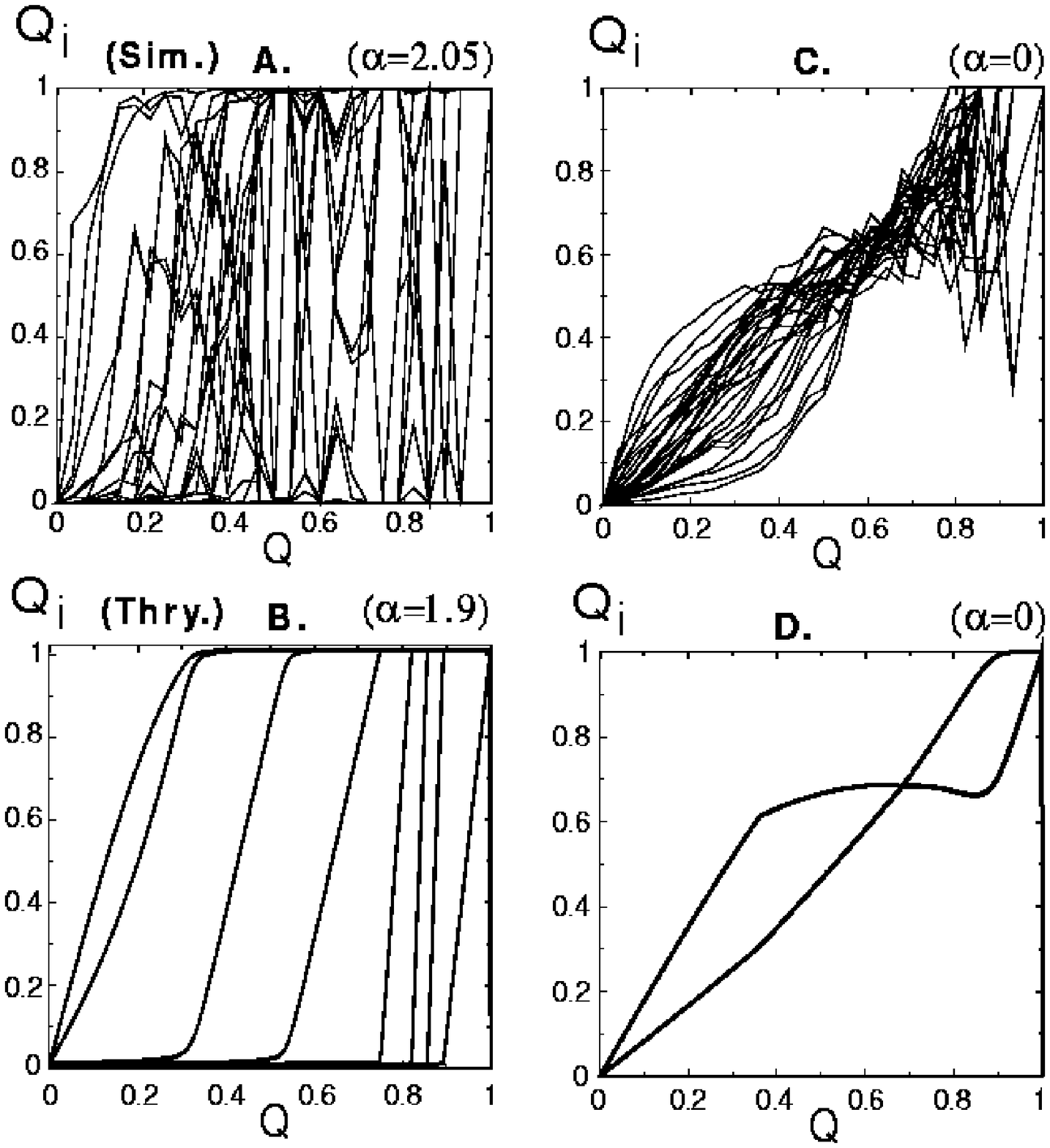,height=8.5cm,width=8.5cm,angle=0}
\caption{}
\label{fig:QivsQroute}
\end{figure}

\begin{figure}[htb]
\hspace{.7cm}
\psfig{file=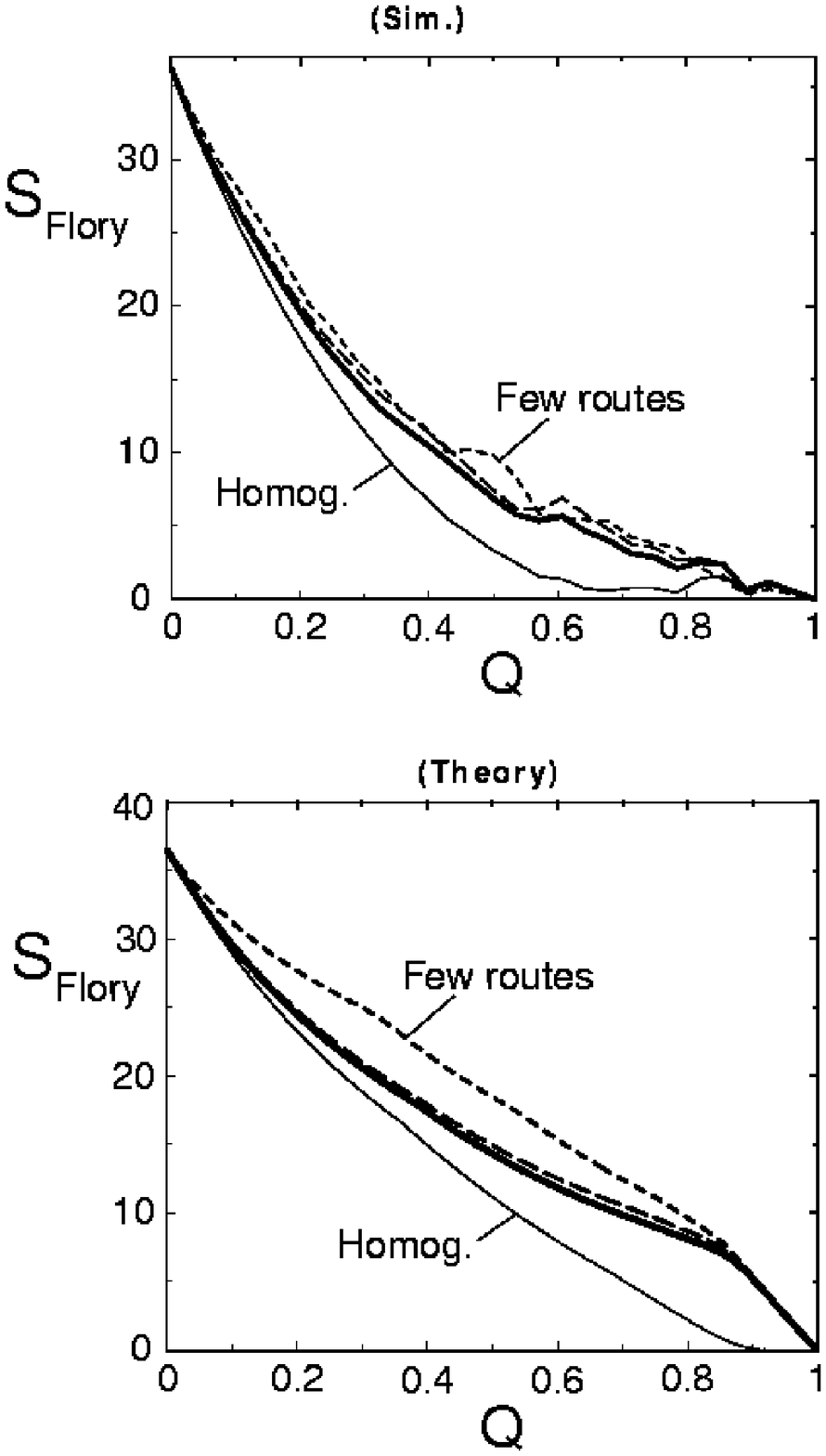,height=14cm,width=8.5cm,angle=0}
\caption{}
\label{fig:stotflory}
\end{figure}

\begin{figure}[htb]
\hspace{.7cm}
\psfig{file=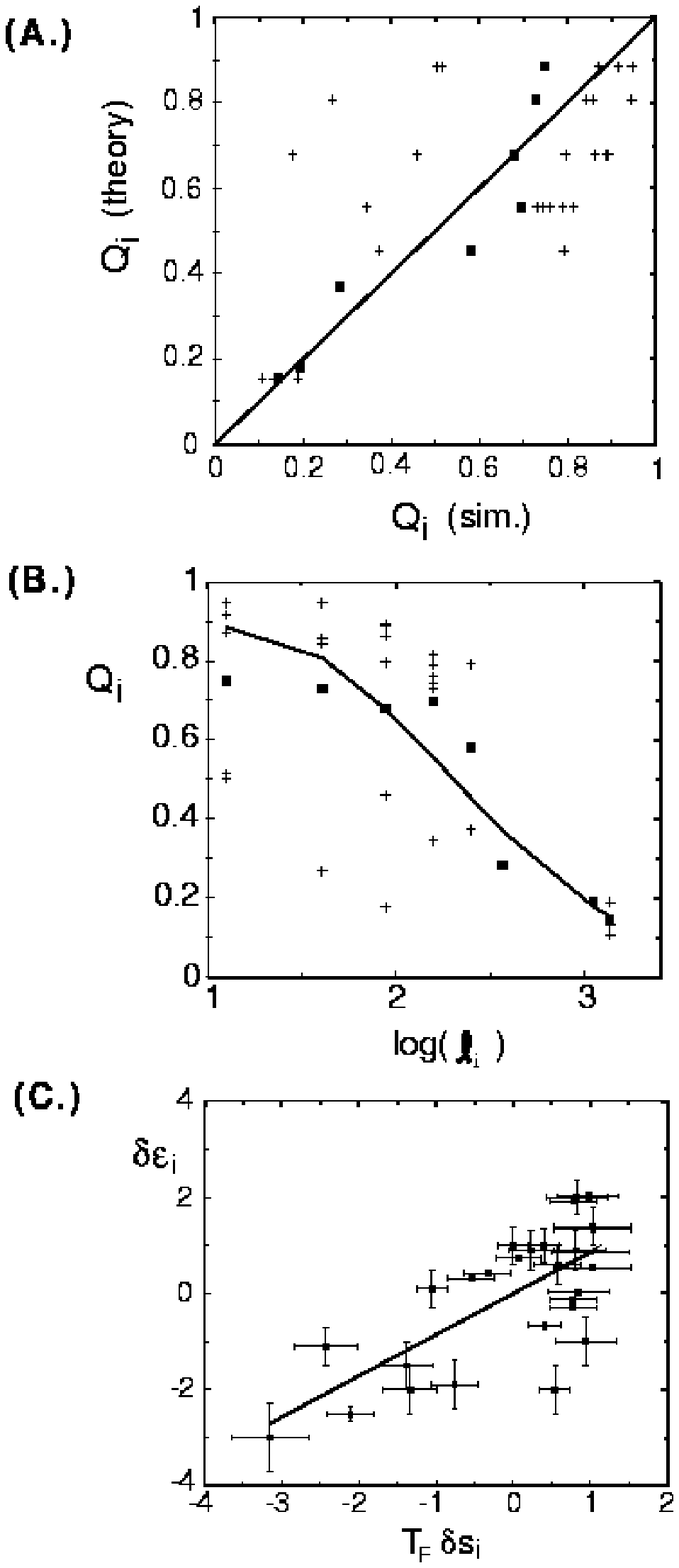,height=18cm,width=8.5cm,angle=0}
\caption{}
\label{fig:simthrytest}
\end{figure}

\begin{figure}[htb]
\hspace{.7cm}
\psfig{file=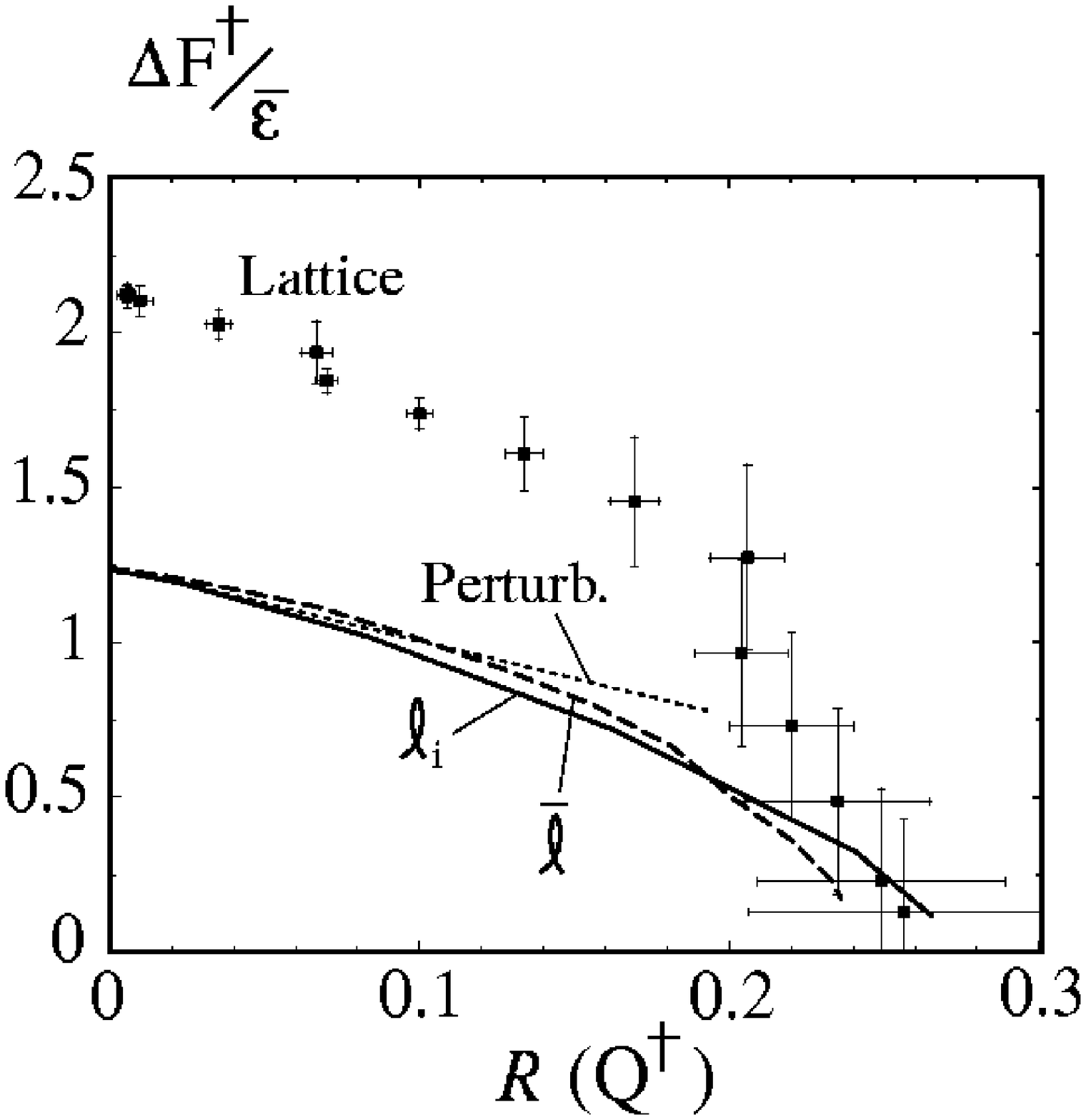,height=9.0cm,width=8.5cm,angle=0}
\caption{}
\label{fig:FvsR}
\end{figure}

\begin{figure}[htb]
\hspace{.7cm}
\psfig{file=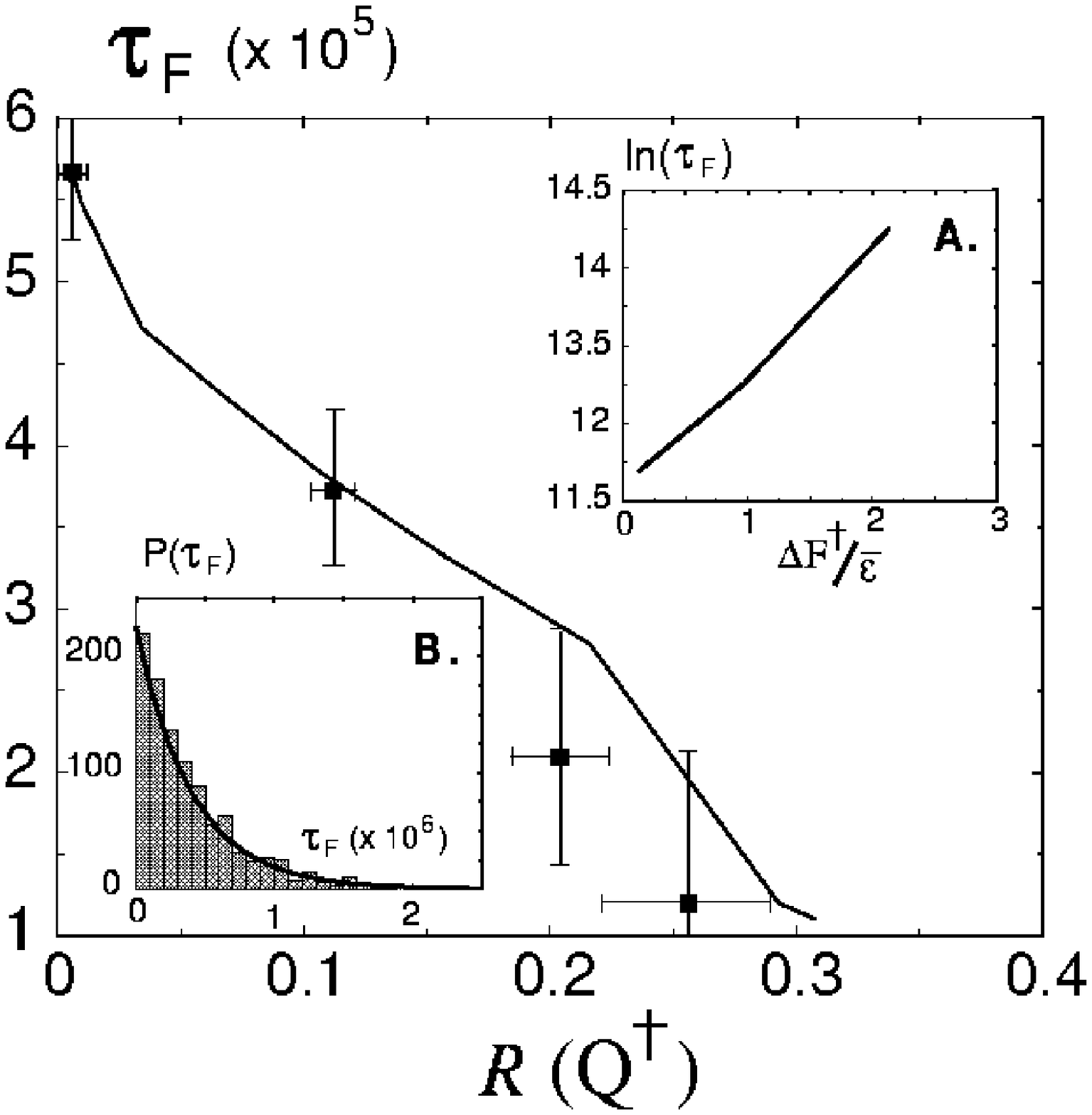,height=9cm,width=8.5cm,angle=0}
\caption{}
\label{fig:tau}
\end{figure}

\begin{figure}[htb]
\hspace{.7cm}
\psfig{file=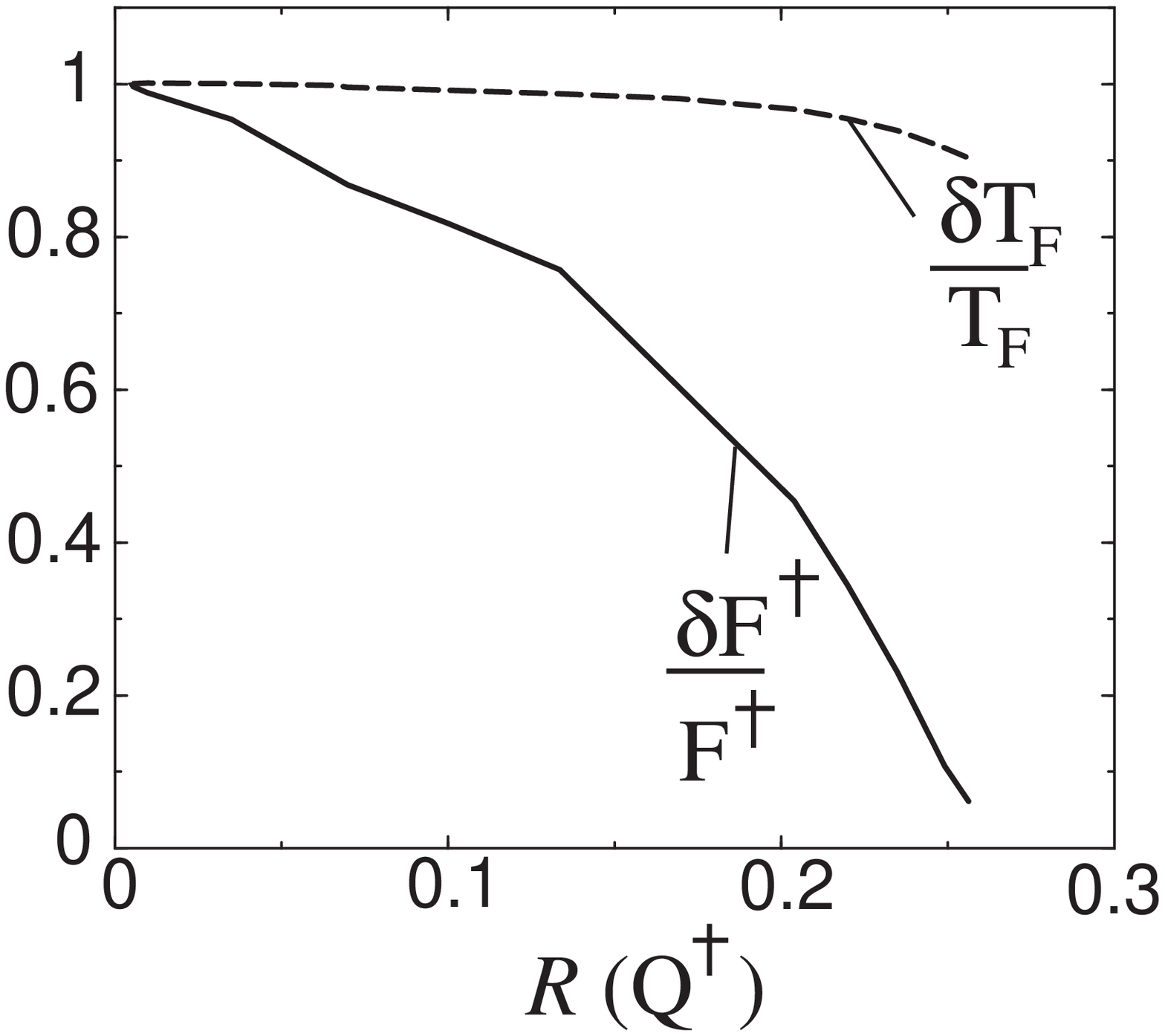,height=7.5cm,width=8.0cm,angle=0}
\caption{}
\label{fig:TTFF}
\end{figure}

\begin{figure}[htb]
\hspace{.7cm}
\psfig{file=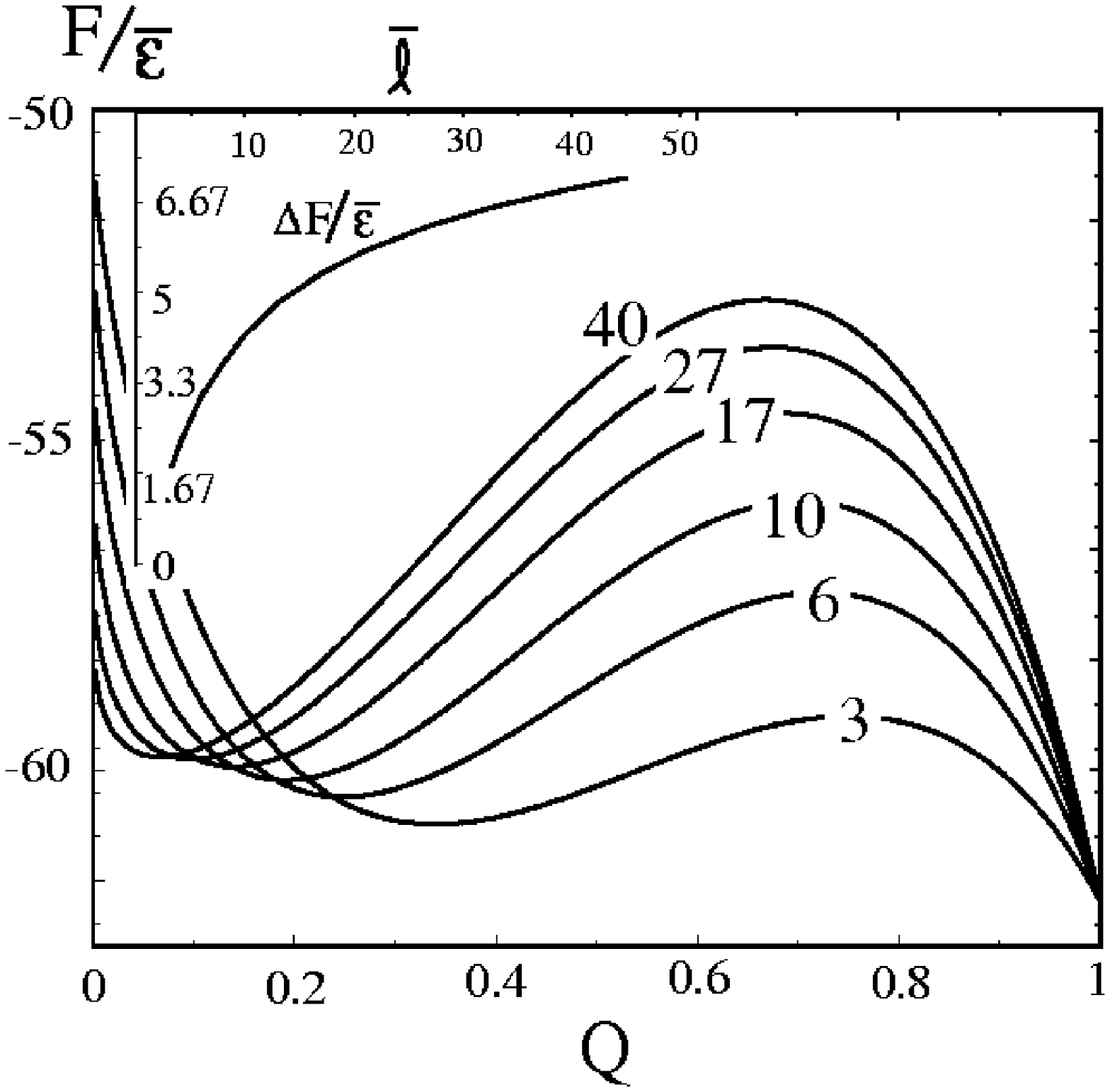,height=8cm,width=8.5cm,angle=0}
\caption{}
\label{fig:LL}
\end{figure}

\begin{figure}[htb]
\hspace{.7cm}
\psfig{file=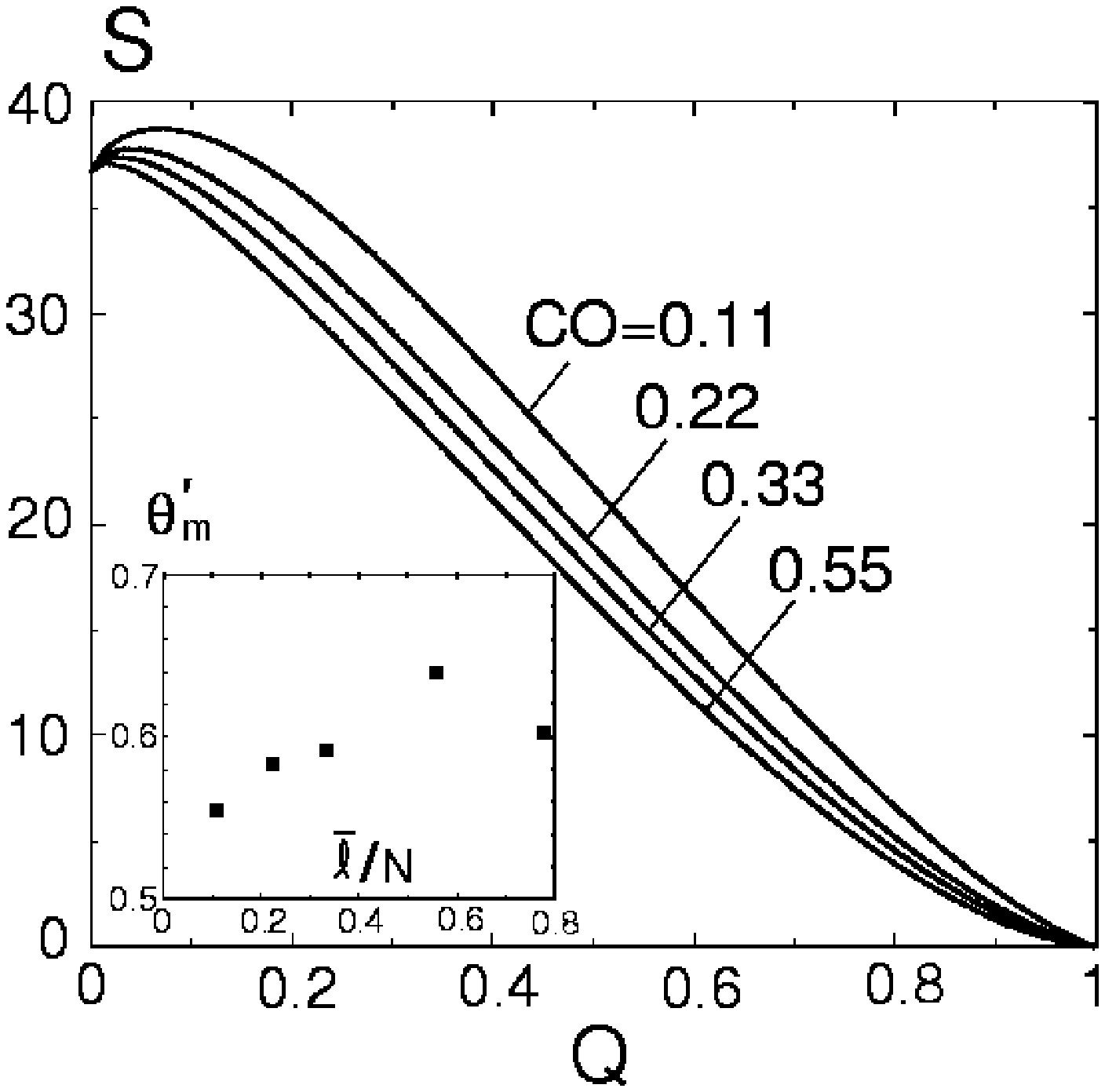,height=8.5cm,width=8.5cm,angle=0}
\caption{}
\label{fig:Sco}
\end{figure}

\begin{figure}[htb]
\hspace{.7cm}
\psfig{file=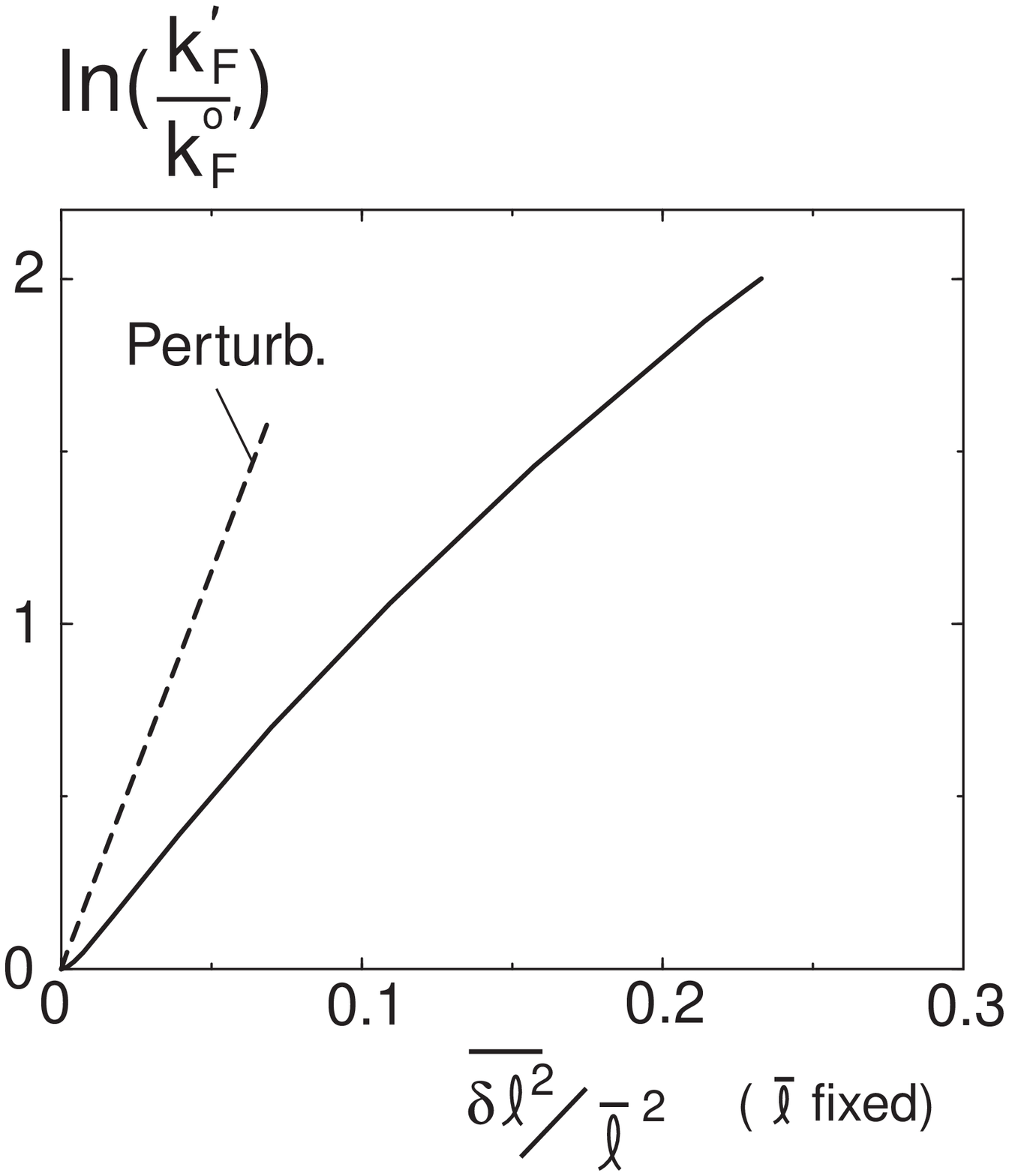,height=9cm,width=8.5cm,angle=0}
\caption{}
\label{fig:sv}
\end{figure}

\begin{figure}[htb]
\hspace{.7cm}
\psfig{file=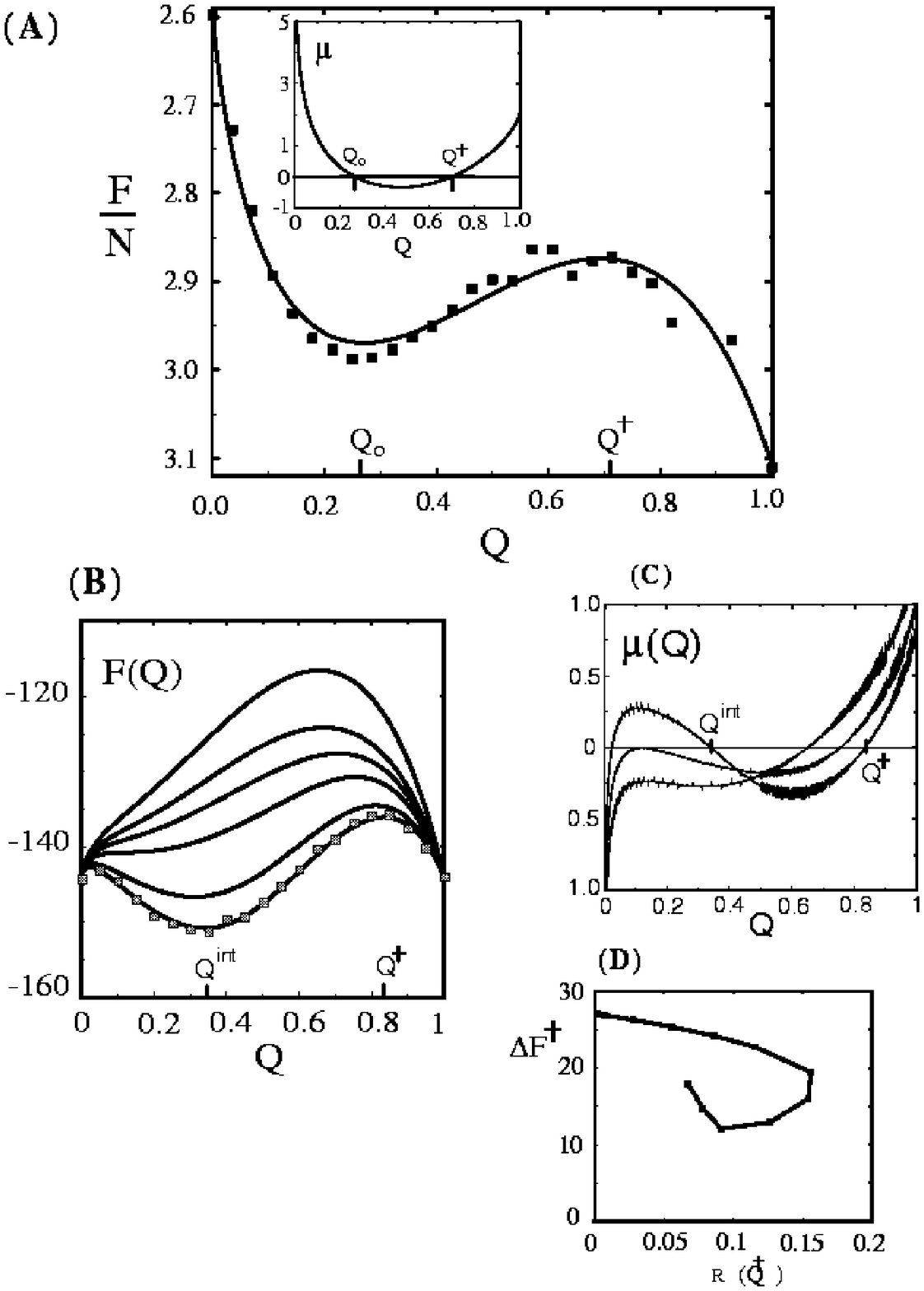,height=13cm,width=8cm,angle=0}
\caption{}
\label{fig:F}
\end{figure}

\begin{figure}[htb]
\hspace{.7cm}
\psfig{file=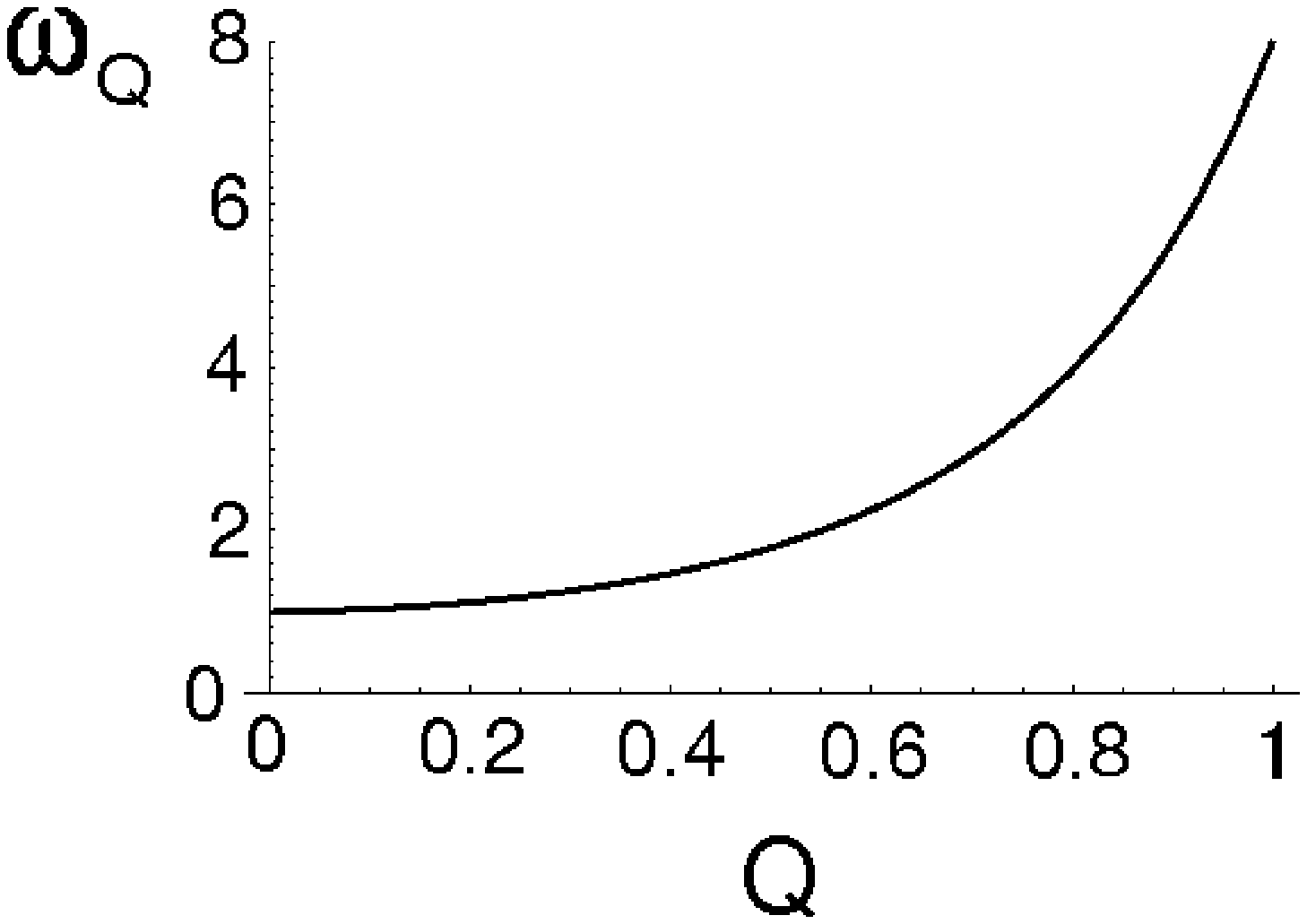,height=7cm,width=8cm,angle=0}
\caption{} 
\label{fig:wq}
\end{figure}

\newpage

\pagestyle{plain}

\begin{thebibliography}{100}

\bibitem{WolynesPG92}
P.~G. Wolynes,  in {\em Spin Glasses and Biology}, edited by D.~L. Stein (World
  Scientific, Singapore, 1992), pp.\ 225--259.

\bibitem{BryngelsonJD95}
J.~D. Bryngelson, J.~N. Onuchic, N.~D. Socci, and P.~G. Wolynes, Proteins {\bf
  21},  167  (1995).

\bibitem{DillKA95:ps}
K.~A. Dill {\it et~al.}, Protein Sci {\bf 4},  561  (1995).

\bibitem{BallKD96}
K.~D. Ball {\it et~al.}, Science {\bf 271},  963  (1996).

\bibitem{DillKA97}
K.~A. Dill and H.~S. Chan, Nat. Struct. Biol. {\bf 4},  10  (1997).

\bibitem{VeitshansT97}
T. Veitshans, D. Klimov, and D. Thirumalai, Folding and Design {\bf 2},  1
  (1997).

\bibitem{Onuchic97}
J.~N. Onuchic, Z. Luthey-Schulten, and P.~G. Wolynes, Annu Rev Phys Chem {\bf
  48},  545  (1997).

\bibitem{PandeVS97:bj}
V.~S. Pande, A.~Y. Grosberg, and T. Tanaka, Biophys J {\bf 73},  3192  (1997).

\bibitem{Dobson98}
C.~M. Dobson, A. Sali, and M. Karplus, Angew Chem Int Ed Engl {\bf 37},  868
  (1998).

\bibitem{GarelT:rev98}
T. Garel, H. Orland, and E. Pitard,  in {\em Spin Glasses and random fields},
  edited by A.~P. Young (World Scientific, River Edge, N.J., 1998).

\bibitem{BrooksCL98:pnas}
C.~L. Brooks, M. Gruebele, J.~N. Onuchic, and P.~G. Wolynes, Proc Nat Acad Sci
  USA {\bf 95},  11037  (1998).

\bibitem{FershtAR99:book}
A.~R. Fersht, {\em Structure and mechanism in protein science}, 1st  ed. (W. H.
  Freeman and Co., New York, 1999).

\bibitem{GruebeleM99}
M. Gruebele, Annu Rev Phys Chem {\bf 50},  485  (1999).

\bibitem{WalesDJ99}
D.~J. Wales and H.~A. Scheraga, Science {\bf 285},  1368  (1999).

\bibitem{OnuchicJN00:apc}
J.~N. Onuchic {\it et~al.}, Adv. Protein Chem. {\bf 53},  87  (2000).

\bibitem{FergusonN99}
N. Ferguson {\it et~al.}, J Mol Biol {\bf 286},  1597  (1999).

\bibitem{KimDE98}
D.~E. Kim, H. Gu, and D. Baker, Proc Nat Acad Sci USA {\bf 95},  4982  (1998).

\bibitem{DalessioPM00}
P.~M. Dalessio and I.~J. Ropson, Biochemistry {\bf 39},  860  (2000).

\bibitem{AbkevichVI94}
V.~I. Abkevich, A.~M. Gutin, and E.~I. Shakhnovich, Biochemistry {\bf 33},
  10026  (1994).

\bibitem{Onuchic96}
J.~N. Onuchic, N.~D. Socci, Z. Luthey-Schulten, and P.~G. Wolynes, Folding and
  Design {\bf 1},  441  (1996).

\bibitem{Klimov98}
D.~K. Klimov and D. Thirumalai, J Mol Biol {\bf 282},  471  (1998).

\bibitem{LiL00}
L. Li, L.~A. Mirny, and E.~I. Shakhnovich, Nature Struct Biol {\bf 7},  336
  (2000).

\bibitem{GoldsteinRA-AMH-92}
R.~A. Goldstein, Z.~A. Luthey-Schulten, and P.~G. Wolynes, Proc Nat Acad Sci
  USA {\bf 89},  4918  (1992).

\bibitem{BryngelsonJD87}
J.~D. Bryngelson and P.~G. Wolynes, Proc Nat Acad Sci USA {\bf 84},  7524
  (1987).

\bibitem{LeopoldPE92}
P.~E. Leopold, M. Montal, and J.~N. Onuchic, Proc Nat Acad Sci USA {\bf 89},
  8721  (1992).

\bibitem{ShakhnovichEI93a}
E.~I. Shakhnovich and A.~M. Gutin, Proc Nat Acad Sci USA {\bf 90},  7195
  (1993).

\bibitem{OnuchicJN95:pnas}
J.~N. Onuchic, P.~G. Wolynes, Z. Luthey-{S}chulten, and N.~D. Socci, Proc Nat
  Acad Sci USA {\bf 92},  3626  (1995).

\bibitem{Bornberg99:pnas}
E. Bornberg-Bauer and H.~S. Chan, Proc Nat Acad Sci USA {\bf 96},  10689
  (1999).

\bibitem{Buchler99:jcp}
N.~E.~G. Buchler and R.~A. Goldstein, J Chem Phys {\bf 111},  6599  (1999).

\bibitem{Plaxco98}
K.~W. Plaxco, K.~T. Simons, and D. Baker, J Mol Biol {\bf 277},  985  (1998).

\bibitem{MunozV99}
V. Munoz and W.~A. Eaton, Proc Nat Acad Sci USA {\bf 96},  11311  (1999).

\bibitem{FershtAR00}
A.~R. Fersht, Proc Nat Acad Sci USA {\bf 97},  1525  (2000).

\bibitem{AlmE99}
E. Alm and D. Baker, Proc Nat Acad Sci USA {\bf 96},  11305  (1999).

\bibitem{ShoemakerWang99}
B.~A. Shoemaker, J. Wang, and P.~G. Wolynes, J Mol Biol {\bf 287},  675
  (1999).

\bibitem{FinkelsteinAV99:pnas}
O.~V. Galzitskaya and A.~V. Finkelstein, Proc Nat Acad Sci USA {\bf 96},  11299
   (1999).

\bibitem{SheaJE99}
J.~E. Shea, J.~N. Onuchic, and C.~L. Brooks, Proc Nat Acad Sci USA {\bf 96},
  12512  (1999).

\bibitem{RiddleDS99}
D.~S. Riddle {\it et~al.}, Nat. Struct. Biol. {\bf 11},  1016  (1999).

\bibitem{DuR99:jcp}
R. Du {\it et~al.}, J Chem Phys {\bf 111},  10375  (1999).

\bibitem{ClementiC00:jmb}
C. Clementi, H. Nymeyer, and J.~N. Onuchic, J Mol Biol {\bf 298},  937  (2000).

\bibitem{ClementiC00:pnas}
C. Clementi, P.~A. Jennings, and J.~N. Onuchic, Proc Nat Acad Sci USA {\bf 97},
   5871  (2000).

\bibitem{TavernaDM00}
D.~M. Taverna and R.~A. Goldstein, Biopolymers {\bf 53},  1  (2000).

\bibitem{MaritanA00}
A. Maritan, C. Micheletti, A. Trovato, and J.~R. Banavar, Nature {\bf 406},
  287  (2000).

\bibitem{PlotkinSS00:pnas}
S.~S. Plotkin and J.~N. Onuchic, Proc Nat Acad Sci USA {\bf 97},  6509  (2000).

\bibitem{ShoemakerBA97}
B.~A. Shoemaker, J. Wang, and P.~G. Wolynes, Proc. Nat. Acad. Sci. USA {\bf
  94},  777  (1997).

\bibitem{ShoemakerBA99}
B.~A. Shoemaker and P.~G. Wolynes, J Mol Biol {\bf 287},  657  (1999).

\bibitem{GutinAM95:pnas}
A.~M. Gutin, V.~I. Abkevich, and E.~I. Shakhnovich, Proc Nat Acad Sci USA {\bf
  92},  1282  (1995).

\bibitem{Radford92}
S.~A. Radford, C.~M. Dobson, and P.~A. Evans, Nature {\bf 358},  302  (1992).

\bibitem{BaiY95}
Y. Bai, T.~R. Sosnick, L. Mayne, and S.~W. Englander, Science {\bf 269},  192
  (1995).

\bibitem{BoczkoEM95}
E.~M. Boczko and C.~L. Brooks, Science {\bf 269},  393  (1995).

\bibitem{Lazaridis97}
T. Lazaridis and M. Karplus, Science {\bf 278},  1928  (1997).

\bibitem{Brookspnas98}
F.~B. Sheinerman and C.~L. Brooks, Proc Nat Acad Sci USA {\bf 95},  1562
  (1998).

\bibitem{FershtAR92}
A.~R. Fersht, A. Matouschek, and L. Serrano, J Mol Biol {\bf 224},  771
  (1992).

\bibitem{HorovitzA92}
A. Horovitz and A. Fersht, J Mol Biol {\bf 224},  733  (1992).

\bibitem{HaoM97}
M.-H. Hao and H. Scheraga, Physica A {\bf 244},  124  (1997).

\bibitem{SorensonJM98}
J.~M. Sorenson and T. Head-Gordon, Folding and Design {\bf 3},  523  (1998).

\bibitem{Klimov98:fd}
D.~K. Klimov and D. Thirumalai, Folding and Design {\bf 3},  127  (1998).

\bibitem{LumK99}
K. Lum, D. Chandler, and J.~D. Weeks, J Phys Chem {\bf 103},  4570  (1999).

\bibitem{TakadaS99:jcp}
S. Takada, Z. Luthey-Schulten, and P.~G. Wolynes, J Chem Phys {\bf 110},  11616
   (1999).

\bibitem{PrinceRB99}
R.~B. Prince, J.~G. Saven, P.~G. Wolynes, and J.~S. Moore, J Am Chem Soc {\bf
  121},  3114  (1999).

\bibitem{KolinskiA96:prot}
A. Kolinski, W. Galazka, and J. Skolnick, Proteins: Struct. Funct. and Genetics
  {\bf 26},  271  (1996).

\bibitem{PlotkinSS97}
S.~S. Plotkin, J. Wang, and P.~G. Wolynes, J Chem Phys {\bf 106},  2932
  (1997).

\bibitem{EastwoodMP00}
M.~P. Eastwood and P.~G. Wolynes, preprint (unpublished).

\bibitem{note:backtop}
The native backbone topology is more precisely specified by a vector
  parameterized by the arc-length along the polymer chain, ${\bf r}(s)$, but it
  is difficult to apply the formalism starting from this description.

\bibitem{Douglas95}
J.~F. Douglas and T. Ishinabe, Phys Rev E {\bf 51},  1791  (1995).

\bibitem{BryngelsonJD89}
J.~D. Bryngelson and P.~G. Wolynes, J Phys Chem {\bf 93},  6902  (1989).

\bibitem{PortmanJprl98}
J.~J. Portman, S. Takada, and P.~G. Wolynes, Phys Rev Lett {\bf 81},  5237
  (1998).

\bibitem{DuR98:jcp}
R. Du {\it et~al.}, J Chem Phys {\bf 108},  334  (1998).

\bibitem{PercusJK82}
J.~K. Percus,  in {\em The liquid state of matter: Fluids, simple and complex},
  edited by E. Montroll and J. Lebowitz (North-Holland, Amsterdam, 1982).

\bibitem{EvansR92}
R. Evans,  in {\em Fundamentals of inhomogeneous fluids}, edited by D.
  Henderson (Dekker, New York, 1992).

\bibitem{Gunton83}
J.~D. Gunton, M.~S. Miguel, and P.~S. Sahni,  in {\em Phase Transitions and
  Critical Phenomena}, edited by C. Domb and J.~L. Lebowitz (Academic Press,
  New York, 1983), Vol.~8, pp.\ 267--466.

\bibitem{BohrHG92}
H.~G. Bohr and P.~G. Wolynes, Phys Rev A {\bf 46},  5242  (1992).

\bibitem{note:kb}
We will generally set Boltzmann's constant $\kB = 1$ in this paper, so
  temperatures have units of energy, and entropies are in units of $\kB$.

\bibitem{note:avg}
Later in the paper we will assume all quantities are thermally equilibrated,
  and the averages will indicate sums over native contacts, e.g. $Q=\left<
  \Qi\right>$. The meaning should be clear from the context.

\bibitem{note:mag}
In this section only $M$ is the total magnetization.

\bibitem{Nymeyer:priv}
We thank H. Nymeyer for helpful discussions on this argument.

\bibitem{DerridaB81}
B. Derrida, Phys Rev B {\bf 24},  2613  (1981).

\bibitem{Karpov94}
V.~G. Karpov, Phys Rev B {\bf 50},  9124  (1994).

\bibitem{Oxtoby96}
V.~G. Karpov and D.~W. Oxtoby, Phys Rev B {\bf 54},  9734  (1996).

\bibitem{LifshitzIM64}
I.~M. Lifshitz, Adv. Phys {\bf 13},  483  (1964).

\bibitem{HalperinBI66}
B.~I. Halperin and M. Lax, Phys. Rev. {\bf 148},  722  (1966).

\bibitem{LangerJS66}
J. Zittartz and J.~S. Langer, Phys. Rev. {\bf 148},  741  (1966).

\bibitem{Landau80}
L.~D. Landau and E.~M. Lifshitz, {\em Statistical Physics}, 3 ed. (Pergamon
  Press, Oxford, 1980).

\bibitem{note:land}
Averaging over native disorder in calculating $\overline{\Qi}$ leaves the
  residual dependence on loop entropy for each contact probability, which is
  then summed using $\sum_i \overline{\Qi} = M Q$.

\bibitem{Wolynes97cap}
P.~G. Wolynes, Proc Nat Acad Sci USA {\bf 94},  6170  (1997).

\bibitem{KolinskiA93:jcp}
A. Kolinski, A. Godzik, and J. Skolnick, J Chem Phys {\bf 98},  7420  (1993).

\bibitem{VendruscoloM99}
M. Vendruscolo, R. Najmanovich, and E. Domany, Phys Rev Lett {\bf 82},  656
  (1999).

\bibitem{Becker35}
R. Becker and W. Doring, Ann. Phys. (Leipzig) {\bf 24},  719  (1935).

\bibitem{Lothe62}
J. Lothe and G.~M. Pound, J Chem Phys {\bf 36},  2080  (1962).

\bibitem{Reiss68}
H. Reiss, J.~L. Katz, and E.~R. Cohen, J Chem Phys {\bf 48},  5553  (1967).

\bibitem{Lothe69}
J. Lothe and G.~M. Pound,  in {\em Nucleation}, edited by A.~C. Zettlemoyer
  (Dekker, New York, 1969), p.\ 109.

\bibitem{Langer67}
J.~S. Langer, Ann. Phys. (N.Y.) {\bf 41},  108  (1967).

\bibitem{Fisher67}
M.~E. Fisher, Physics {\bf 3},  255  (1967).

\bibitem{UngerC84}
C. Unger and W. Klein, Phys Rev B {\bf 29},  2698  (1984).

\bibitem{Cahn58}
J.~W. Cahn and J.~E. Hilliard, J Chem Phys {\bf 28},  258  (1958).

\bibitem{Langer69}
J.~S. Langer, Ann. Phys. (N.Y.) {\bf 54},  258  (1969).

\bibitem{PlotkinSS96}
S.~S. Plotkin, J. Wang, and P.~G. Wolynes, Phys Rev E {\bf 53},  6271  (1996).

\bibitem{PandeVS97:fd}
V.~S. Pande, A.~Y. Grosberg, and T. Tanaka, Folding and Design {\bf 2},  109
  (1997).

\bibitem{FernandezA00:jcp}
A. Fern\'{a}ndez, K.~S. Kostov, and R.~S. Berry, J Chem Phys {\bf 112},  5223
  (2000).

\bibitem{note:resum}
This term contains in principle a resummation of all the moments of a virial
  expansion of the entropy $ {\cal S}_{\mbox{\tiny V}} ( \{ \Qi^f \} | \{ \Qi^o
  \} ) = \sum_i s_i^o \times ( \Qi^f - \Qi^o ) + \sum_{i,j} s_{i,j}^o \times
  (\Qi^f - \Qi^o ) \times (\Qj^f - \Qj^o ) + \cdots $ which is a slowly
  convergent expansion consisting of 2-body, 4-body and higher order terms.

\bibitem{FloryPJ56:jacs}
P.~J. Flory, J Am Chem Soc {\bf 78},  5222  (1956).

\bibitem{GutinA94}
A. Gutin and E. Shakhnovich, J Chem Phys {\bf 100},  5290  (1994).

\bibitem{FinkelsteinAV97:molbiol}
A.~V. Finkelstein and A.~Y. Badretdinov, Molecular Biology {\bf 31},  391
  (1997).

\bibitem{ChanHS90}
H.~S. Chan and K.~A. Dill, J Chem Phys {\bf 92},  3118  (1990).

\bibitem{DillKA93:pnas}
K.~A. Dill, K.~M. Fiebig, and H.~S. Chan, Proc Nat Acad Sci USA {\bf 90},  1942
   (1993).

\bibitem{Hao94}
M.-H. Hao and H.~A. Scheraga, J.\ Phys.\ Chem. {\bf 98},  4940  (1994).

\bibitem{note:lag}
We could equivalently have explicitly invoked this by introducing the Lagrange
  constraint $\sum_i \Qi = M Q$ into the problem from the outset.

\bibitem{Itzhaki95}
L.~S. Itzhaki, D.~E. Otzen, and A.~R. Fersht, J Mol Biol {\bf 254},  260
  (1995).

\bibitem{NymeyerH00:pnas}
H. Nymeyer, N.~D. Socci, and J.~N. Onuchic, Proc Nat Acad Sci USA {\bf 97},
  634  (2000).

\bibitem{note:note}
Note that by the structure of eq.~(\ref{leff}), eq~(\ref{eq:el}) is satisfied
  for all $Q$, i.e. when the energies are tuned to extremize $\Delta
  F^{\ddag}$, the $\Qi$ are all equal essentially for all $Q$. Thus in the
  model there is full symmetry in the ordering of the protein. However, when
  the entropy is exactly counted, and also when the entropy cutoff mentioned
  after eq.~(\ref{eq:smfflory}) is present, $\Qi(Q) = Q$ only at the barrier
  peak. See fig.~\ref{fig:QivsQroute}C and~\ref{fig:QivsQroute}D.

\bibitem{SaliA94:nat}
A. {\u{S}}ali, E. Shakhnovich, and M. Karplus, Nature {\bf 369},  248  (1994).

\bibitem{SocciND95:jcp}
N.~D. Socci and J.~N. Onuchic, J Chem Phys {\bf 103},  4732  (1995).

\bibitem{AbkevichVI95}
V.~I. Abkevich, A.~M. Gutin, and E.~I. Shakhnovich, J Mol Biol {\bf 252},  460
  (1995).

\bibitem{SocciND96:jcp}
N.~D. Socci, J.~N. Onuchic, and P.~G. Wolynes, J Chem Phys {\bf 104},  5860
  (1996).

\bibitem{PlotkinSS00:un}
S.~S. Plotkin, submitted to Proteins: Struct. Funct. and Genetics
  (unpublished).

\bibitem{WallFT51}
F.~T. Wall and P.~J. Flory, J Chem Phys {\bf 19},  1435  (1951).

\bibitem{DeamRT76}
R.~T. Deam and S.~F. Edwards, Phil. Trans. R. Soc. A {\bf 280},  317  (1976).

\bibitem{WangPlot97}
J. Wang, S.~S. Plotkin, and P.~G. Wolynes, J. Phys. I France {\bf 7},  395
  (1997).

\bibitem{TakadaS97:pnas}
S. Takada, J.~J. Portman, and P.~G. Wolynes, Proc Nat Acad Sci USA {\bf 94},
  23188  (1997).

\bibitem{VigueraAR96}
A.~R. Viguera, V. Villegas, F.~X. Aviles, and L. Serrano, Folding and Design
  {\bf 2},  23  (1996).

\bibitem{KimDE00}
D.~E. Kim, C. Fisher, and D. Baker, J Mol Biol {\bf 298},  971  (2000).

\bibitem{Munoz96:Rev}
V. Munoz and L. Serrano, Folding and Design {\bf 1},  R71  (1996).

\bibitem{HagenSJ96:pnas}
S.~J. Hagen, J.~A. Hofrichter, A. Szabo, and W.~A. Eaton, Proc Nat Acad Sci USA
  {\bf 93},  11615  (1996).

\bibitem{BrownBM99}
B.~M. Brown and R.~T. Sauer, Proc Nat Acad Sci USA {\bf 96},  1983  (1999).

\bibitem{Wolynes97nsb}
P.~G. Wolynes, Nature Struct Biol {\bf 4},  871  (1997).

\bibitem{FiebigKM92}
K.~M. Fiebig and K.~A. Dill, J Chem Phys {\bf 98},  3475  (1992).

\bibitem{SerranoL98nsb}
J.~C. Martinez, M.~T. Pisabarro, and L. Serrano, Nature Struct Biol {\bf 5},
  721  (1998).

\bibitem{BakerD98nsb}
V.~P. Grantcharova, J.~V. Santiago, D. Baker, and D.~S. Riddle, Nature Struct
  Biol {\bf 5},  714  (1998).

\bibitem{Nymeyer:thank}
We are grateful to H. Nymeyer for providing the lattice data for this plot.

\end{thebibliography}

\end{document}